\documentclass{aastex61}

\shorttitle{Optimal stellar photometry for MCAO}
\shortauthors{Turri et al.}

\begin{document}
\title{Optimal stellar photometry for multi-conjugate adaptive optics systems\\ using science-based metrics}

\author[0000-0002-6451-6239]{P. Turri}
\affiliation{Department of Physics and Astronomy, University of Victoria, 3800 Finnerty Road, Victoria, BC V8P 5C2, Canada}
\affiliation{NRC Herzberg -- Astronomy and Astrophysics, 5071 West Saanich Road, Victoria, BC V9E 2E7, Canada}

\author{A. W. McConnachie, P. B. Stetson}
\affiliation{NRC Herzberg -- Astronomy and Astrophysics, 5071 West Saanich Road, Victoria, BC V9E 2E7, Canada}

\author{G. Fiorentino}
\affiliation{INAF -- Osservatorio Astronomico di Bologna, Via Ranzani 1, I-40127 Bologna, Italy}

\author{D. R. Andersen}
\affiliation{NRC Herzberg -- Astronomy and Astrophysics, 5071 West Saanich Road, Victoria, BC V9E 2E7, Canada}

\author{G. Bono}
\affiliation{Dipartimento di Fisica -- Universit\`{a} di Roma Tor Vergata, Via della Ricerca Scientifica 1, I-00133 Roma, Italy}
\affiliation{INAF -- Osservatorio Astronomico di Roma, Via Frascati 33, I-00040 Monte Porzio Catone (RM), Italy}

\author{D. Massari}
\affiliation{INAF -- Osservatorio Astronomico di Bologna, Via Ranzani 1, I-40127 Bologna, Italy}

\author{J.-P. V{\'e}ran}
\affiliation{NRC Herzberg -- Astronomy and Astrophysics, 5071 West Saanich Road, Victoria, BC V9E 2E7, Canada}

\correspondingauthor{P. Turri}
\email{turri@uvic.ca}

\begin{abstract}

We present a detailed discussion of how to obtain precise stellar photometry in crowded fields using images from multi-conjugate adaptive optics (MCAO) systems, with the intent of informing the scientific development of this key technology for the Extremely Large Telescopes. We use deep J and K$_\mathrm{s}$ exposures of NGC 1851 taken with the Gemini Multi-Conjugate Adaptive Optics System (GeMS) on Gemini South to quantify the performance of the instrument and to develop an optimal strategy for stellar photometry using PSF-fitting techniques. We judge the success of the various methods we employ by using science-based metrics, particularly the width of the main sequence turn-off region. We also compare the GeMS photometry with the exquisite HST data in the visible of the same target. We show that the PSF produced by GeMS possesses significant spatial and temporal variability that must be accounted for during the analysis. We show that the majority of the variation of the PSF occurs within the ``control radius'' of the MCAO system and that the best photometry is obtained when the PSF radius is chosen to closely match this spatial scale. We identify photometric calibration as a critical issue for next generation MCAO systems such as those on TMT and E-ELT. Our final CMDs reach K$_\mathrm{s}\sim$22---below the main sequence knee---making it one of the deepest for a globular cluster available from the ground. Theoretical isochrones are in remarkable agreement with the stellar locus in our data from below the main sequence knee to the upper red giant branch.
\end{abstract}


\section{Introduction}

The extremely large telescopes (ELT)---the European Extremely Large Telescope (E-ELT), the Thirty Meter Telescope (TMT), and the Giant Magellan Telescope (GMT)---will reach unprecedented depths and spatial resolutions on account of their enormous apertures. To fully exploit the potential ``$D^{4}$'' benefit of these facilities, taking advantage of both depth and resolution, requires adaptive optics (AO; see \cite{bib:davies12} for a comprehensive review; see \cite{bib:boyer16} for discussion relating to TMT and \cite{bib:diolaiti16} for discussion relating to E-ELT). This technology improves the Strehl ratio of the point-spread function (PSF), a measurement of the peak of the PSF relative to the theoretical diffraction-limited value. It does this by correcting the wavefront distortions in the science light caused by the atmosphere. Adaptive optics has a better performance at longer wavelengths and so they generally operate at near-infrared (NIR) wavelengths.

In classic adaptive optics, optical aberrations are first measured in the direction of a point-like source above the turbulence by using a wavefront sensor. Natural guide stars (NGS) are the best sources to use since they are distant and point-like. However, there can be not enough suitable (i.e., bright) NGSs in a given field of view, leading to the use of artificial laser guide star (LGS). These are created by a laser beam launched from the proximity of the telescope in the direction of the science target. A return of photons toward the telescope aperture is produced by either Rayleigh scattering from molecules in the atmosphere or by excitement of sodium atoms at high altitudes. Since the LGS wavefront sensors are blind to the tip-tilt aberration \citep{bib:rigaut92}, NGSs are still necessary but they can be fainter because they are required only to determine tip and tilt. A control system acquires the data from the wavefront sensors, measures the wavefront distortions and calculates the correction that the AO system has to apply. A deformable mirror then injects an equal and opposite distortion to correct the image.

The main shortcoming of classical AO is the limited size of the corrected field, typically of the order of a several arcseconds. A single guide star can probe only the cylinder of atmosphere in its direction; the more distant a target is from it, the less the guide star probes the relevant aberrations and so the less effective the correction. This effect, known as anisoplanatism, is improved by multi-conjugate adaptive optics (MCAO). Here, multiple guide stars are employed to probe the atmosphere in different directions and reconstruct the atmospheric distortion in a larger volume. Multiple deformable mirrors, optically conjugated to different altitudes, apply the correction to the layers of turbulence associated to those heights. While an MCAO system can work using only NGSs, such as the MCAO demonstrator MAD on VLT \citep{bib:marchetti03,bib:ferraro09,bib:moretti09,bib:bono10,bib:fiorentino11}, the use of LGSs is advantageous because it alleviates the requirements for an asterism of bright stars close to the scientific target, thus increasing significantly the sky coverage.

The Gemini Multi-Conjugate Adaptive Optics System (GeMS) on Gemini South is the first facility class MCAO instrument, and had first light in 2012. High resolution imaging of extended crowded stellar fields has generally been a prerogative of space telescopes, but GeMS now routinely enables this capability from the ground. We have observed the Galactic globular clusters NGC 1851, NGC 2808, NGC 5904, NGC 6652, NGC 6681, and NGC 6723 in the J and K$_\mathrm{s}$ bands. These clusters were selected as southern hemisphere targets which had preexisting, high quality, space-based imaging from the HST ACS Survey of Galactic Globular Clusters of \cite{bib:sarajedini07}. First results for NGC 1851 are presented in \cite{bib:turri15} and results for NGC 2808 and NGC 6681 are published in \cite{bib:massari16a,bib:massari16b}.

In addition to the study of their stellar populations and dynamics, a major goal of our observing program is to better understand how best to exploit MCAO data for precision photometry and astrometry, in order to prepare for science observations with the ELTs. The large number of relatively bright point sources distributed over the field of view allows our globular clusters to be used as a training ground for measurement techniques on MCAO data. The success of MCAO as a technology relies on its ability to produce science-ready astronomical measurements---particularly photometry and astrometry---and so it is important to use science-based metrics to judge the success or otherwise of any techniques. 

In this paper, we examine the critical question of how to extract the most precise photometry from MCAO data, using our observations of NGC 1851. This cluster is known for the presence of multiple stellar populations from both photometry and spectroscopy \citep{bib:calamida07,bib:milone08,bib:lee09,bib:han09,bib:cummings14,bib:simpson17}. In particular, \cite{bib:milone08} has observed a split subgiant branch with two populations. Examination of the quality of the color magnitude diagrams will allow us to determine the utility of the strategies that we employ to obtain the photometry. Since NGC 1851 is affected by a very low color excess \citep{bib:harris96}, we can ignore the effect of differential reddening on the width of the sequences. While our conclusions are specifically valid for GeMS, they are also expected to be of relevance for any MCAO dataset, and it is our expectation that such studies will prove useful for the development of the MCAO systems and observing strategies for the TMT and E-ELT. A subsequent contribution will present similar details for astrometric measurements, noting that the capacity of GeMS to produce precise astrometry has already been shown in \cite{bib:massari16b}.

The format of this paper is as follows: Section~\ref{sec:observations} discusses the details of our observations of NGC 1851; Section~\ref{sec:images} presents a summary of the image processing; Section~\ref{sec:performance} overviews the photometric performance of GeMS/GSAOI, especially regarding the uniformity of the shape of the PSF; Section~\ref{sec:photometry} presents an overview of our photometric analysis procedure and a detailed discussion of the steps taken to optimize the process to obtain the most precise photometry; Section~\ref{sec:cmds} presents our final color-magnitude diagrams and a brief discussion of their most prominent features; Section~\ref{sec:conclusion} summarises our main conclusions.

\section{Observations}\label{sec:observations}

GeMS, the Gemini Multi-Conjugate Adaptive Optics System \citep{bib:neichel14,bib:rigaut14}, is the first facility-class MCAO system. It started science operations at the end of 2012 on the 8-meter Gemini South Telescope on Cerro Pach{\'o}n (Chile). It measures the aberrations caused by the volume of atmosphere in the direction of the observed field using five laser beams, tuned to the sodium D2 line at 589 nm, that are launched from the back of the secondary mirror. By exciting the layer of atmospheric neutral sodium at 90--100 km \citep{bib:neichel13}, the system generates five LGSs arranged in a 60\arcsec\, square asterism with the fifth star in the center. A wavefront sensor (WFS) for each LGS measures the wavefront distortions along these five directions. Since the LGS WFSs are blind to tip-tilt aberrations \citep{bib:rigaut92}, 1--3 bright natural guide stars (NGSs) are required in a field of 2\arcmin\, to determine this low-order mode, each with their own WFS. All wavefront measurements are combined to calculate the commands given to the two deformable mirrors that provide the correction. The deformable mirrors are optically conjugated to 0 and 9 km to correct for the turbulence at the ground and at high altitudes, respectively. The instrument was originally designed with a third deformable mirror conjugated at 4.5 km but, following the failure of multiple actuators one of them, it was commissioned with only two \citep{bib:rigaut14}. When the deformable mirror replacement will be implemented in the system, the quality of the correction is expected to increase in consequence of the lower tomographic error \citep{bib:neichel14}.

The corrected wavefront is imaged by GSAOI, the Gemini South Adaptive Optics Imager \citep{bib:mcgregor04,bib:carrasco12}. This camera has four Teledyne HAWAII-2RG NIR detectors \citep{bib:blank12} arranged in a 2 x 2 array, each chip with 2040 x 2040 active pixels. The pixel scale is 0.0196\arcsec px$^\mathrm{-1}$ and the total field of view is 82.5 x 82.5\arcsec\, (with a gap of 2.5\arcsec\, between the chips). The detectors are read out using Fowler sampling \citep{bib:fowler90}, which consists of a series of nondestructive reads of the pixels at the beginning and at the end of each exposure. The two sets are averaged and their difference constitutes the signal accumulated. The larger the number of samples taken, the lower the read noise.

\begin{figure}
\centering
\includegraphics[width=0.6\textwidth]{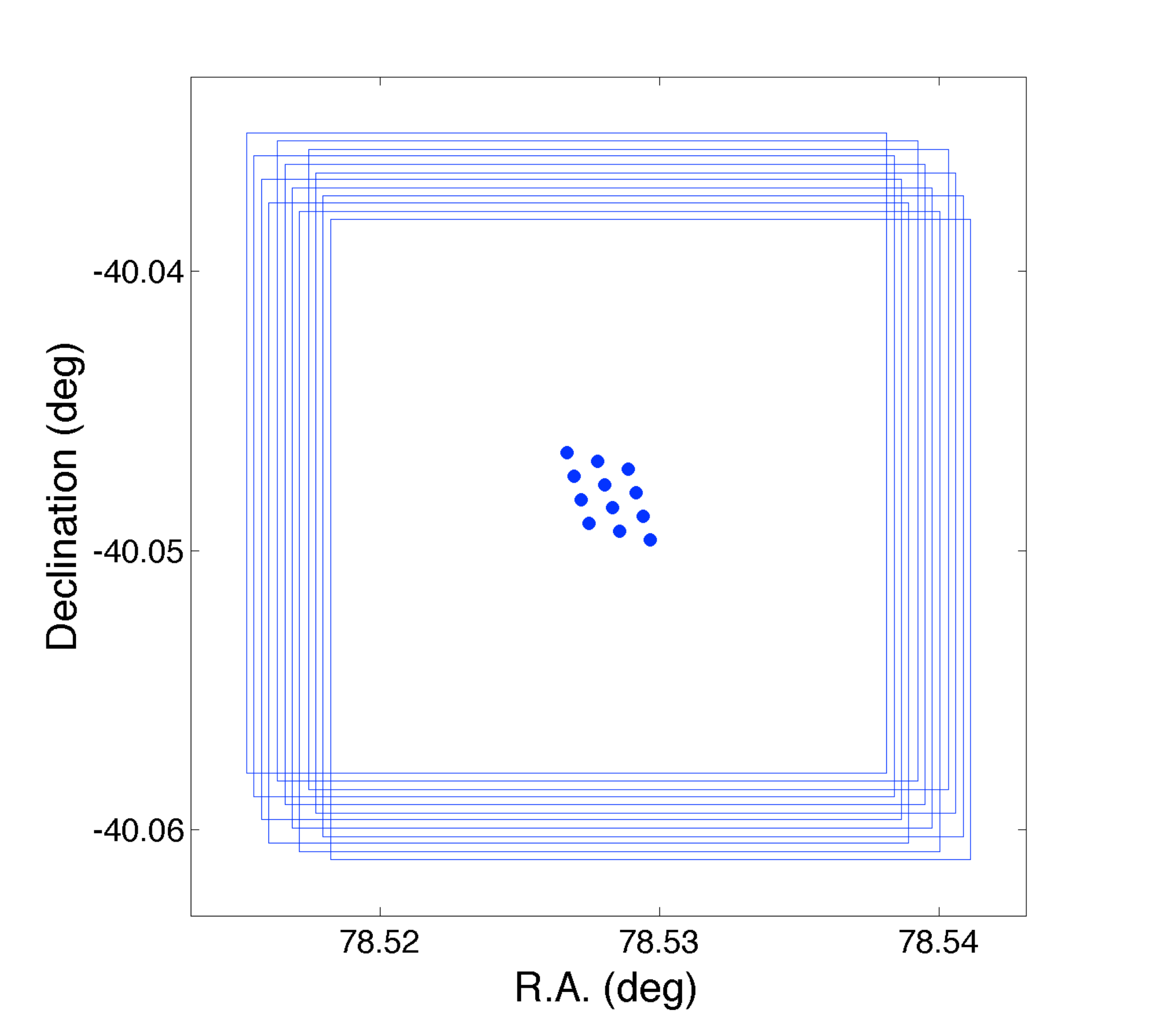}
\caption{The dithering pattern for the 12 long exposures in the K$_\mathrm{s}$ band. The same pattern was used for the 17 J band long exposures, with 5 of the pointings repeated.}\label{fig:dithering}
\end{figure}

\begin{figure}
\centering
\includegraphics[width=0.4\textwidth]{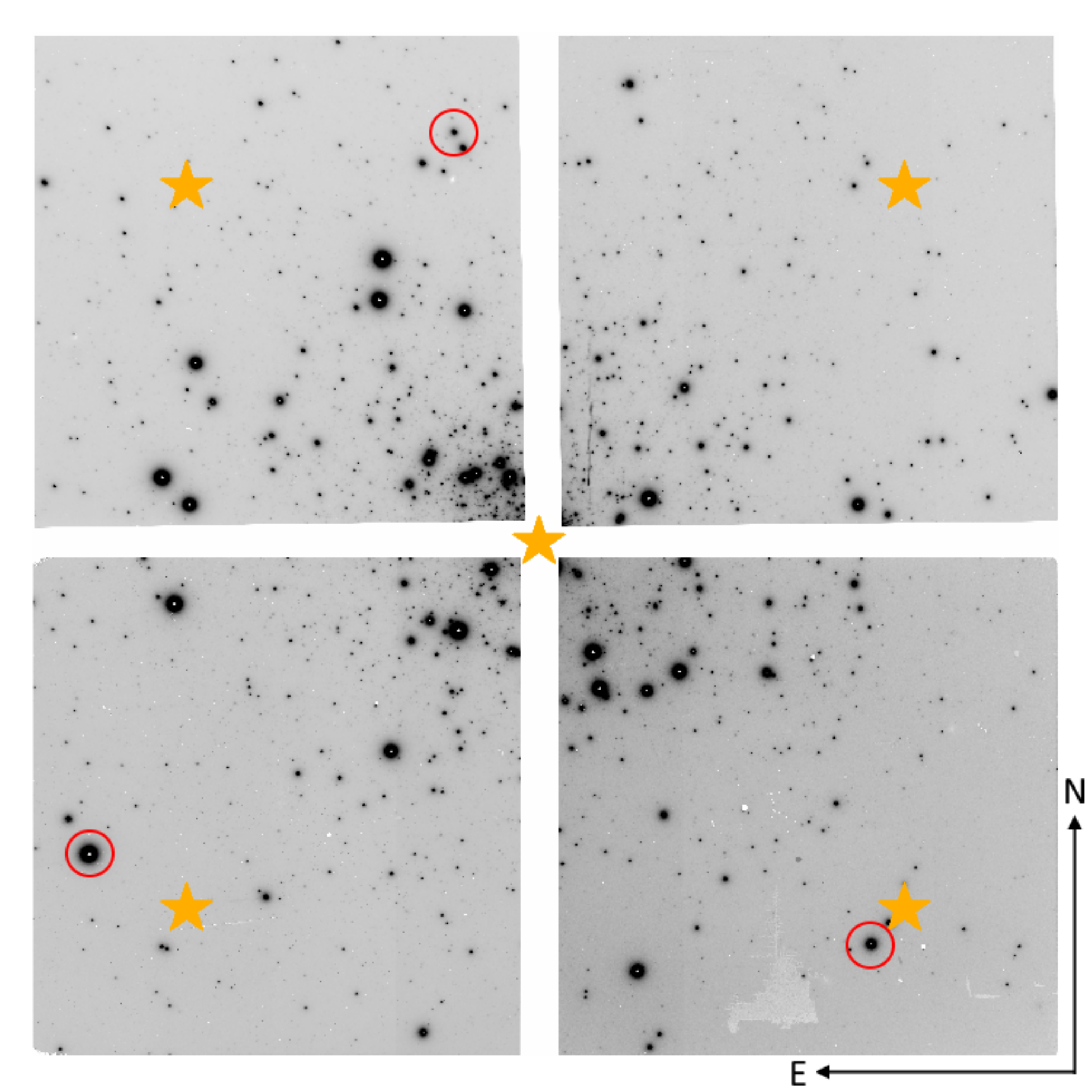}
\caption{Example of a processed K$_\mathrm{s}$ exposure of 160 seconds of NGC 1851 with GeMS/GSAOI. The orange star symbols show the position of the five LGSs. The three NGSs used are circled in red.\label{fig:mosaic}}
\end{figure}

During the GeMS/GSAOI Science Verification period, we observed the Galactic globular cluster NGC 1851 (05$^\mathrm{h}$14$^\mathrm{m}$06$^\mathrm{s}$.76, -40\arcdeg 02\arcmin 47\arcsec .6) with the NIR broad band filters J ($\lambda_{c}=1.24$ \micron) and K$_\mathrm{s}$ ($\lambda_{c}=2.16$ \micron) on the nights of 2012 December 31, 2013 January 27, and 2013 January 30 (Gemini program GS-2012B-SV-406, PI: McConnachie). Sub-exposures of 160 seconds were used and dithered in a pattern of 3\arcsec\, steps to cover the gap between the four chips (Figure~\ref{fig:dithering}). Seventeen exposures were taken in the J band and twelve in the K$_\mathrm{s}$. Two shorter exposures were also taken in each band to obtain non-saturated photometry of the brightest stars. Table~\ref{tab:exposures} provides a summary of our exposures. We used the ``faint object'' read mode with four Fowler samples per set (read noise of 13.6 $\mathrm{e^-}$). Three natural guide stars in the field of view have been used for the tip-tilt correction, with magnitudes of 13.3, 13.7, and 14.4 in the R band, all at least one magnitude brighter than the faint limit at 15.5. Unfortunately, reliable measurements of the seeing conditions at the telescope site during individual observations were not recorded. However, the seeing median values were estimated from DIMM measurements 1.1\arcsec\, and 0.8\arcsec, for, respectively, the J and K$_\mathrm{s}$ exposures, classified at or above the 85\% -ile image quality for Cerro Pach{\'o}n. One of the J band images (exposure \#10) has not been used in the data analysis because of its particularly poor MCAO correction. A single K$_\mathrm{s}$ band exposure is shown in Figure~\ref{fig:mosaic} with the positions of the natural and laser guide stars highlighted.

\begin{table}
\centering
\caption{Observing log of NGC 1851 with GeMS}\label{tab:exposures}
\begin{tabular}{ccccc}
\hline
\hline
Band & \multicolumn{1}{p{2cm}}{\centering Exposure\\ time (s)} & Exposures & Date & Notes\\
\hline
J & 41.1 & 1 & 2013 Jan 27 & \\
J & 90 & 1 & 2013 Jan 27 & \\
J & 160 & 10 & 2013 Jan 27 & Exposure \#10 discarded\\
J & 160 & 7 & 2013 Jan 30 & \\
K$_\mathrm{s}$ & 21.5 & 1 & 2012 Dec 31 & \\
K$_\mathrm{s}$ & 90 & 1 & 2012 Dec 31 & \\
K$_\mathrm{s}$ & 160 & 12 & 2012 Dec 31 & \\
\hline
\end{tabular}
\end{table}

\section{Image processing}\label{sec:images}

Prior to performing the photometric analysis, we process the raw images of every chip of every exposure independently, using the Gemini/GSAOI IRAF package correcting for bad pixels, linearity and flat fielding.

\begin{figure}
\centering
\includegraphics[width=0.4\textwidth]{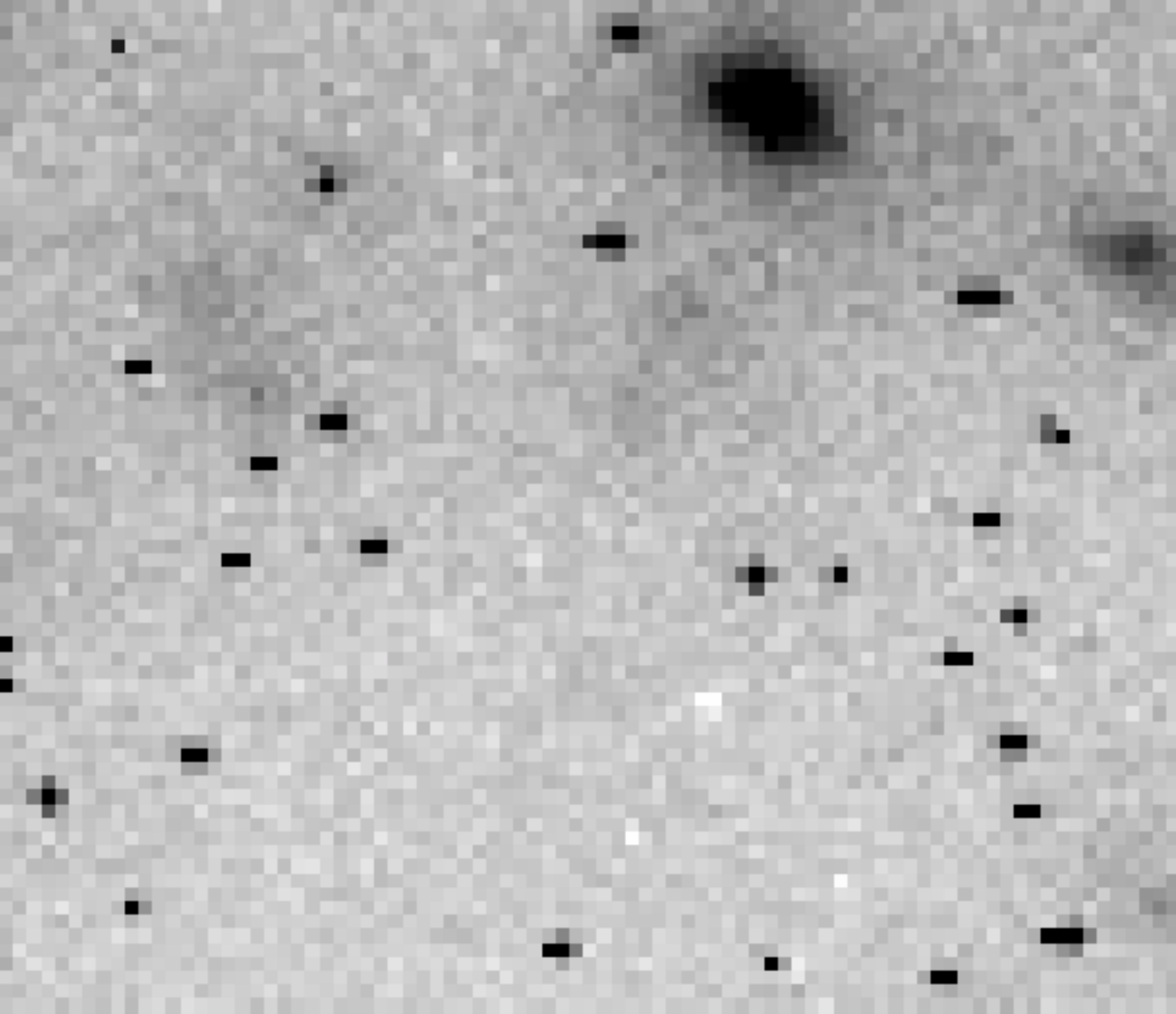}
\caption{Example of a raw J band image with a large concentration of hot (black) pixels. The cross-like pattern visible around them is due to the effect of inter-pixel capacitance discussed in the text. Dead (white) pixels are also visible.\label{fig:capac}}
\end{figure}

\begin{itemize}
\item \textit{Bad pixels ---} The GSAOI detector has a large number of both dead and bright pixels. The defective elements have been flagged and their values excluded from the photometric analysis using the bad pixel map provided by the Gemini IRAF package. Most of the hot pixels also cause an uncorrected leak of signal towards adjacent pixels that have lower counts, producing a ``cross-like'' pattern as shown in Figure~\ref{fig:capac}. These are evident especially in the J band exposures due to the lower sky emission. This effect is caused by inter-pixel capacitance \citep{bib:finger06}, an additional voltage on a pixel during its reading that is generated by the large difference in collected charges with respect to an adjacent one.

\item \textit{Linearity ---} HAWAII-2RG chips deviate appreciably at high signal levels from a linear response because of the change in the reverse bias voltage caused by the accumulated charges \citep{bib:figer04}. To compensate for this behavior, the values of every pixel are corrected with a quadratic polynomial whose coefficients were measured during the GSAOI commissioning \citep{bib:carrasco12}.

\item \textit{Flat field ---} To create our flat fields in each band, we first produced a master dome flat and a master twilight flat by median-combing individual exposures using average sigma clipping. The dome flats have the benefit of a high signal to noise, although their spectral signature and uniformity is inferior to the sky flats. The latter are therefore preferred but also have a lower signal to noise than the dome flats. Thus, we combine the dome and twilight master flats together to create a ``super'' flat in the following way. We first divide the master twilight by the master dome and we apply on the result a top-hat median circular filter to keep only the large-scale spatial variation. This creates a low-pass filtered image that is then multiplied by the dome master flat that has information on the pixel-to-pixel difference in response. The resulting ``super'' flat has the high signal-to-noise of the dome exposures and the uniform illumination and spectral signature of the twilight exposures, making it our choice for the flat fielding of the science exposures.
\end{itemize}

\begin{figure}
\plottwo{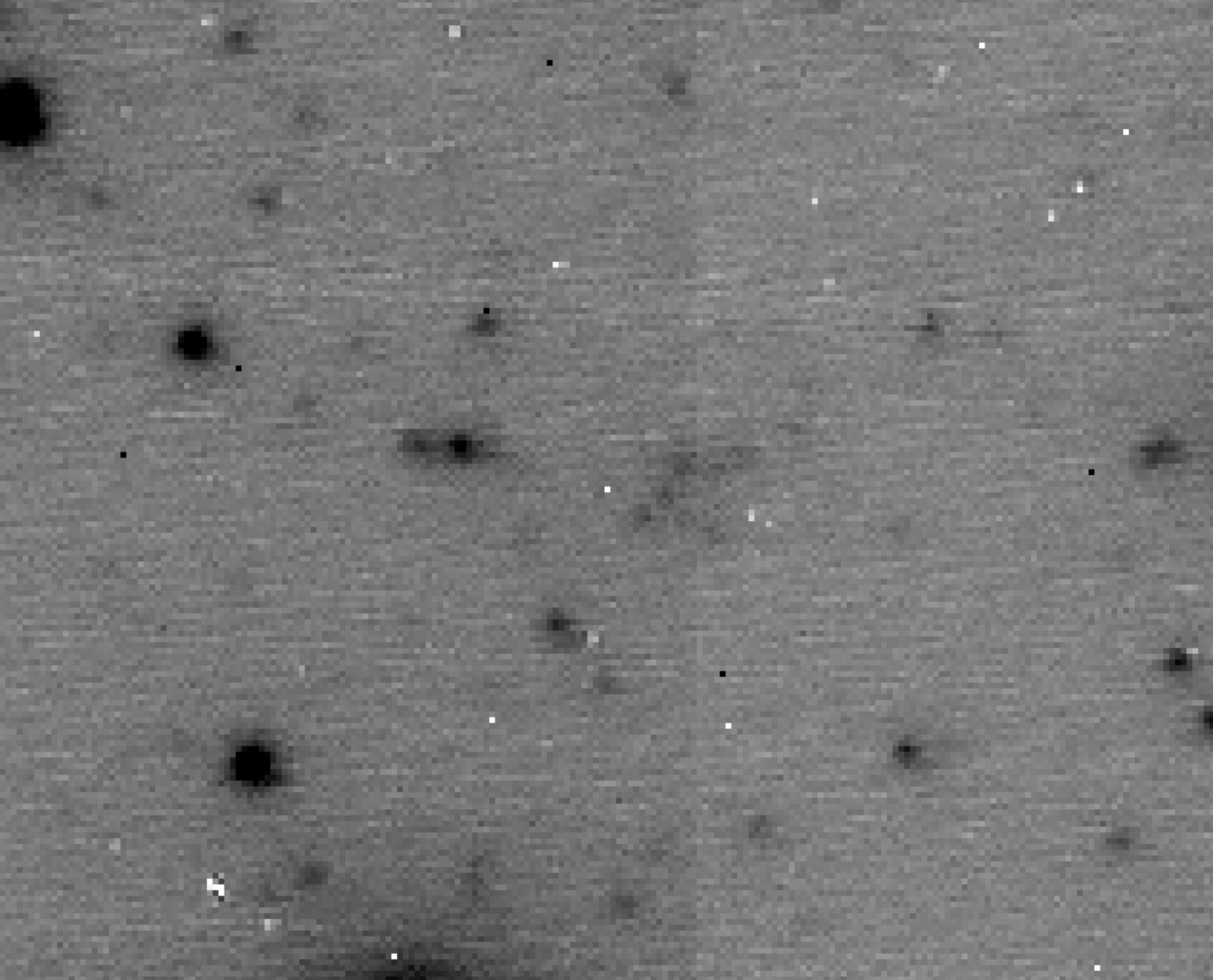}{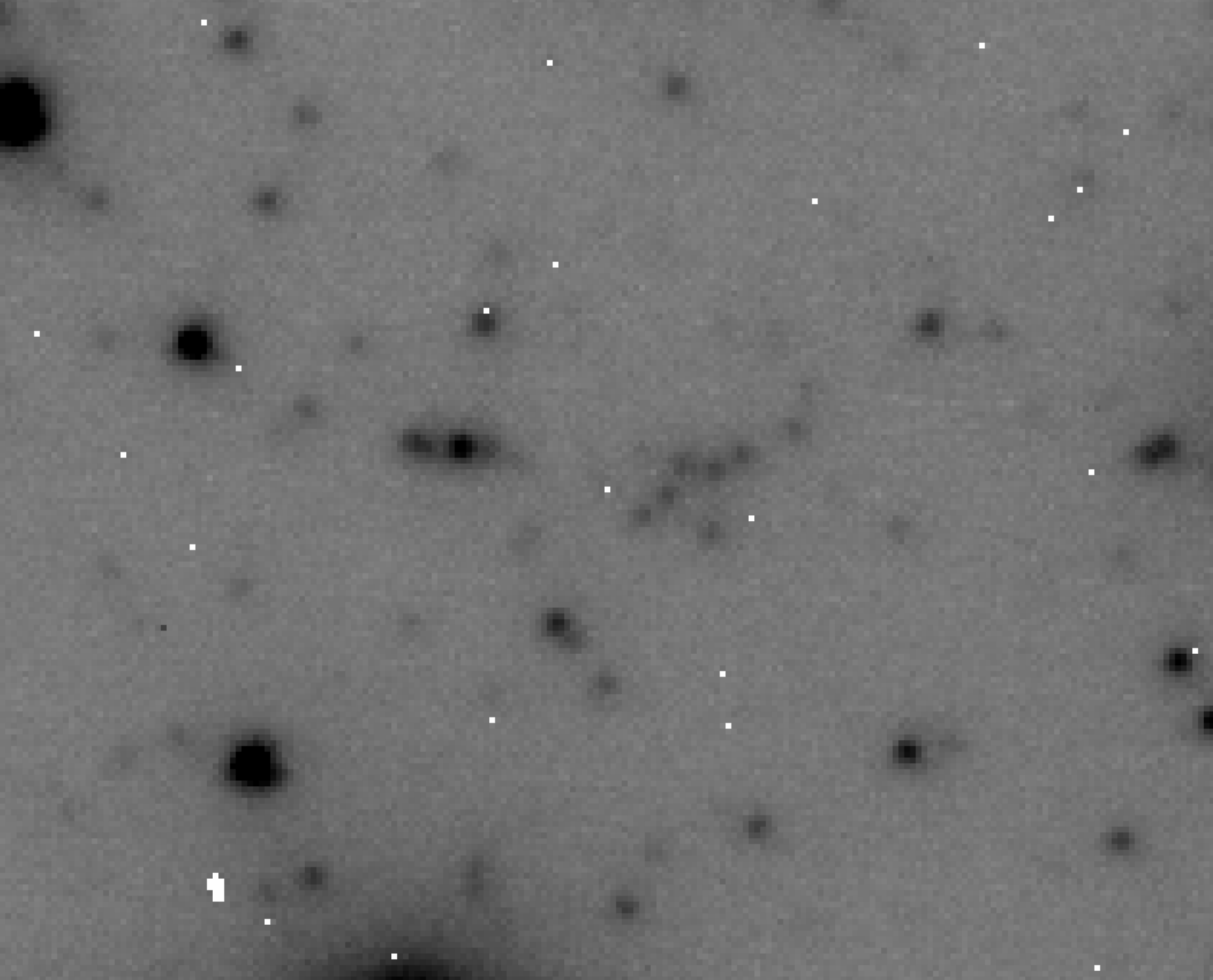}
\caption{Example region (4\arcsec\, x 3.3\arcsec ) of a K$_\mathrm{s}$ exposure before (\emph{left}) and after (\emph{right}) the image processing.\label{fig:iraf}}
\end{figure}

We note that the detectors have a very low dark current (0.01 $\mathrm{e^-}$ s$^\mathrm{-1}$ px$^\mathrm{-1}$). The scale of this effect is of the order of one electron per pixel in our longest exposures; hence, no dark frame has been subtracted. We also highlight that we do not apply sky frames that contain the sky and thermal signature on large and medium spatial scales. The focus of our analysis is point source photometry and, on the spatial scale of a PSF, the flux from the sky is expected to be constant. Applying a sky frame introduces additional systematic uncertainties and contributes to the photon noise. Instead, we evaluate the sky at the position of each star using an annulus around them during the profile fitting, as discussed in Section~\ref{sec:photometry}. An example region (4\arcsec\, x 3.3\arcsec ) of one of the K$_\mathrm{s}$ exposures before and after processing is shown in Figure~\ref{fig:iraf}.

\section{Photometric performance of GeMS}\label{sec:performance}

The purpose of MCAO is to provide diffraction-limited imaging over the largest field of view possible. In the case of a perfect correction, the PSFs across the field would have the same size, shape, and flux concentration. Globular clusters provide an ideal testing ground of MCAO systems in this regard, with a large number of stars to examine. We now characterize the quality and stability of the MCAO correction in our exposures to inform the more detailed photometric measurements that are described in Section~\ref{sec:photometry}.

\begin{figure}
\plottwo{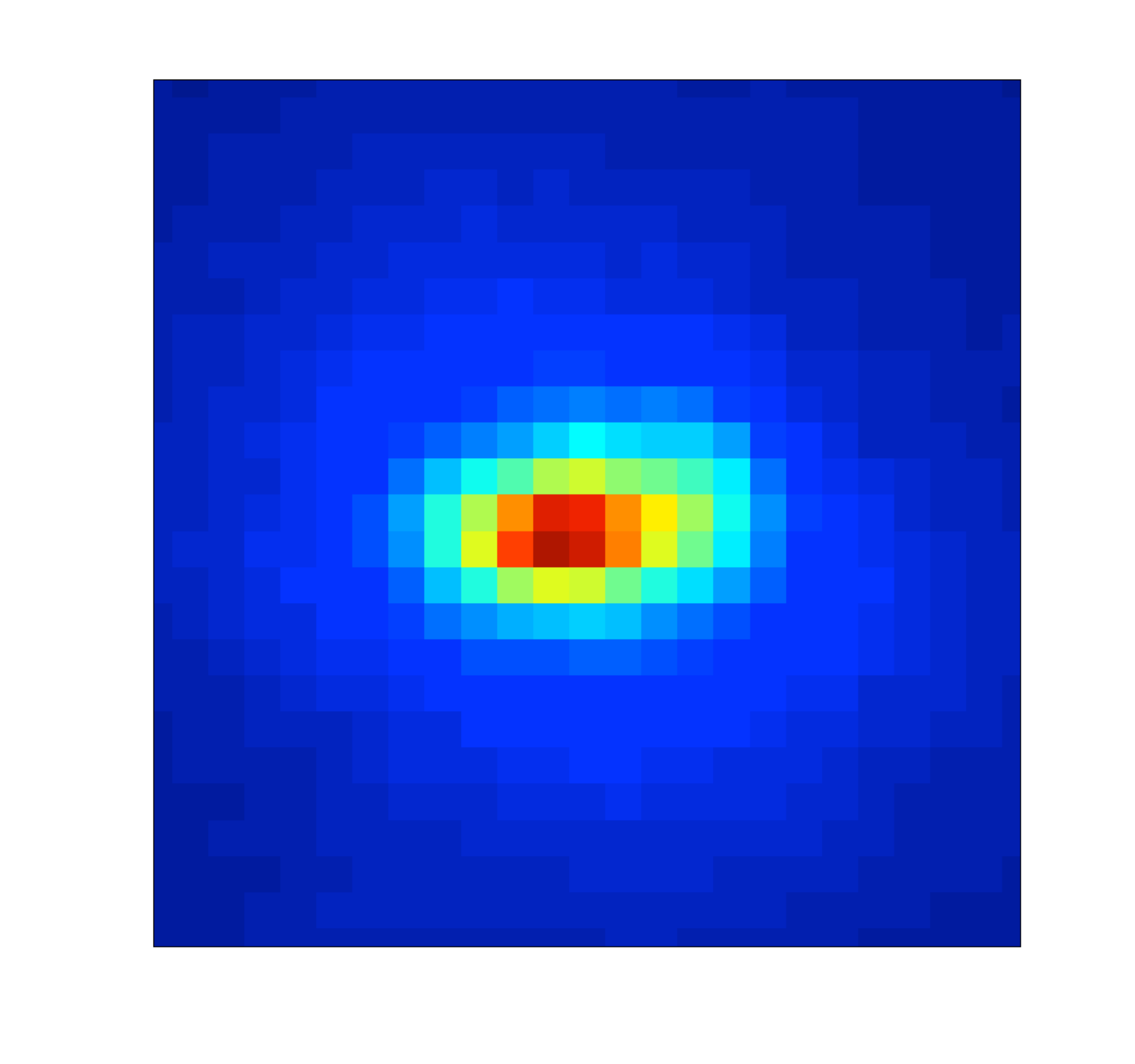}{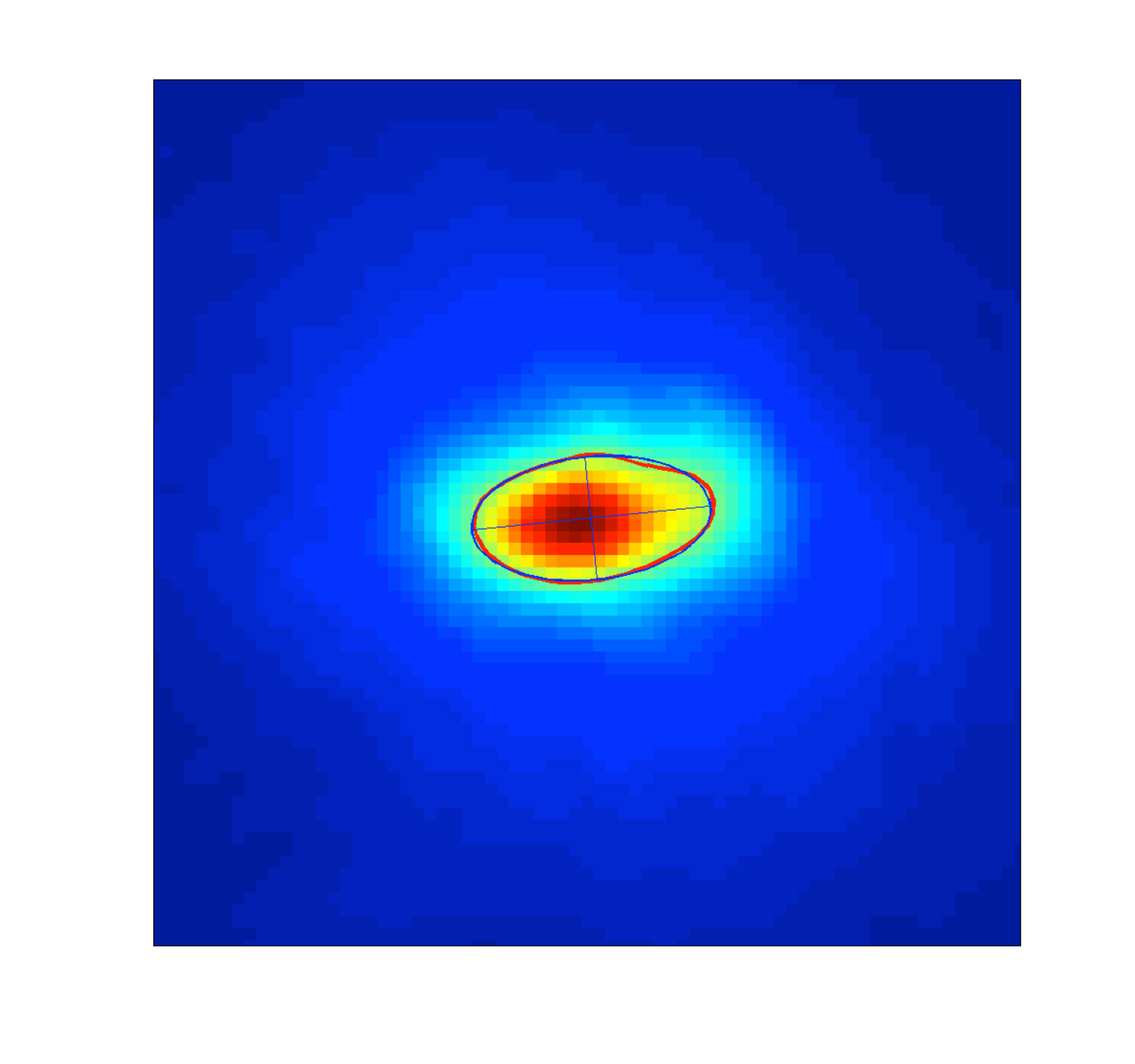}
\caption{Image of a PSF star in the J band that shows significant elongation. The left and right images are before and after the interpolation, respectively. The red contour in the right panel shows the half-maximum values and the blue contour shows the ellipse fitted to them.\label{fig:interp}}
\end{figure}

We use the large number of bright and isolated PSF stars (see discussion in Section~\ref{sec:psfestimate}), distributed across the field of view. We subtract all other stars from the images (using the technique described in Section~\ref{sec:photometry}) and we apply a median filter on the bad pixels to remove them from the images. An example of a stellar image is shown in the left panel of  Figure~\ref{fig:interp}. The stellar images are then interpolated onto a finer grid by zero-padding their Fourier transform by a factor of 3, an example of which is shown in the right panel of Figure~\ref{fig:interp}.

The Strehl ratio (SR) is the ratio of the peak flux of the actual PSF to the peak flux of the diffraction-limited PSF. To measure it, we adopt the peak value of the interpolated image as the peak of the PSF. The diffraction-limited PSF is generated by the Fourier transform of the Gemini South telescope aperture, which we model as a circle with a diameter of 8.1 m and a central circular obstruction of 1 m. We omit the spider of the secondary mirror since the change in the PSF peak by including this more complex structure is negligible. The diffraction peak is normalized to the flux of the star using the instrumental magnitudes measured by PSF fitting by the techniques in Section~\ref{sec:photometry}. However, the PSF model we use here has a radius of 45 pixels that is larger than the radius used in Section~\ref{sec:psfrad} This is so that we include in each measurement the majority of flux from the star (see Figure~\ref{fig:psfradstat}). This improves the accuracy of the flux, at a cost to the precision, by avoiding larger systematic errors.

\begin{figure}
\plottwo{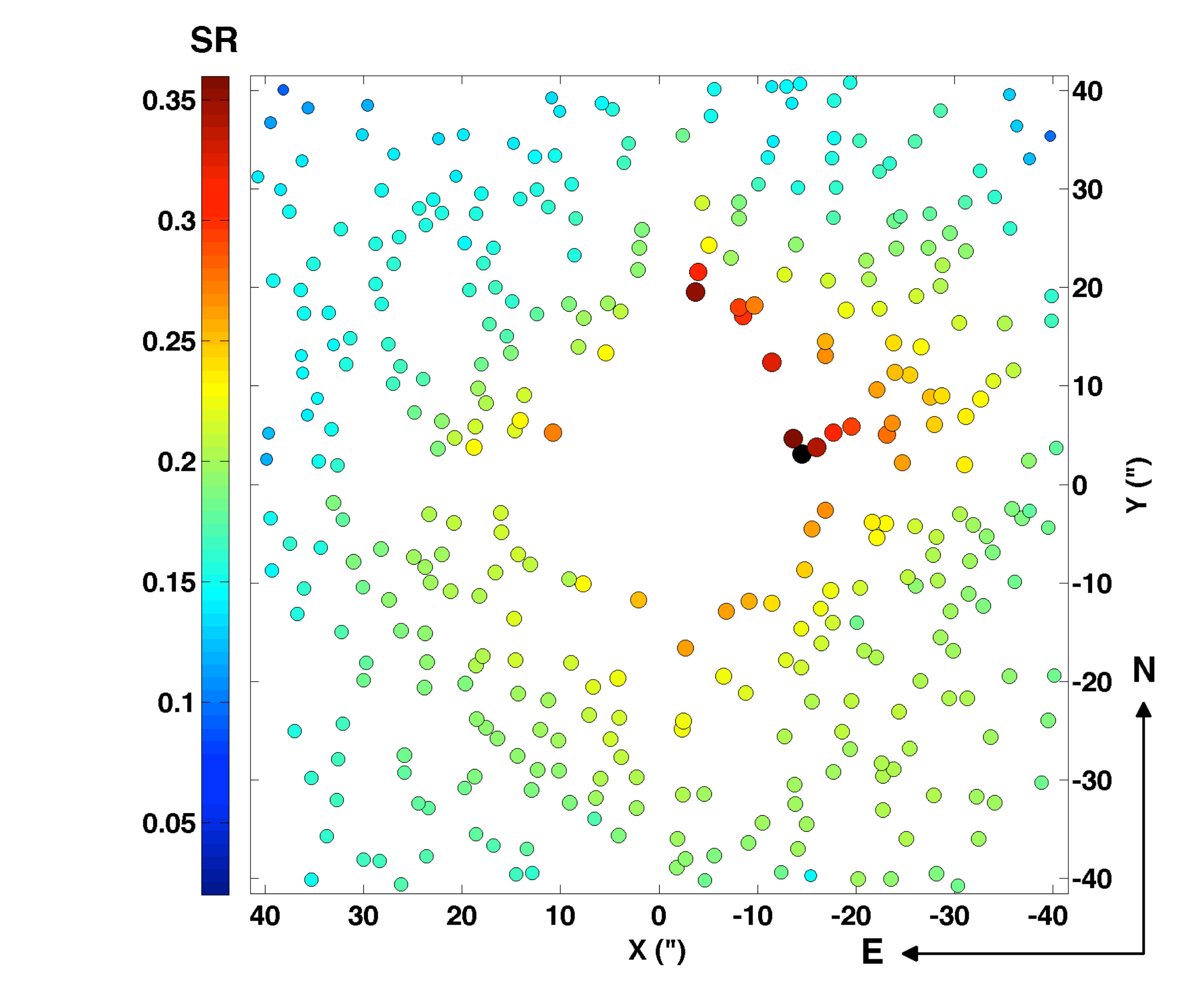}{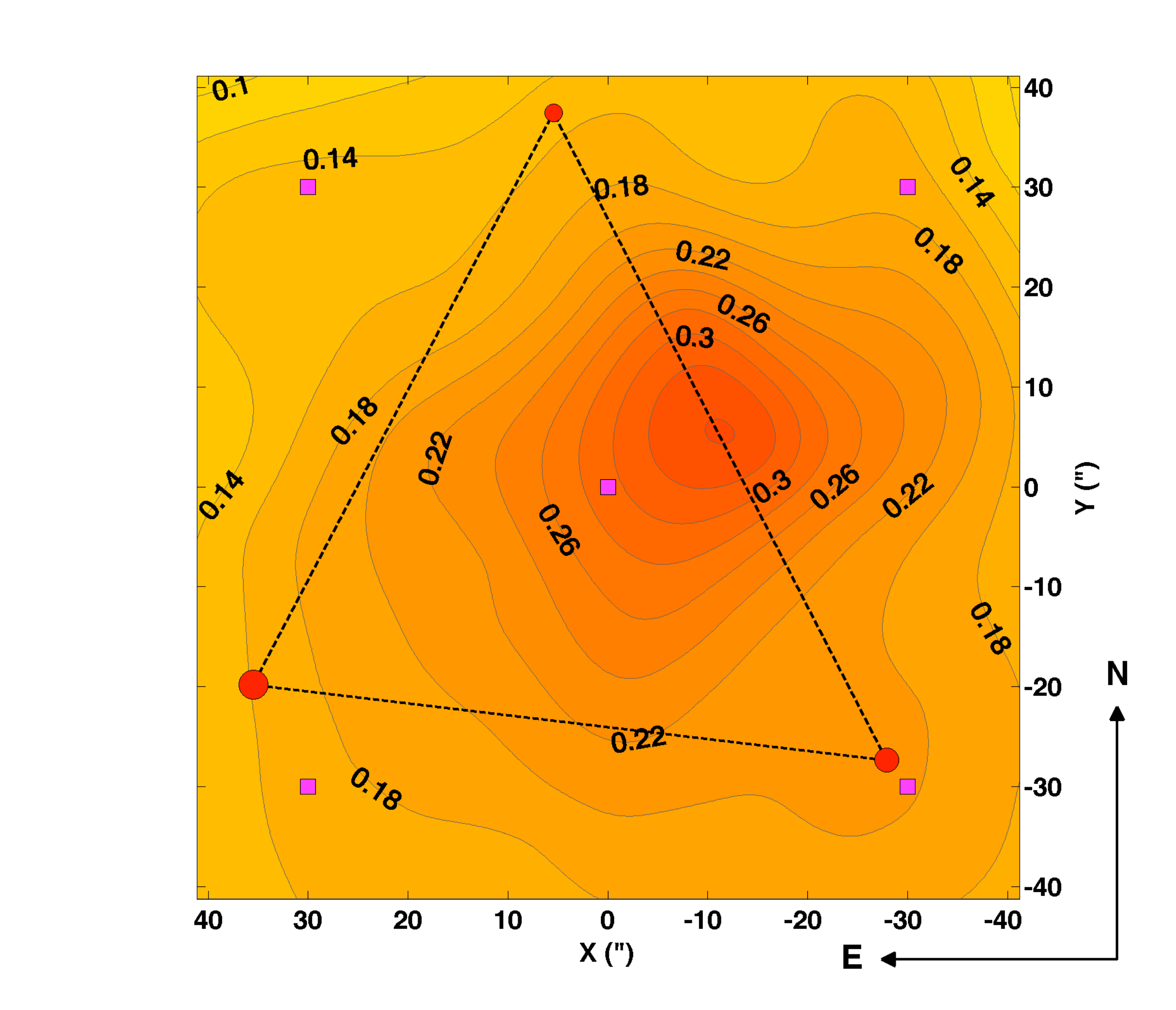}
\caption{Strehl ratios of the PSF stars in a K$_\mathrm{s}$ long exposure (\emph{left}) and their interpolation witha smoothing spline (\emph{right}). The position of the NGSs and LGSs in the field of view are indicated by the red circles and the magenta squares, respectively.\label{fig:spline}}
\end{figure}

We provide a simple quantification of the two dimensional shape of the PSF by fitting an ellipse to the contour that defines the half-maximum of the PSF. An illustration of this is shown in the right hand panel of Figure~\ref{fig:interp}. The FWHM is then defined from the geometric mean of the semi-major ($a$) and semi-minor ($b$) axes, such that FWHM $=2\sqrt{ab}$, and the ellipticity is defined as $e=1-b/a$. We also measure the position angle of each ellipse.  This allows us to create spatial performance maps of the MCAO correction, an example of which is shown in Figure~\ref{fig:spline} for one of the long K$_\mathrm{s}$ exposures (exposure 10). The left panel shows the distribution of raw SR measurements for each of the PSF stars. Note that there is a lack of PSF stars in the central region of the image due to extreme crowding. The right panel shows the same performance map but interpolated across the full field using a smoothing spline and shown as a contour map. 

\begin{figure}
\centering
\includegraphics[width=0.5\textwidth]{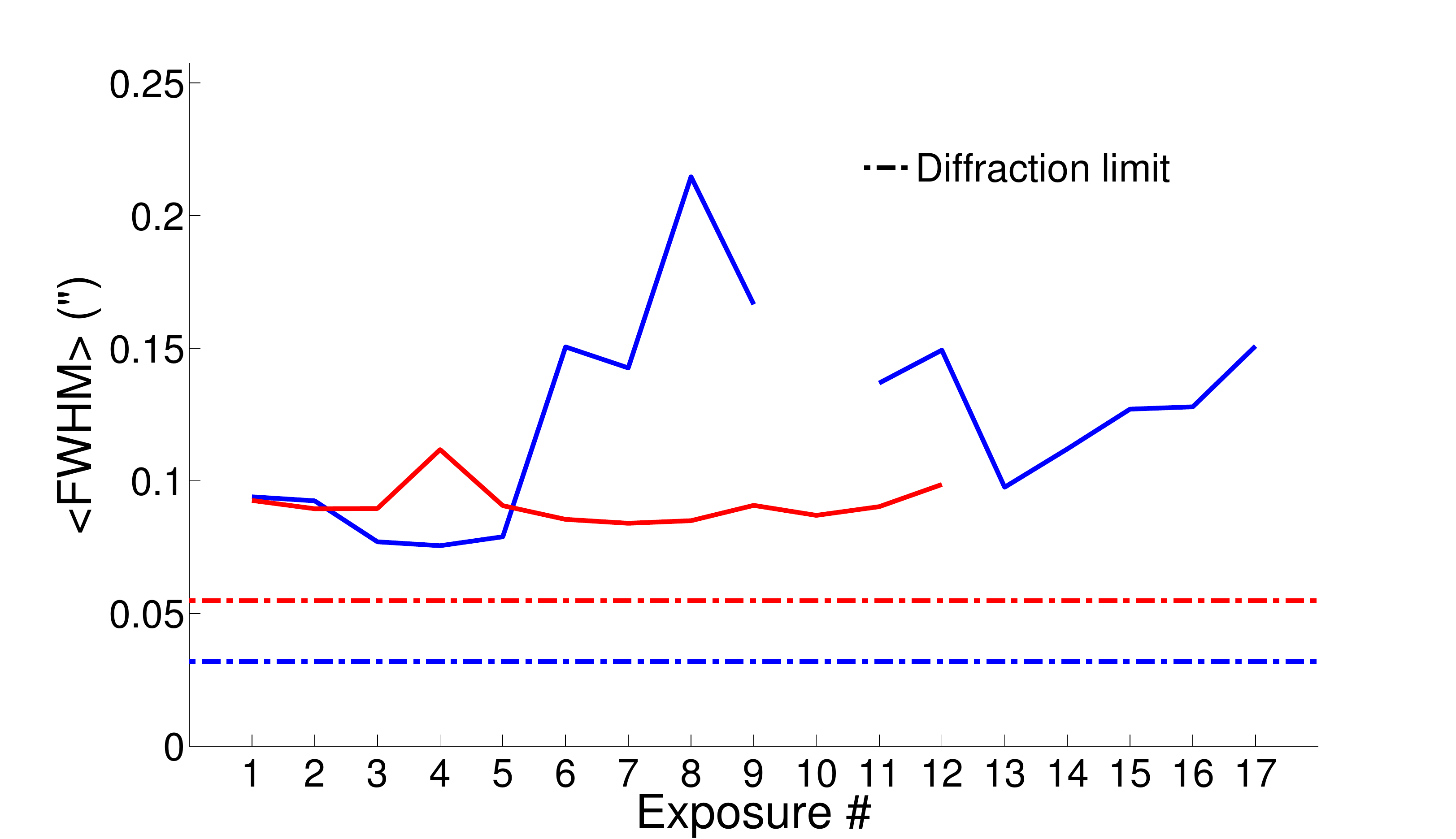}
\includegraphics[width=0.5\textwidth]{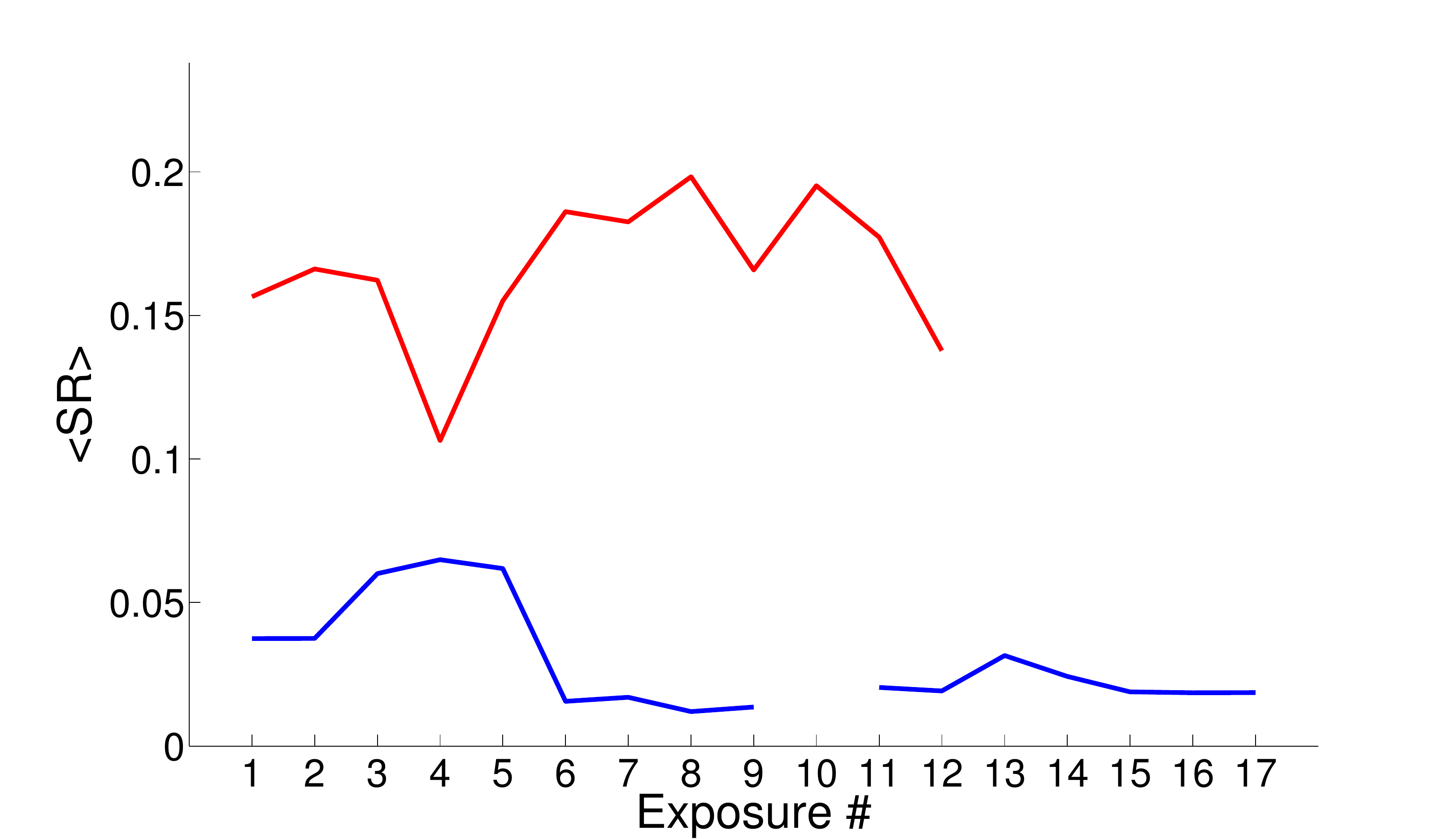}
\includegraphics[width=0.5\textwidth]{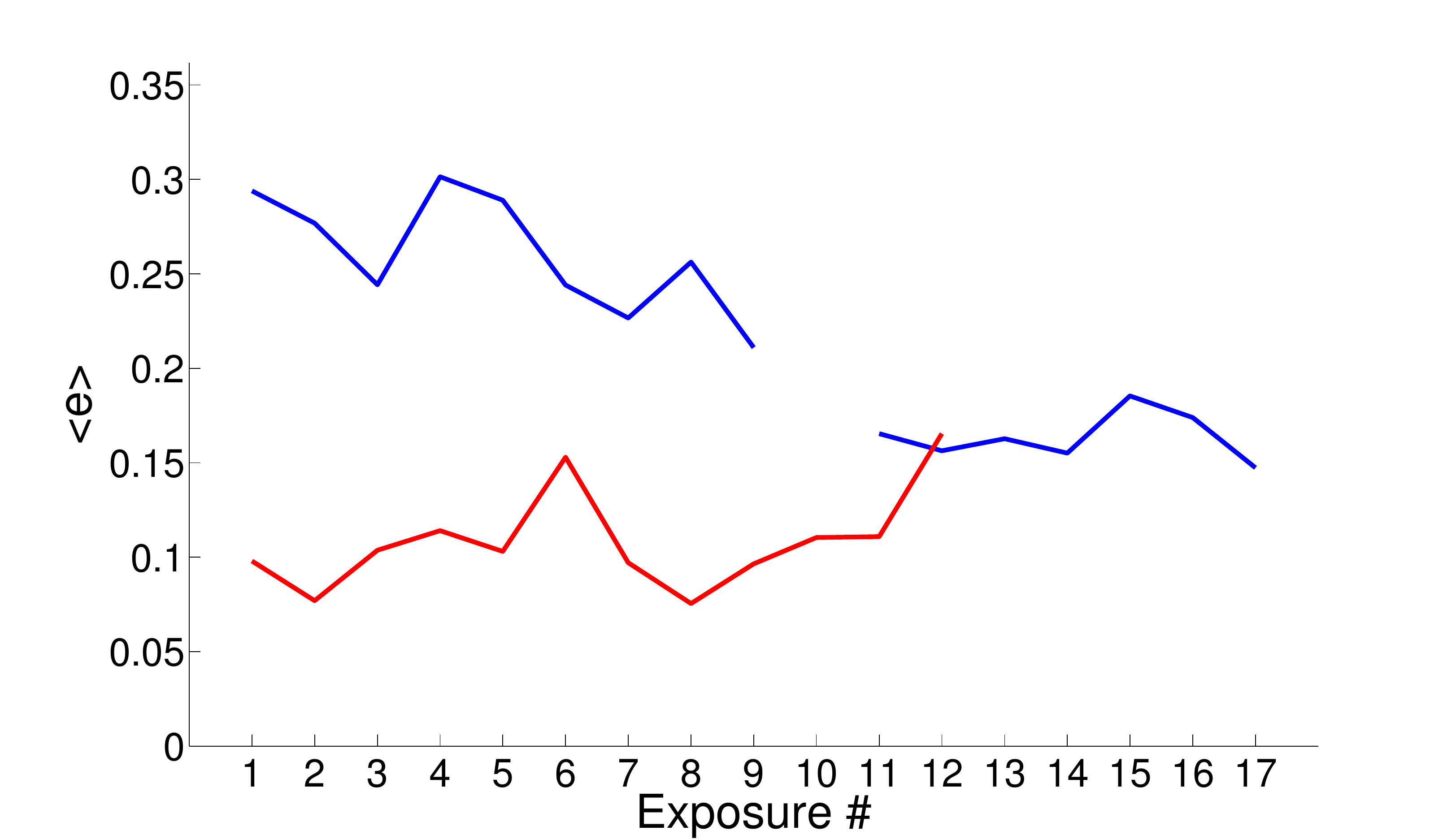}
\caption{Average FWHM \emph{(top panel)}, Strehl ratio \emph{(middle panel)}, and ellipticity \emph{(bottom panel)} of the PSF stars in each long exposure in J and K$_\mathrm{s}$ bands (blue and red, respectively). The diffraction-limit FWHM values are shown in the top panel for comparison. We have not analyzed exposure \#10 in the J band due to a particularly poor MCAO correction. These exposures are numbered sequentially in the order they were observed; all K$_\mathrm{s}$ band exposures were observed on the same night, whereas the J band exposures were observed on two different nights, with the first ten observed on the first night.\label{fig:perfstats}}
\end{figure}

The average values of the FWHM, Strehl ratio, and ellipticity of each long exposure image for both the J and K$_\mathrm{s}$ bands are shown in Figure~\ref{fig:perfstats}. These exposures are numbered sequentially in the order that they were taken. The average FWHM in the J band is $0.13\pm 0.04\arcsec$ and in the K$_\mathrm{s}$ band is $0.09\pm 0.01\arcsec$. For the Strehl ratio, the average value for the J band is $0.03\pm 0.02$, whereas for the K$_\mathrm{s}$ band it is $0.17\pm 0.03$. It is expected that adaptive optics performs in K$_\mathrm{s}$ band better than in J since the aberrated wavefront has a larger spatial and temporal coherence at longer wavelengths \citep{bib:davies12}, and we note that stars in the J band appear on average more elongated than in K$_\mathrm{s}$. We do not see a correlation between the average ellipticity and FWHM or Strehl ratio in our observations.

\begin{figure}
\gridline{\fig{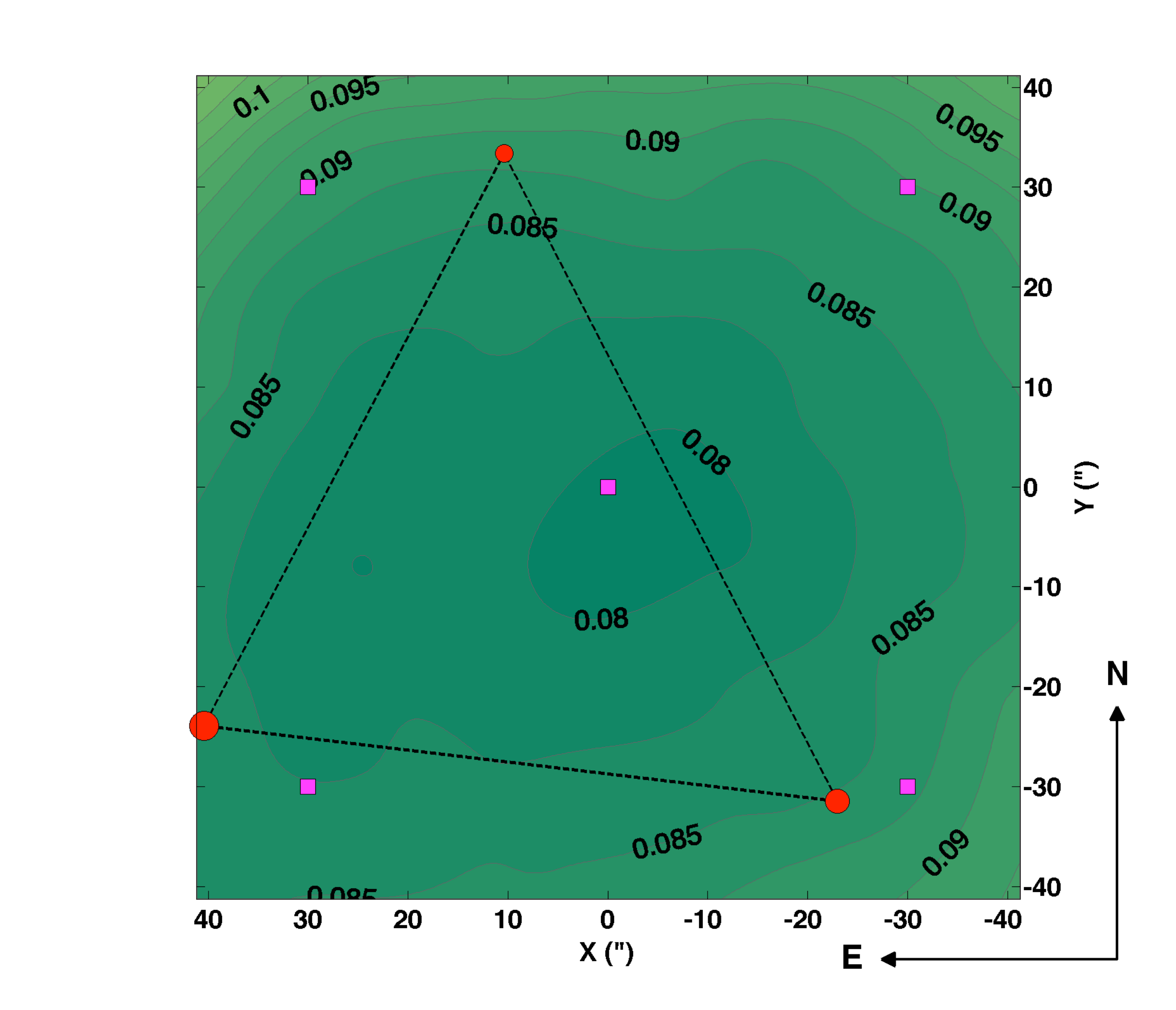}{0.45\textwidth}{FWHM (\arcsec)} \fig{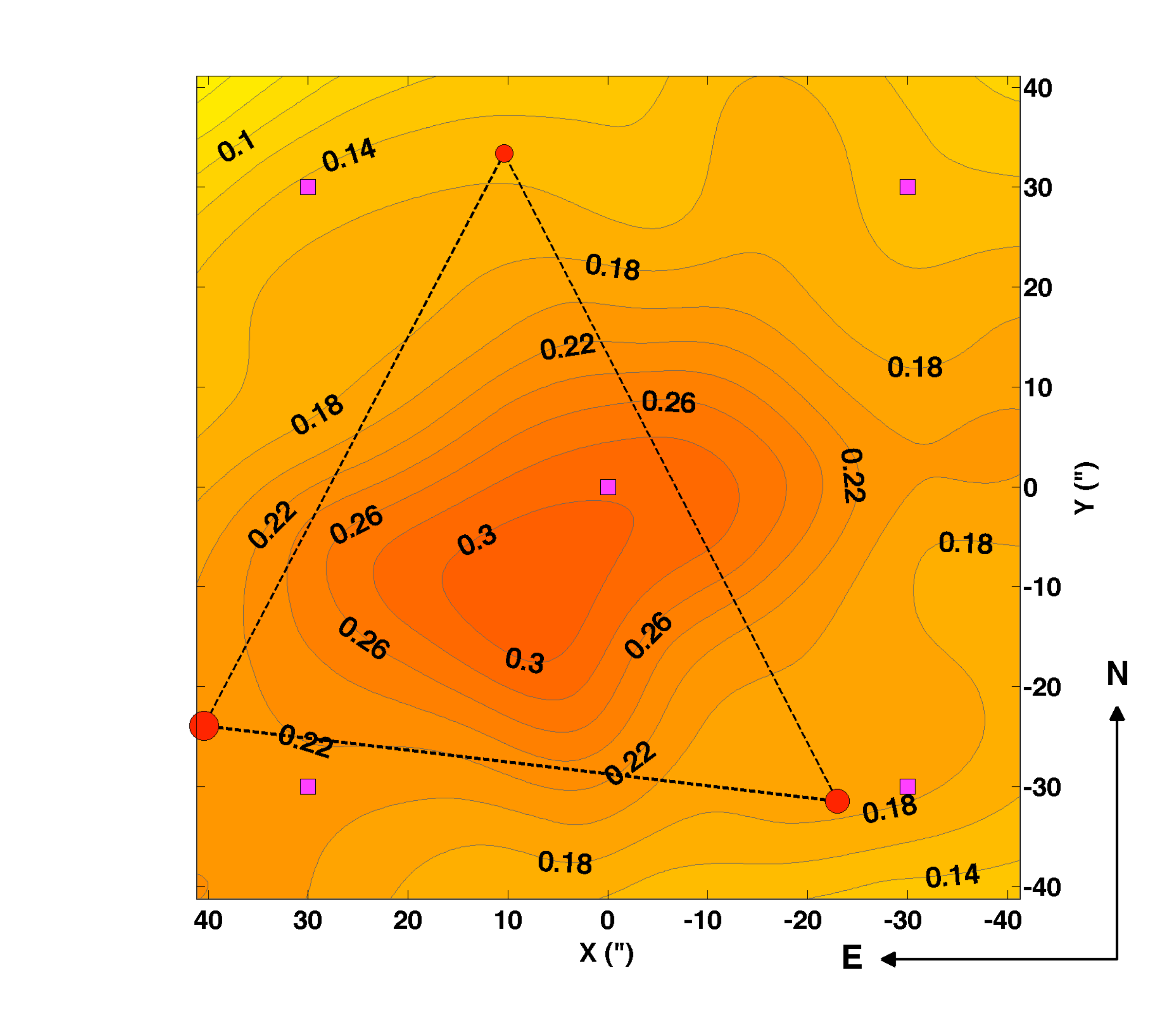}{0.45\textwidth}{Strehl ratio}}
\gridline{\fig{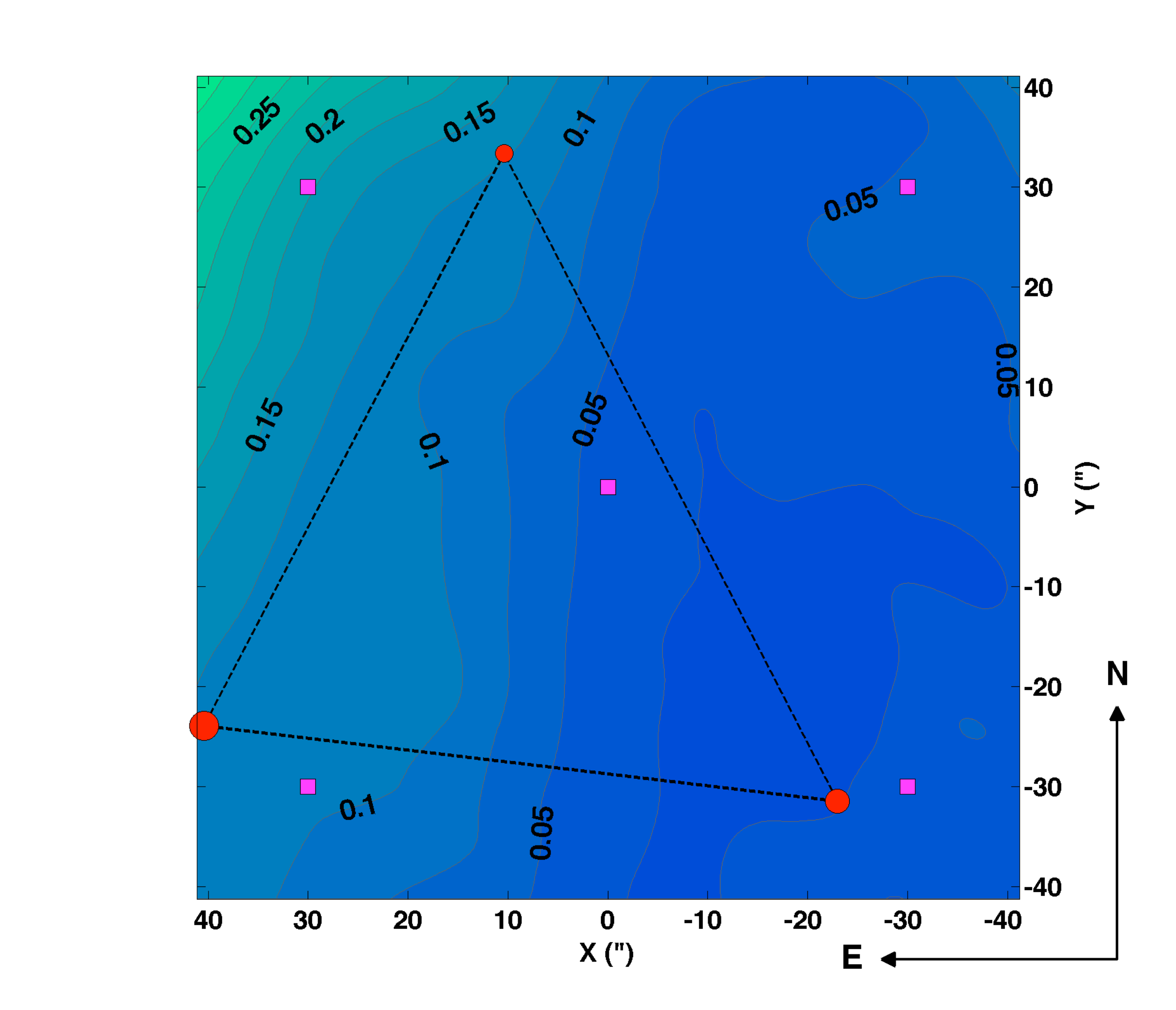}{0.45\textwidth}{Ellipticity} \fig{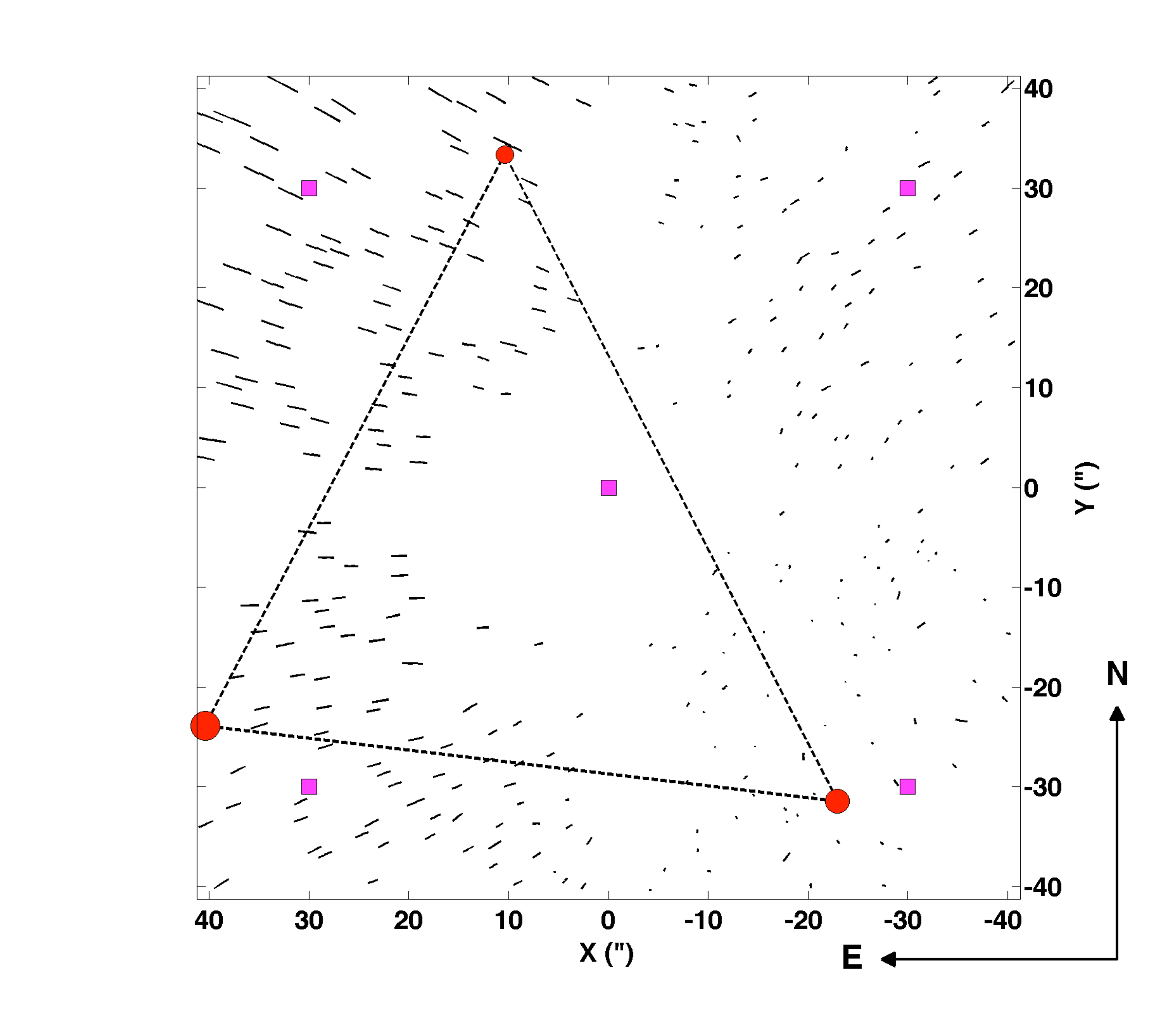}{0.45\textwidth}{Position angle}}
\caption{Performance maps of a K$_\mathrm{s}$ long exposure, based on the measuremnt of its PSF stars. The map of the position angles shows the orientation of the ellipses fitting the PSF stars' contours at half maximum; the length of the segments is proportional to the ellipticity.\label{fig:perfmaps}}
\end{figure}

\begin{figure}
\gridline{\fig{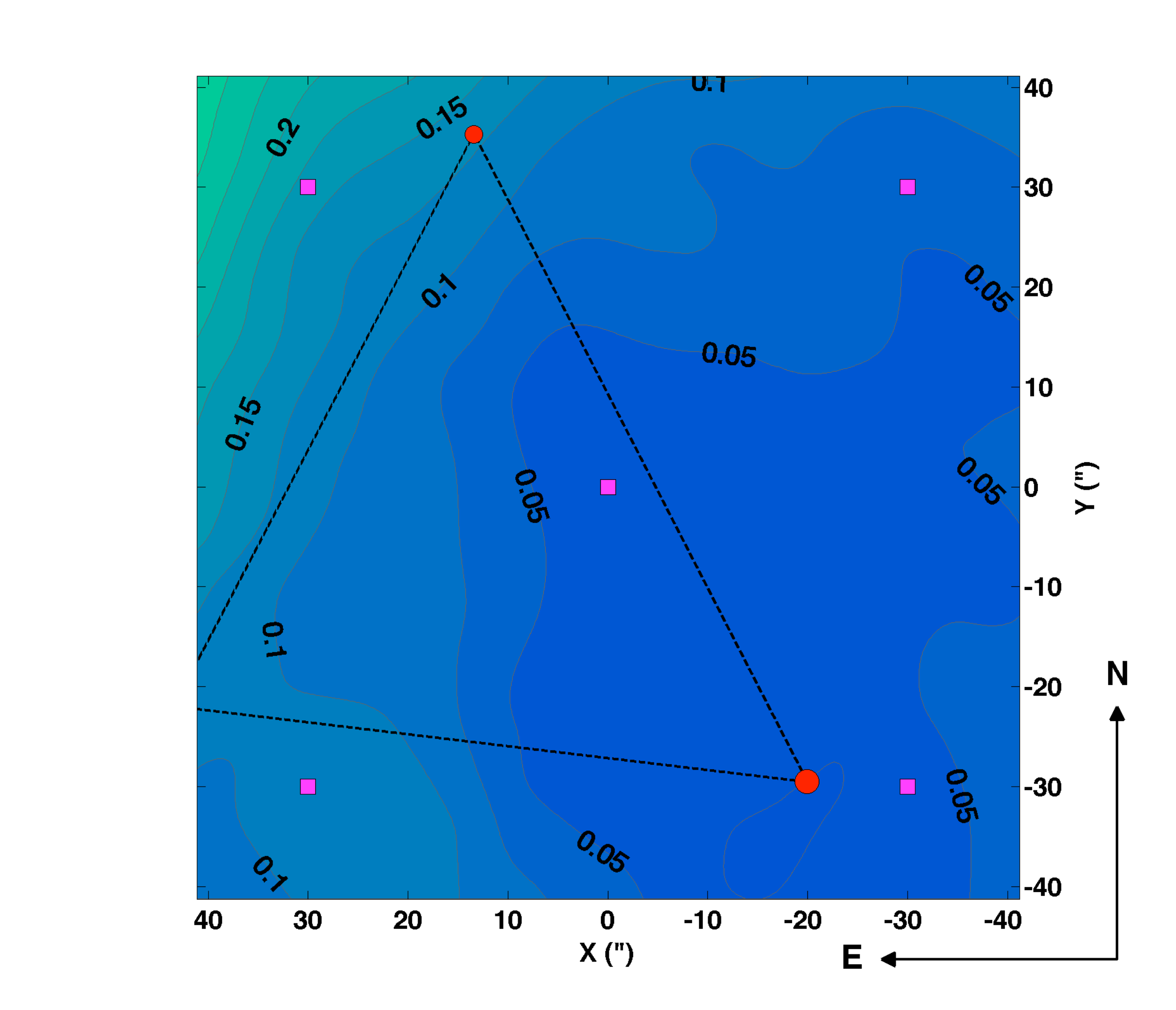}{0.33\textwidth}{} \fig{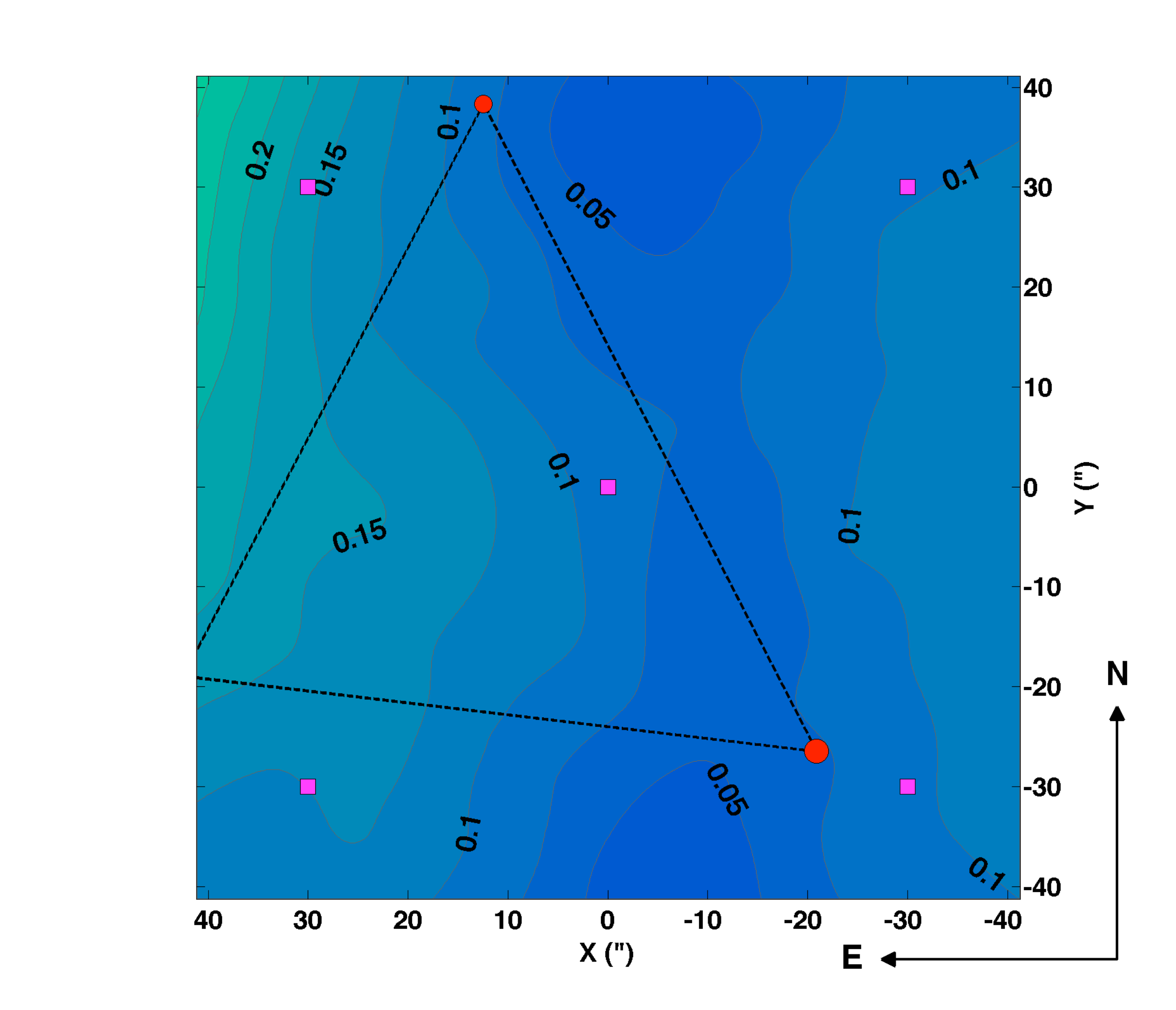}{0.33\textwidth}{} \fig{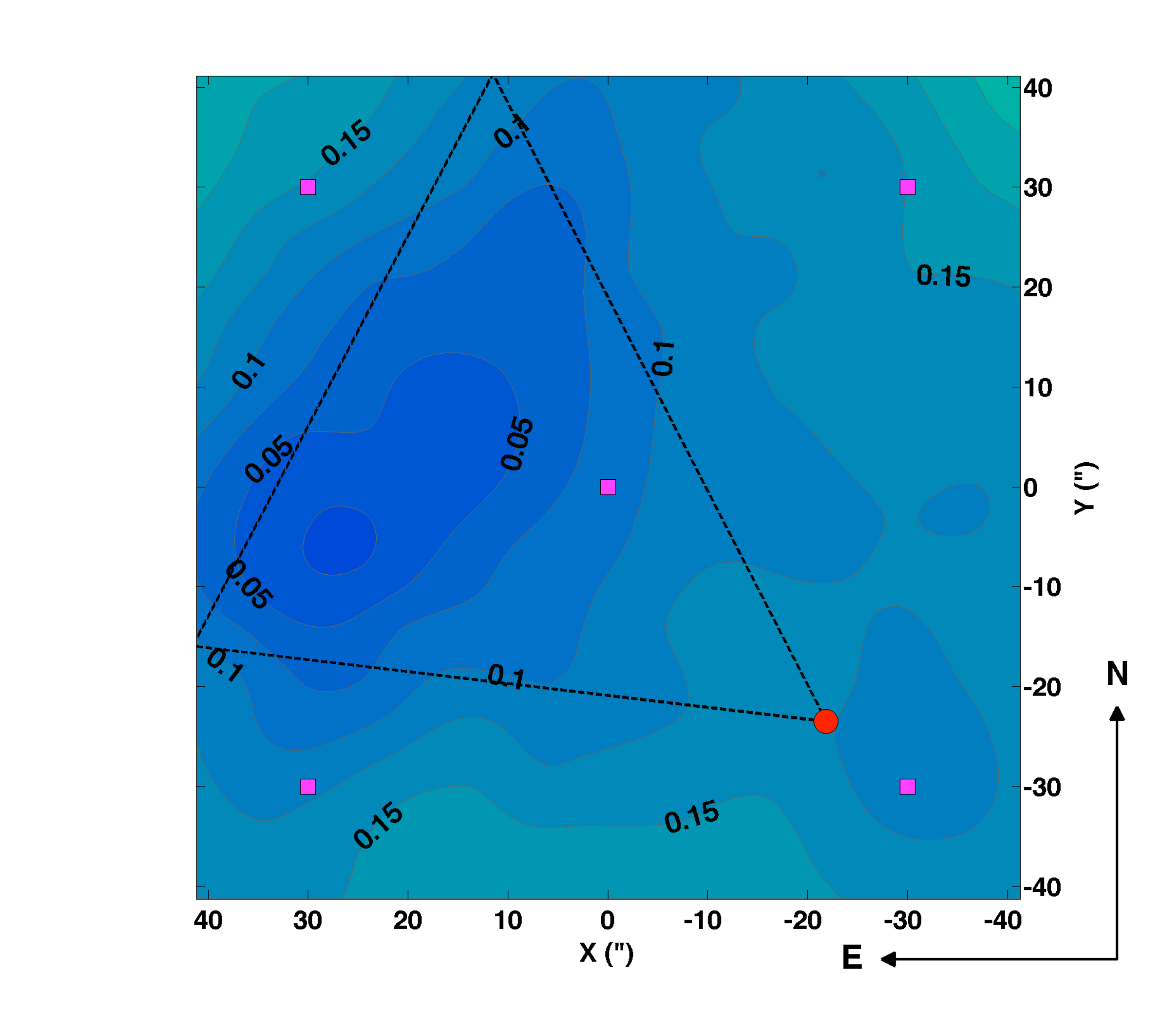}{0.33\textwidth}{}}
\caption{Maps of the ellipticities of three consecutive K$_\mathrm{s}$ long exposures.\label{fig:elliptmaps}}
\end{figure}

In Figure~\ref{fig:perfmaps}, we show the performance maps for the best of our exposure (\#8 in the K$_\mathrm{s}$ band), which has the smallest FWHM, the largest Strehl ratios, and the most circular PSFs. The first three panels show the interpolated maps of the FWHM, SR, and ellipticity. The fourth panel indicates the shape of each FWHM ellipse, where the angle of the dash indicates the position angle and the length indicates the ellipticity (short dashes are more circular ellipses and thus the position angle is not as well measured as for longer dashes). In all four panels, it is clear that the performance is relatively uniform but certainly not constant across the field. In line with expectations, the best values of FWHM and SR are near the center of the LGS asterism. But we also notice that the shape of the PSF, described by its ellipticity, has a more complex spatial pattern and does not perform better in the center (thus our best exposure has smaller but more elliptical PSFs in the center of the asterism compared to elsewhere). Indeed, we find that the behavior of the ellipticity and position angles varies abruptly in consecutive exposures, as demonstrated in Figure~\ref{fig:elliptmaps} for three consecutive K$_\mathrm{s}$ band exposures. Note that this does not correlate with the movement of the NGS asterism caused by the dithering of the telescope.

One of the problems in delivering a better performance was the incorrect conjugation in altitude of the optical elements in the LGS wavefront sensors, as documented in a Gemini Observatory internal report (G. Sivo, private communication). This was identified and corrected in the year after our observations, to reduce the size and ellipticity of the PSF delivered.

Despite these few issues, our MCAO observations show a substantial improvement from seeing-limited images. Our estimation of the performance of GeMS are consistent with previous findings \citep{bib:neichel14,bib:saracino15,bib:massari16a,bib:massari16b,bib:saracino16}. We now demonstrate that they can produce unprecedented levels of photometric precision in a crowded field from the ground.

\section{Photometric analysis}\label{sec:photometry}

It is well know that precision photometry in a crowded stellar field cannot be achieved by aperture photometry due to crowding, where starlight from one star compromises the measurement of the flux of any nearby star. A more robust method is profile fitting photometry, especially if the PSF can be estimated empirically on the images. This is the case for our observations of NGC 1851, where we have a large number of bright point-like sources in every part of the field of view. For the ensuing photometric analysis we have used the DAOPHOT II suite of programs \citep{bib:stetson87,bib:stetson88,bib:stetson94}. We begin by providing a broad overview of the process that we use to obtain our photometry prior to discussing how certain critical stages of our analysis are optimized for best performance.

\subsection{Overview of the photometry procedure}

\subsubsection{Creation of the master source catalog}

\begin{figure}
\plottwo{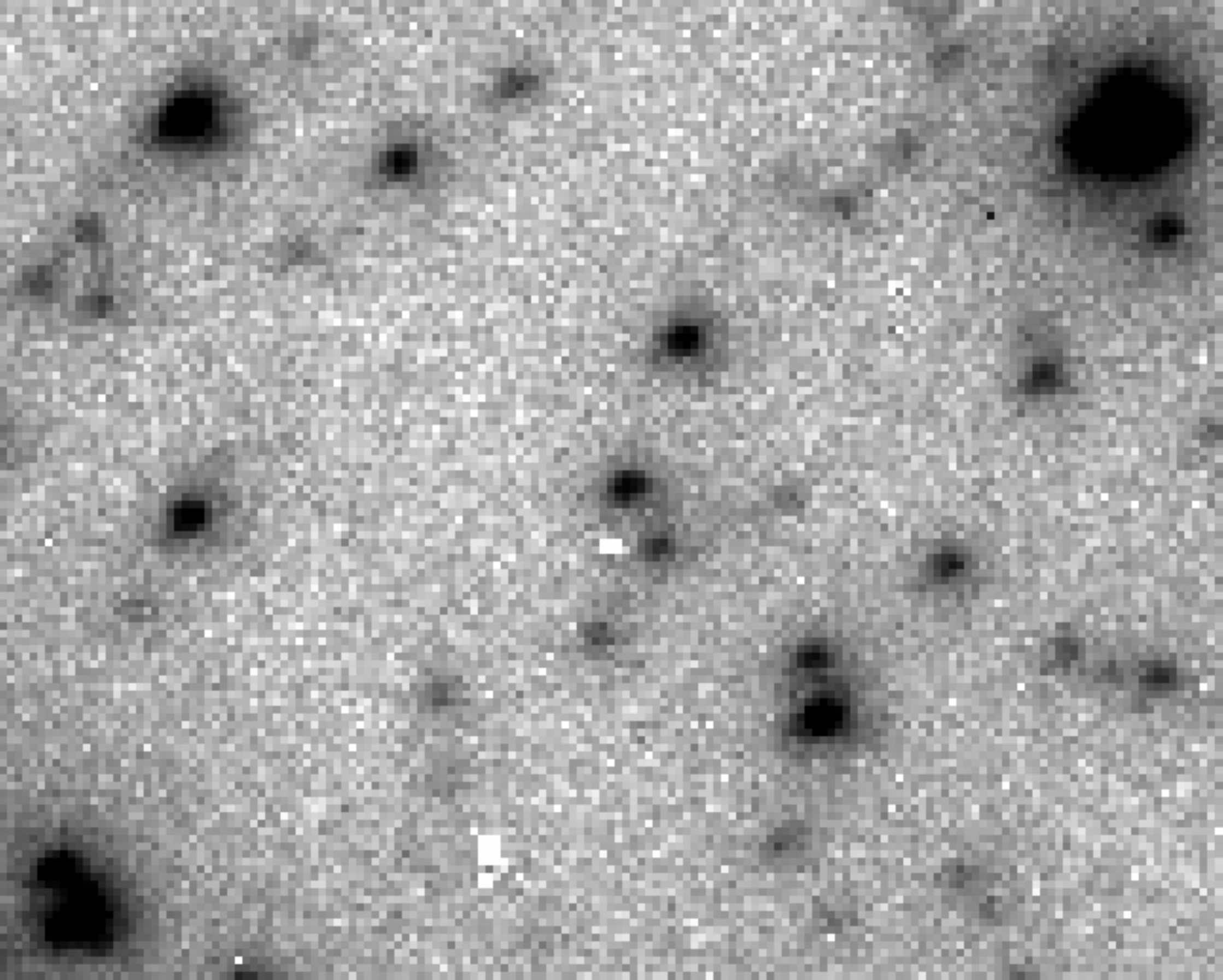}{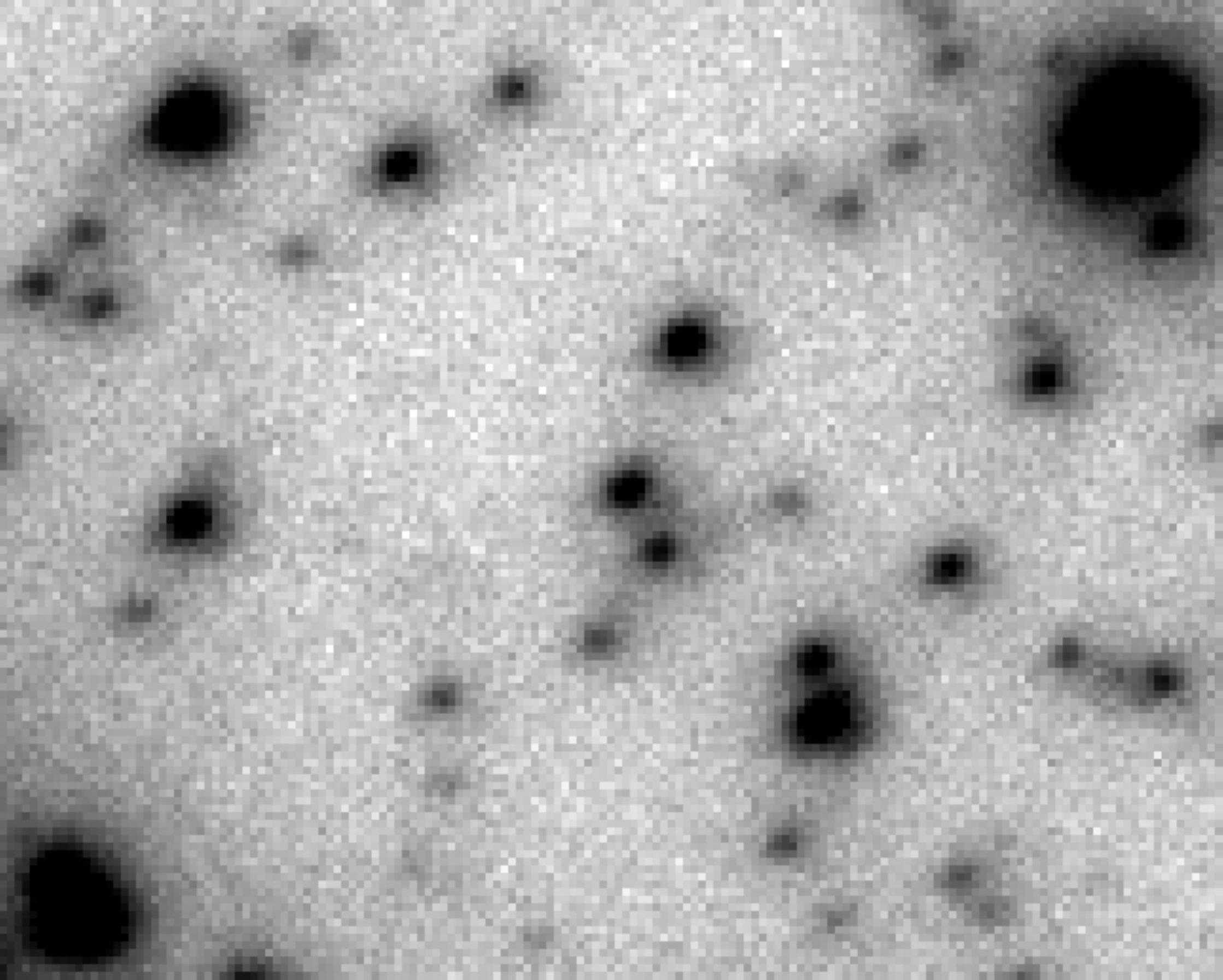}
\caption{Detail of a long exposure (\emph{left}) and stacked image (\emph{right}) in K$_\mathrm{s}$ band.\label{fig:stack}}
\end{figure}

\begin{itemize}
\item The stars are first detected (using the FIND routine) by convolving each individual image with a truncated Gaussian profile, lowered to have a zero integral and that has a FWHM similar to that of the real PSF. Local maxima in the convolved images are identified as stars if they are above a chosen noise threshold ($3.5\sigma$ for J and $4\sigma$ for K$_\mathrm{s}$) and within a range of ``sharpness'' and ``roundness'' parameters, selected to discriminate a star from extended objects, cosmic rays or artifacts of the detector.

\item For each catalog corresponding to each of the original images, we transform the coordinate system to a master frame of reference (using the DAOMASTER program). The master frame is defined by the seeing-limited catalog used for the photometric calibration (Section \ref{sec:calibration}). A polynomial transformation of order 3 is used.

\item We additionally stack the long exposure images (using MONTAGE) by assigning to each pixel of the stacked image the median value of its nearest pixel from each of the transformed frames, and identifying all sources (using FIND). While the resulting image has a larger PSF than any of the single exposures (due to the different PSFs of each image, see Section~\ref{sec:performance}, and the nearest-pixel approach instead of interpolation), these stacked images allow us to detect and measure a large number of faint stars that are not detected in the individual exposures (Figure~\ref{fig:stack}).

\item The master catalog is created in each band by combining all detections found in each of the short exposures, the stacked images, and in at least three of the dithered long exposures. The short exposures are primarily used to find the brightest unsaturated stars, the stacked images are primarily for the faintest stars, and the long exposures are for all the other sources, with the requirement of three independent detections to remove most spurious objects.
\end{itemize}

\subsubsection{Estimating the PSF}\label{sec:psfestimate}

\begin{figure}
\centering
\plottwo{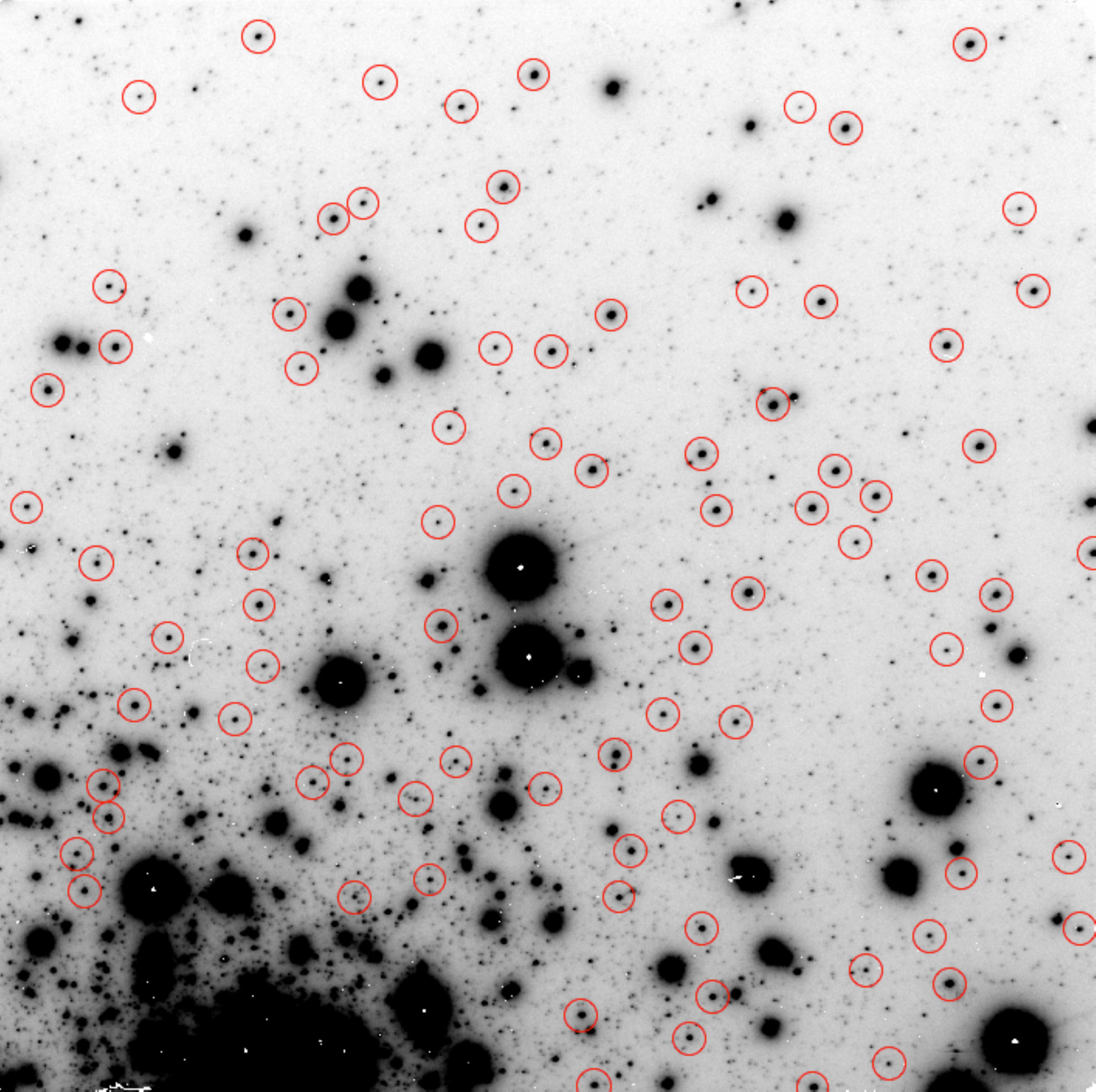}{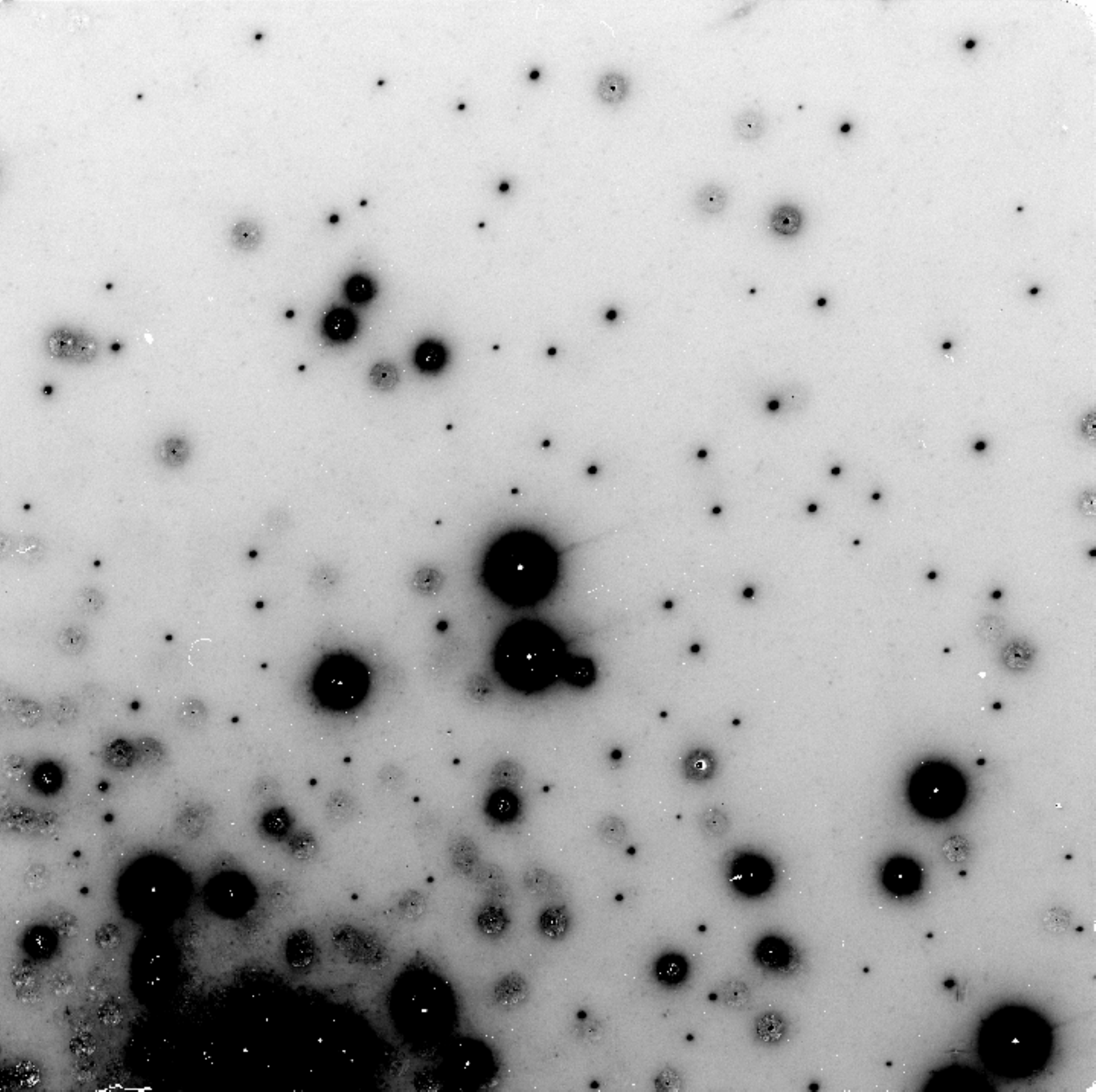}
\caption{\emph{Left}: selection of the PSF stars used for one of the K$_\mathrm{s}$ images. \emph{Right}: cleaned image to measure the two-iterations PSF.\label{fig:psf_stars}}
\end{figure}

\begin{itemize}
\item We select 500 ``PSF stars'' in the entire field of view, approximately 90 in each quadrant of the detector (left panel of Figure~\ref{fig:psf_stars}), that are used to model the PSF independently for every chip of every exposure. These are stars that are bright but not saturated and that do not have obvious close companions. They are chosen to be distributed as uniformly as possible across the observed field of view, except in the cluster core where many saturated stars cause large flux gradients that render reliable background estimates difficult. The significant number of PSF stars ensures that the photon noise in the PSF model is reduced and that systematic effects on individual PSF stars (such as detector cosmetics or contamination from unrecognized companion stars) are minimized. A constant background is estimated for each PSF star as the modal value of the pixels in a surrounding annulus and is then subtracted from each PSF star. We choose the inner edge of this sky region to start from the pixel immediately outside the PSF model and the outer edge to contain approximately 2000 pixels in the annulus.

\item The PSF is modeled in two parts. The first component is a simple analytic function, for which we have chosen a Lorentzian, that has residuals to the actual PSF smaller than the other available options (Gaussian, Moffat and ``Penny''). The second component is a two dimensional grid of empirical corrections calculated from the residuals of the PSF stars after subtracting the analytic component. This look-up table can be set to have either the same values everywhere in the field of view or to vary spatially. In the spatially varying case, each element of the table is associated with a set of coefficients that describe a bivariate polynomial in x and y, the position on the field of view. In Section~\ref{sec:performance}, we demonstrated that there is significant spatial variation in the PSF in our images, and so we have chosen to use a cubic polynomial, the highest degree of variability available in DAOPHOT. The possibility to calculate such a high variability is given by the large number of PSF stars available in the images.
\end{itemize}

\subsubsection{Profile fitting}

Photometric measurements are obtained for objects in each image, including the stacked images, by fitting the appropriate PSF model to each object in the master source catalog (using ALLSTAR). At each iteration, the flux of a star is calculated by first subtracting the profiles of its neighbors using their positions and magnitudes derived from the previous iteration (starting with estimates from aperture photometry). The star's background (as estimated from the previous iteration) is then subtracted and the instrumental magnitude is measured by fitting the PSF model to the star's central pixels within three FWHMa, where the S/N is higher. The sky beneath the star is then updated by removing the star itself and measuring the values in an annulus starting at two pixels from the peak (to avoid the large residuals of the core) out to the edge of the PSF model.

\subsubsection{Photometric calibration}\label{sec:calibrationover}

The photometric calibration of the instrumental magnitudes is performed with respect to the 2MASS photometric system. However, we can not use the 2MASS Point Source Catalog directly for the calibration since it has only about 150 stars in our field that are, for the most part, too bright for our exposures. Therefore, we use an auxiliary catalog that has been obtained by one of the authors (PBS) from 256 seeing-limited archival images of NGC 1851 (median FWHM of 1.32\arcsec) taken with the NEWFIRM instrument on the Blanco telescope on Cerro Tololo (Chile), has a depth intermediate between our Gemini images and the 2MASS catalog and is calibrated to 2MASS. By matching the stars in the auxiliary catalog to the corresponding stars in our Gemini dataset, we are able to determine the zeropoints of each chip in each exposure in both bands by using the technique described in Section~\ref{sec:calibration}.

\subsubsection{Quality control criteria}\label{sec:quality}

The final step is to clean the calibrated J and K$_\mathrm{s}$ catalogs of spurious objects, extended objects, or poor photometry. To do this, we use the various photometric parameters calculated by DAOPHOT. Specifically, we use: $\chi$, the goodness of the photometric fit that is measured by the ratio between the fitting residuals and the expected scatter caused by noise only; sharpness, which is derived from the difference in width between the object and the PSF (objects that are too small are usually bad pixels or cosmic rays, objects that are too large are usually galaxies); the magnitude error, which is a combination of the estimated random noise and the fitting residuals (weighted toward the bigger of the two).

\begin{figure}
\gridline{\fig{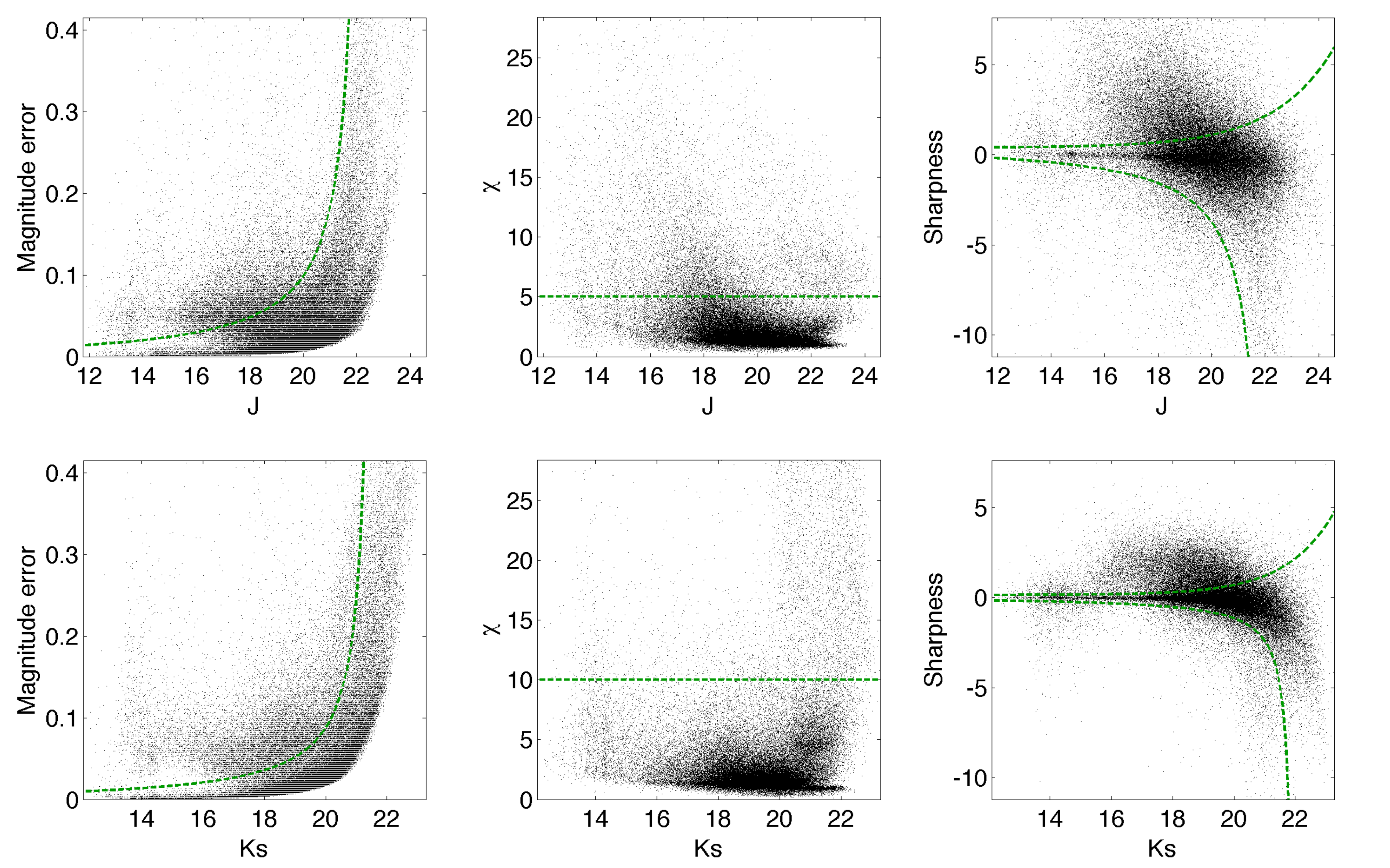}{0.55\textwidth}{} \fig{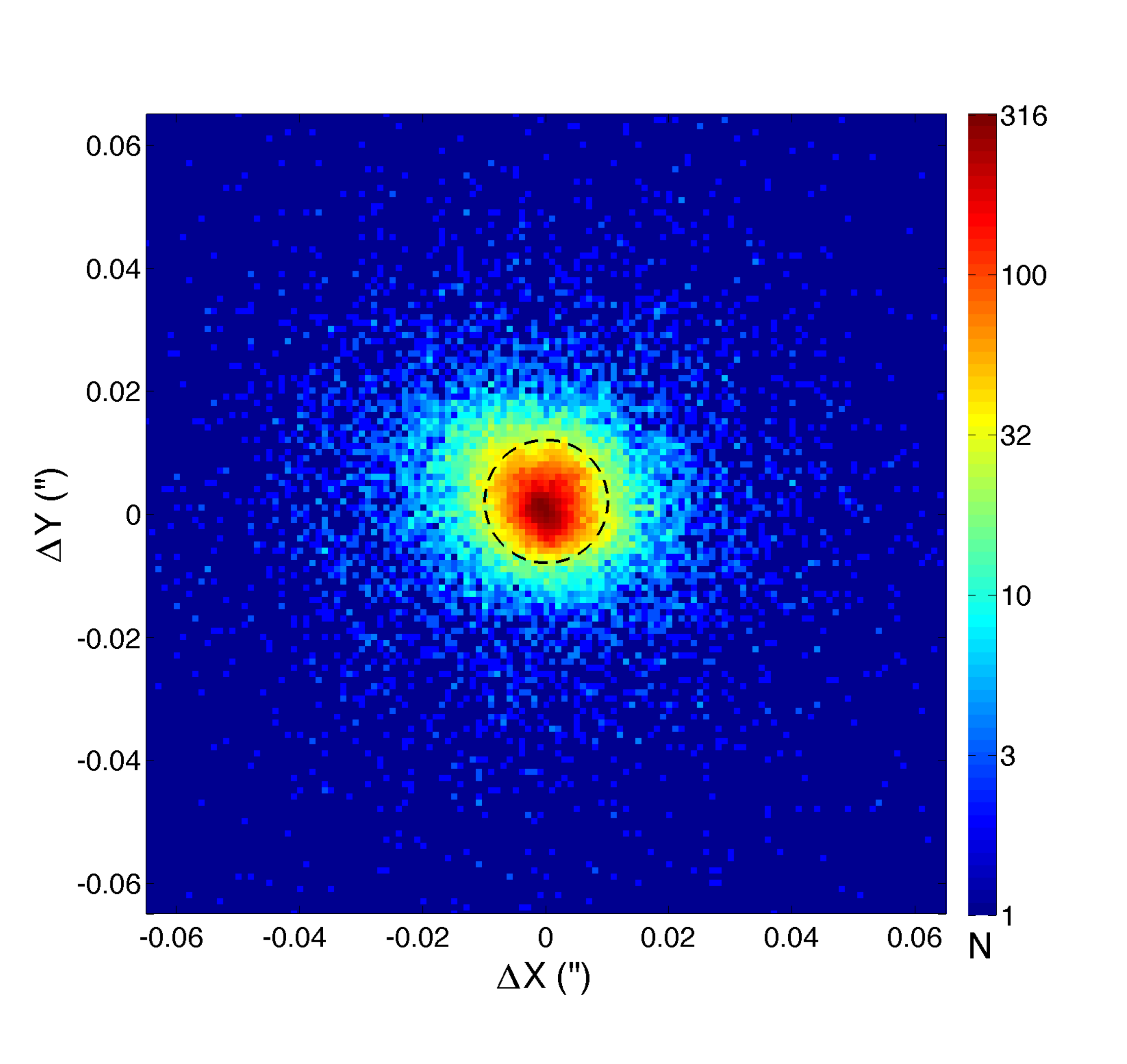}{0.35\textwidth}{}} 
\caption{\emph{Left:} Distribution of the three quality control parameters discussed in the text, for both the J and K$_\mathrm{s}$ bands. Green lines show the cuts used to remove poor detections, spurious objects and poor photometry. \emph{Right:} Two dimension distribution (on a logarithmic scale) showing the difference in position of matched objects between the J and K$_\mathrm{s}$ bands. We only keep objects that lie within the dashed circle with a radius of 0.01\arcsec .\label{fig:cmdclean}}
\end{figure}

The left panels of Figure~\ref{fig:cmdclean} show the cuts applied to these three parameters in both bands. The right panel of Figure~\ref{fig:cmdclean} shows the two dimensional distribution of the difference in position of matched objects between the two bands. We remove from the final catalog those stars with more than 0.01\arcsec\, of discrepancy, corresponding to half a pixel. Thus, our final catalog contains only high quality detections in both bands.

\subsection{Optimization of the photometry}

The preceding subsection provided a high level overview of the methodology for obtaining the final photometry from our processed images. We have followed a standard procedure for PSF photometry that has been implemented on optical and near IR observations of stellar fields for many years. However, science quality MCAO observations of crowded fields are a relatively new capability and a major focus of this work has been to understand what modifications and subtleties are required at various stages in the process to have confidence that the final photometry is optimal. In particular, our analysis in Section~\ref{sec:performance} demonstrates that the images still have significant distortions present that change the shape of the PSF across the field and between exposures in a complex fashion. What follows is a discussion of what we have identified as critical steps and decisions in the method described in the previous subsection, to obtain the highest quality photometry. Specifically:

\begin{enumerate}
\item Size of the PSF model
\item Variability of the PSF
\item Selection of the sky annuli
\item Minimizing contamination light in the PSF stars
\item Independent profile fitting
\end{enumerate}

Most of these items are judged by science-based metrics based on the quality of the final color magnitude diagram, instead of using parameters like FWHM and Strehl ratio that are difficult to measure and do not translate easily into an impact for the science. Calibration is discussed independently in Section~\ref{sec:calibration}. In this discussion we show results stemming from the analysis in the K$_\mathrm{s}$ band, but we note similar results with the J filter.

\subsubsection{Size of the PSF model}\label{sec:psfrad}

The PSF model in DAOPHOT is defined within a radius (termed the ``PSF radius'') measured from the centroid. In the case of no variation in the PSF across the field, this aperture will contain the same fraction of flux for every star. In this way, the flux measured within the PSF radius for each star can be translated into a total flux via an additional term in the photometric zeropoint measured during the calibration (Section~\ref{sec:calibration}). In the case of a spatially variable PSF, the PSF radius should ideally contain most of the spatially variable component, so that the flux within the aperture is still a constant fraction of the total. Ideally, the PSF radius should also be large enough to include most of the real PSF. In this way, when the star is subtracted during the profile fitting, all of its flux is removed from the image and it does not contaminate the profile of nearby stars.

Most of the spatial variability in the PSF of MCAO images is caused by an inhomegeneous correction applied over the field of view. The maximum spatial frequency of the aberrated wavefront that an MCAO system can correct is $f_{c}=(2d)^{-1}$, where $d$  is the pitch of the highest density of actuators in the deformable mirrors projected on to the aperture of the telescope. For GeMS it is $d=0.51\,m$  \citep{bib:bec08}. The MCAO system behaves as a high-pass filter for the wavefront, such that only light within the ``control radius'', $r_{c}=\lambda /(2d)$, of any PSF is affected by the correction, leaving the wings of the profile mostly unchanged with respect to the seeing-limited PSF. Since the actuators are actually arranged on a square grid, the PSF region influenced by the adaptive optics for GeMS is a square with a side of $2r_{c}$. For GeMS, $r_{c}=22\,px$ in K$_\mathrm{s}$ band and the PSF radius used for our images should therefore, in principle, be at least this value.

\begin{figure}
\gridline{\fig{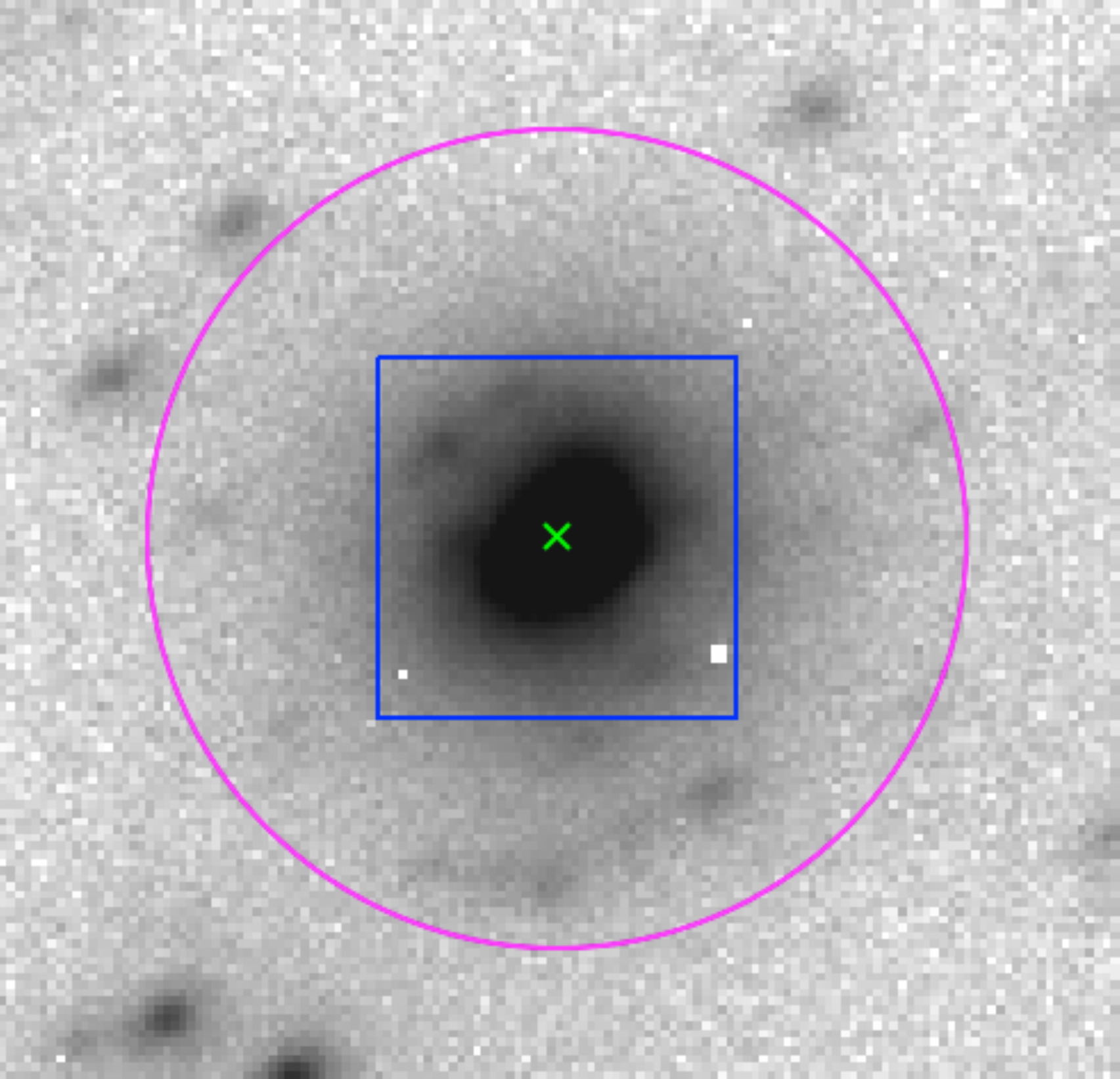}{0.3\textwidth}{} \fig{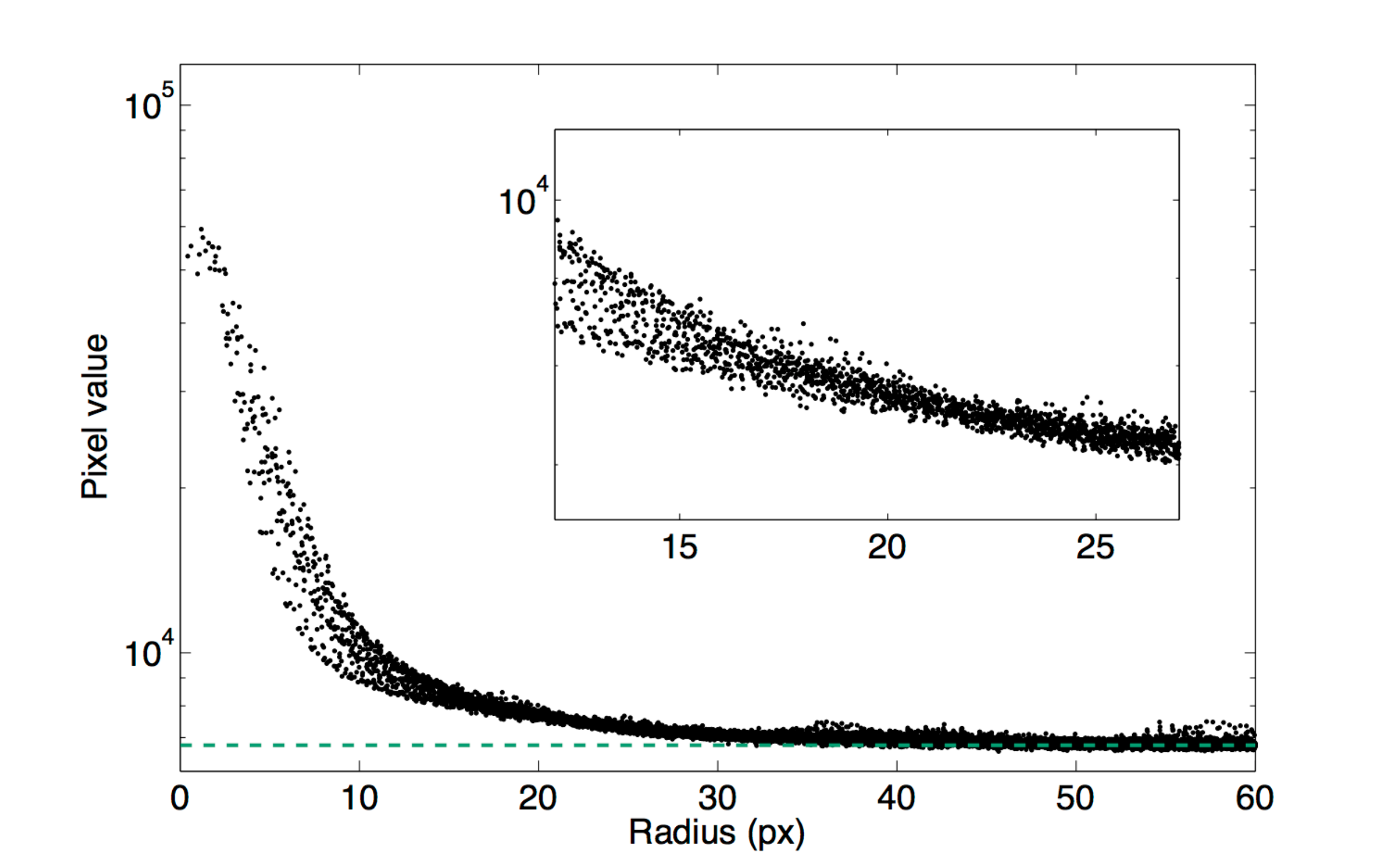}{0.65\textwidth}{}}
\caption{\emph{Left}: a bright star in a K$_\mathrm{s}$ long exposure. The centroid is marked with a green cross. The blue square superimposed has sides of $2r_{c}=44\,px$ and is the region within which the adaptive optics correction is applied. Shown for reference is also a magenta circle with a radius of 50 pixels. \emph{Right}: radial profile of the same star, where each point is a pixel from the image in the left panel. The estimated background level is indicated by the green dashed line. The inset panel shows the detail of the radial profile around the control distance $r_{c}=22\,px$.\label{fig:rad}}
\end{figure}

Examination of a typical PSF in our images suggests that this argument is valid. Figure~\ref{fig:rad} presents an example of a stellar radial profile from our data. The left panel shows the two dimensional image, with the centroid marked by the green cross and a blue square superimposed with sides of $2r_{c}=44\,px$. Shown for reference is also a circle with a radius of 50 pixels. The right panel shows the radial profile of the PSF, where each dot corresponds to a pixel in the image. Firstly, we see that the profile does not level off to a constant value until around 50 pixels. Secondly, the spread in the profile at small radius is significant and increases with decreasing radius. This is to be expected if the PSF is significantly distorted (e.g., has significant ellipticity). As the inset panel shows, however, the spread in the points becomes broadly constant at a radius $\sim$20 pixels. This is in line with expectations from our argument above and confirms that the asymmetric part of the PSF is mostly within the control radius.

\begin{figure}
\gridline{\fig{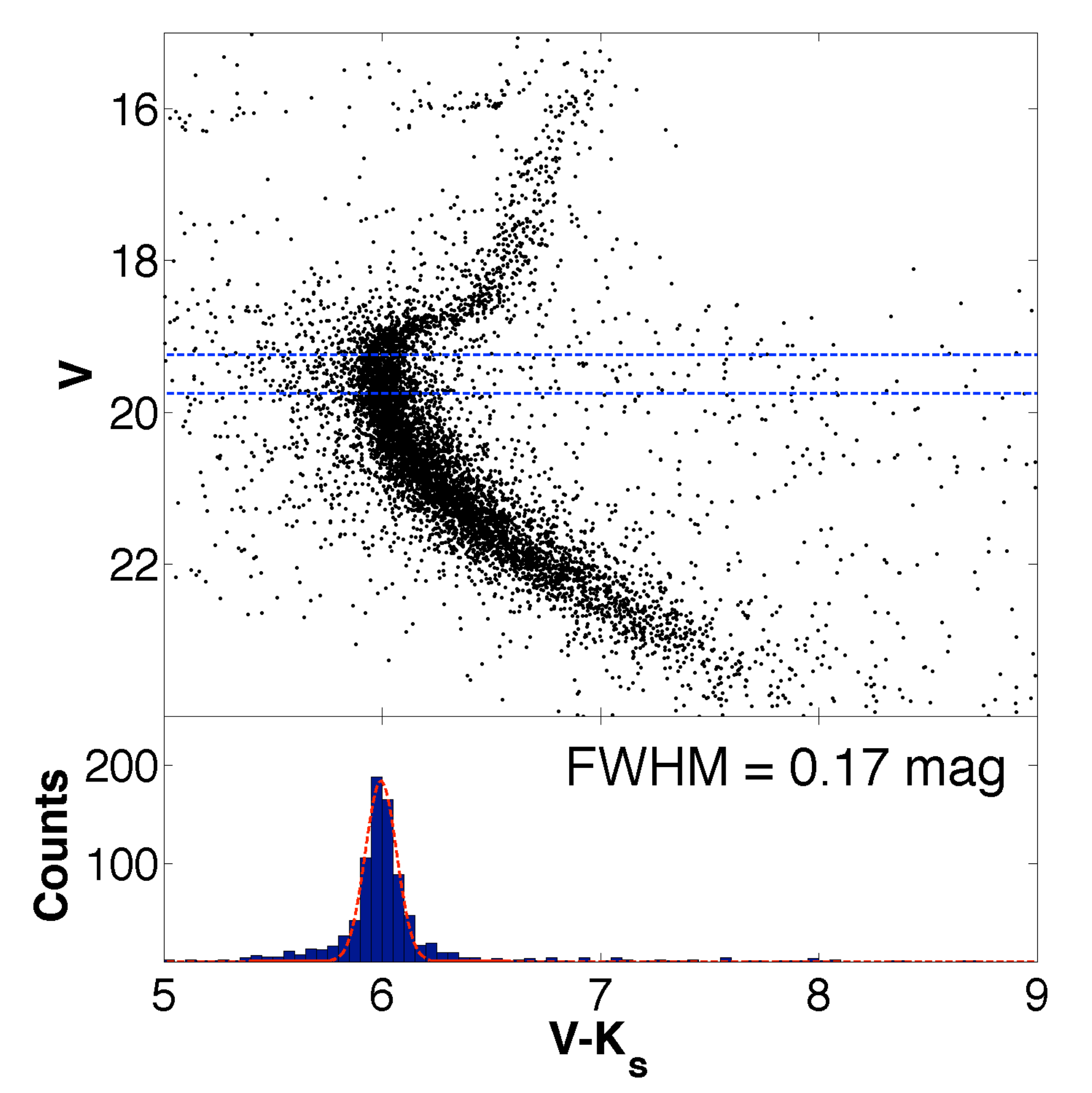}{0.3\textwidth}{PSF radius: 8 px} \fig{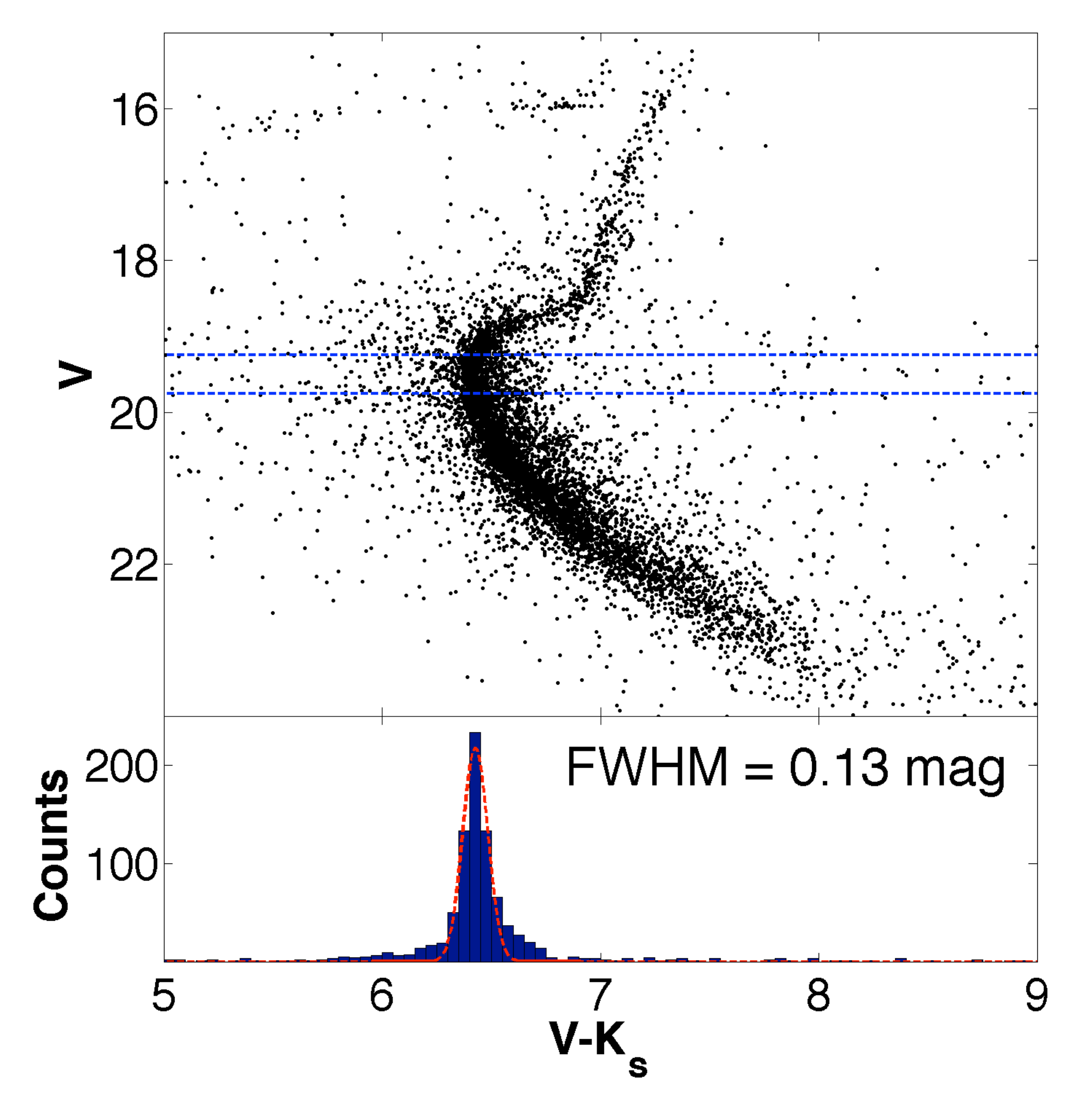}{0.3\textwidth}{PSF radius: 16 px} \fig{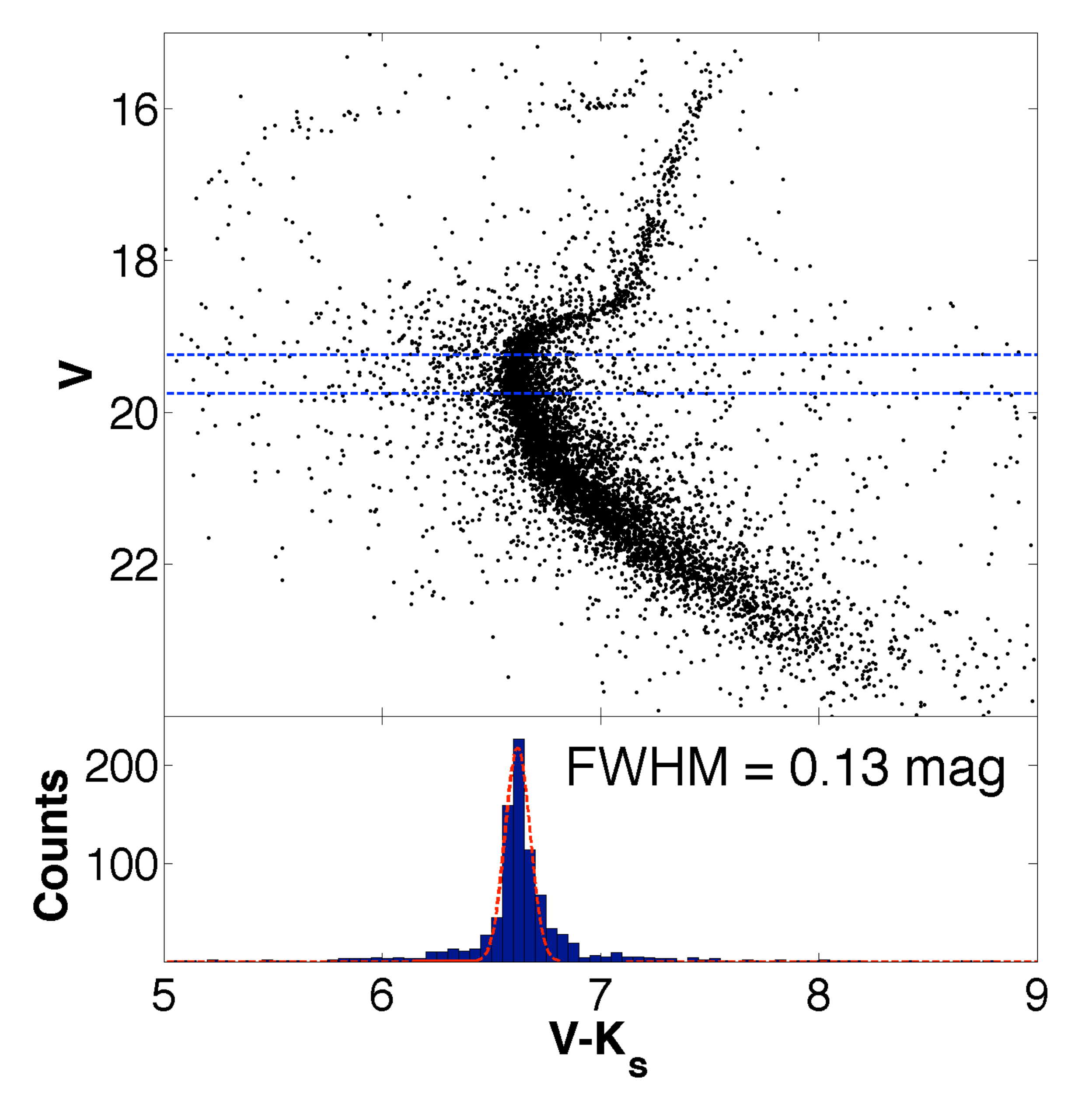}{0.3\textwidth}{PSF radius: 24 px}}
\gridline{\fig{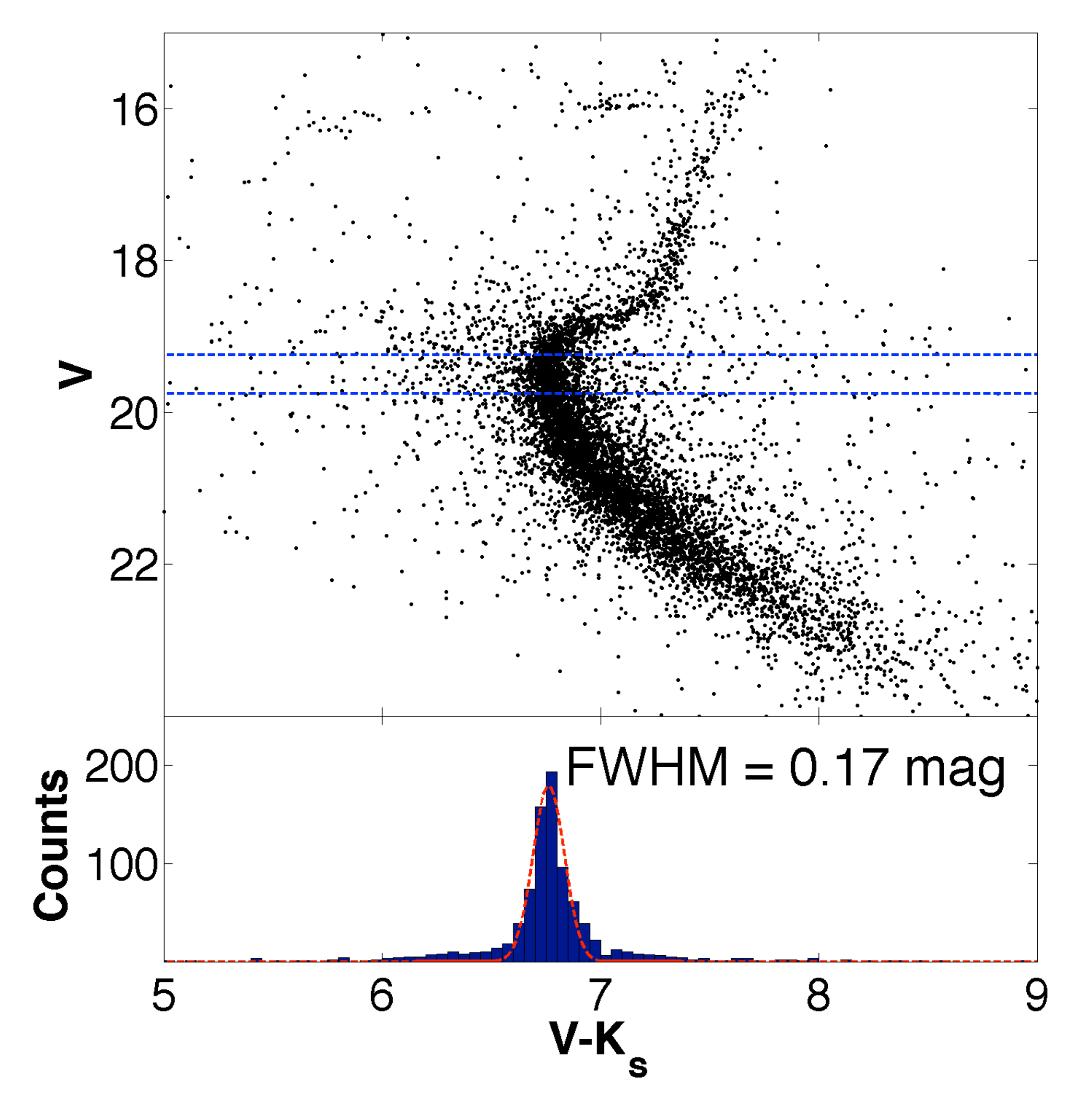}{0.3\textwidth}{PSF radius: 32 px} \fig{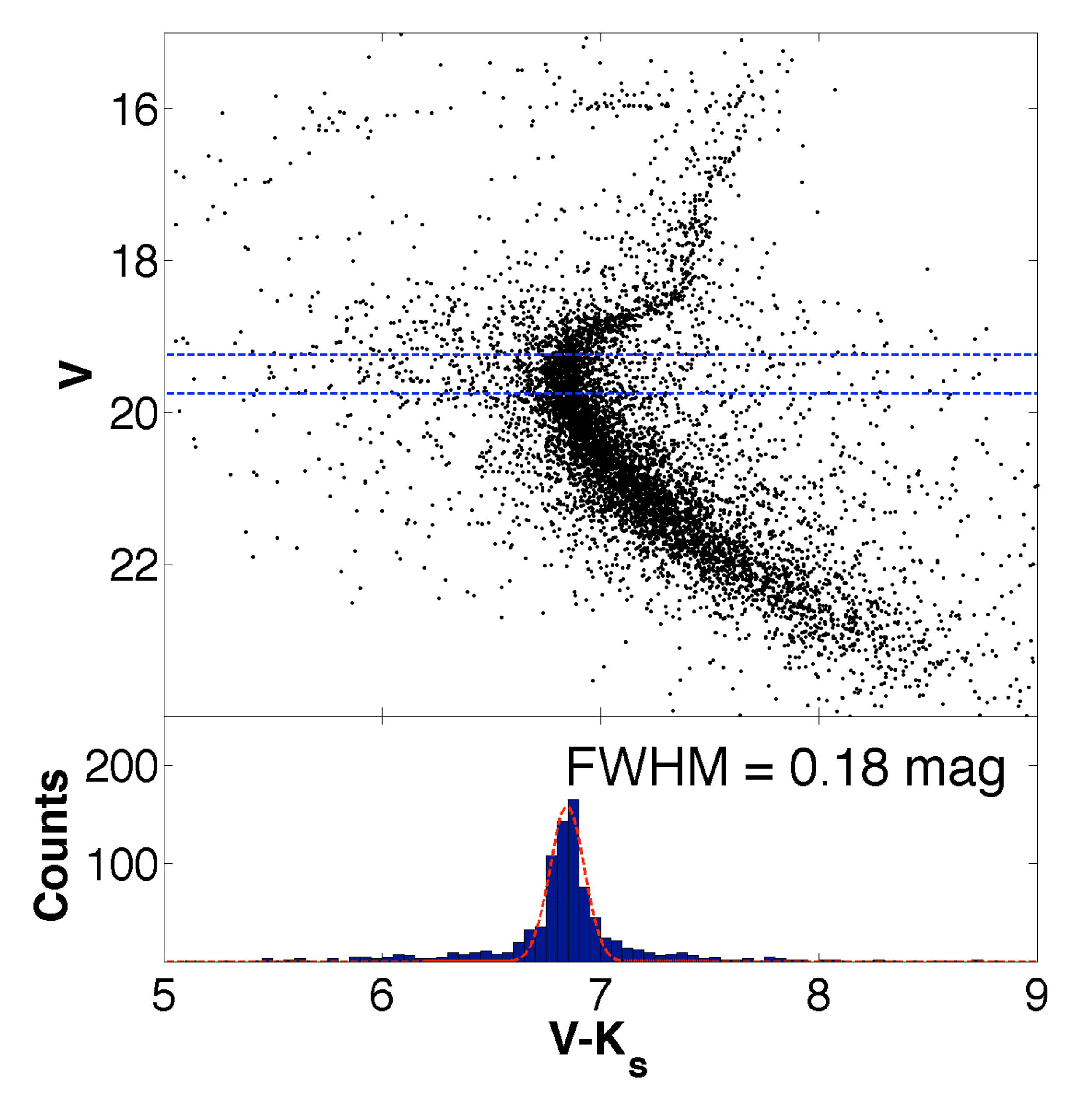}{0.3\textwidth}{PSF radius: 40 px} \fig{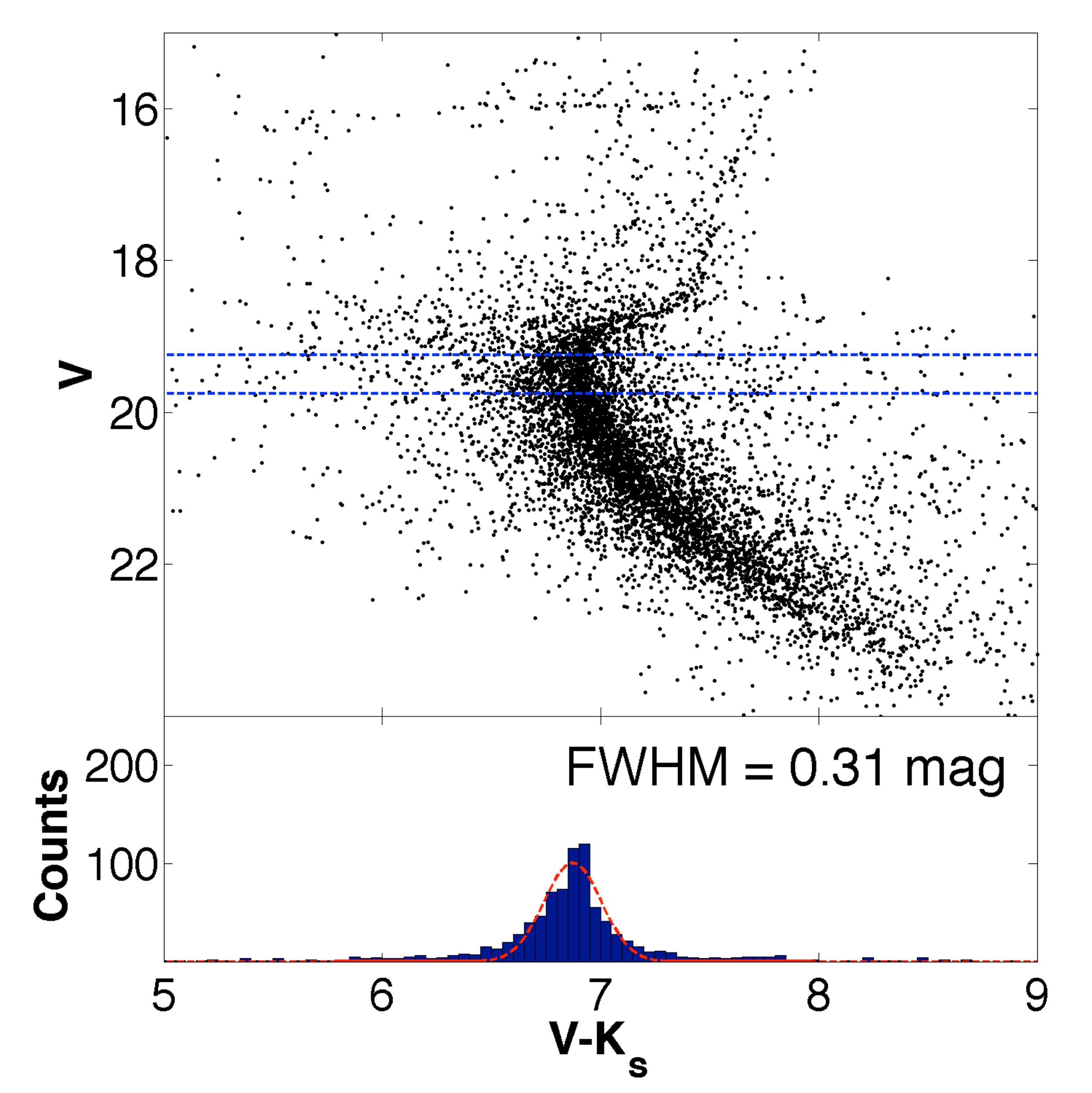}{0.3\textwidth}{PSF radius: 48 px}}
\caption{Uncalibrated CMDs of a K$_\mathrm{s}$ image using different PSF radii and the HST V photometry. The two blue horizontal lines include the stars around the brightness of the MSTO used to measure the width of the sequence. The lower panels show the histogram of stellar colors in this region and the red line is the best fit Gaussian.\label{fig:psfradcmd}}
\end{figure}

We have tested the effect of modifying the PSF radius on the resulting photometry from our images. We have produced several different photometric catalogs, where each version uses a different value for the PSF radius. All other steps and parameters in the photometric analysis are identical. We have combined the resulting instrumental magnitudes with the high precision photometry from the HST/ACS catalogue of NGC 1851 in the V band (F606W) presented in \cite{bib:sarajedini07}. Since this catalog is extremely precise, examination of the relative quality of the resulting optical-NIR CMDs allows us to develop a science-based metric to judge the quality of our NIR photometry. Figure~\ref{fig:psfradcmd} shows the CMDs corresponding to each version of our K$_\mathrm{s}$ photometry (a similar analysis was made for the J band). We use the width of the main sequence turn-off (MSTO), in the region where the sequence is nearly vertical, as a metric on the quality of the photometry. We quantify the width of this sequence via the FWHM of a Gaussian curve fitted to the histogram of the color distribution of the stars. Photometric errors will act to increase the measured width of the main sequence, even in the presence of multiple populations (as is the case for NGC 1851, see Section~\ref{sec:cmds}), thus higher quality photometry will produce a smaller FWHM. We note that Figure~\ref{fig:psfradcmd} uses uncalibrated K$_\mathrm{s}$ magnitudes. Therefore, each CMD is shifted horizontally with respect to the others since the instrumental magnitudes are ``fainter'' when a smaller PSF radius is used.

\begin{figure}
\centering
\includegraphics[width=0.6\textwidth]{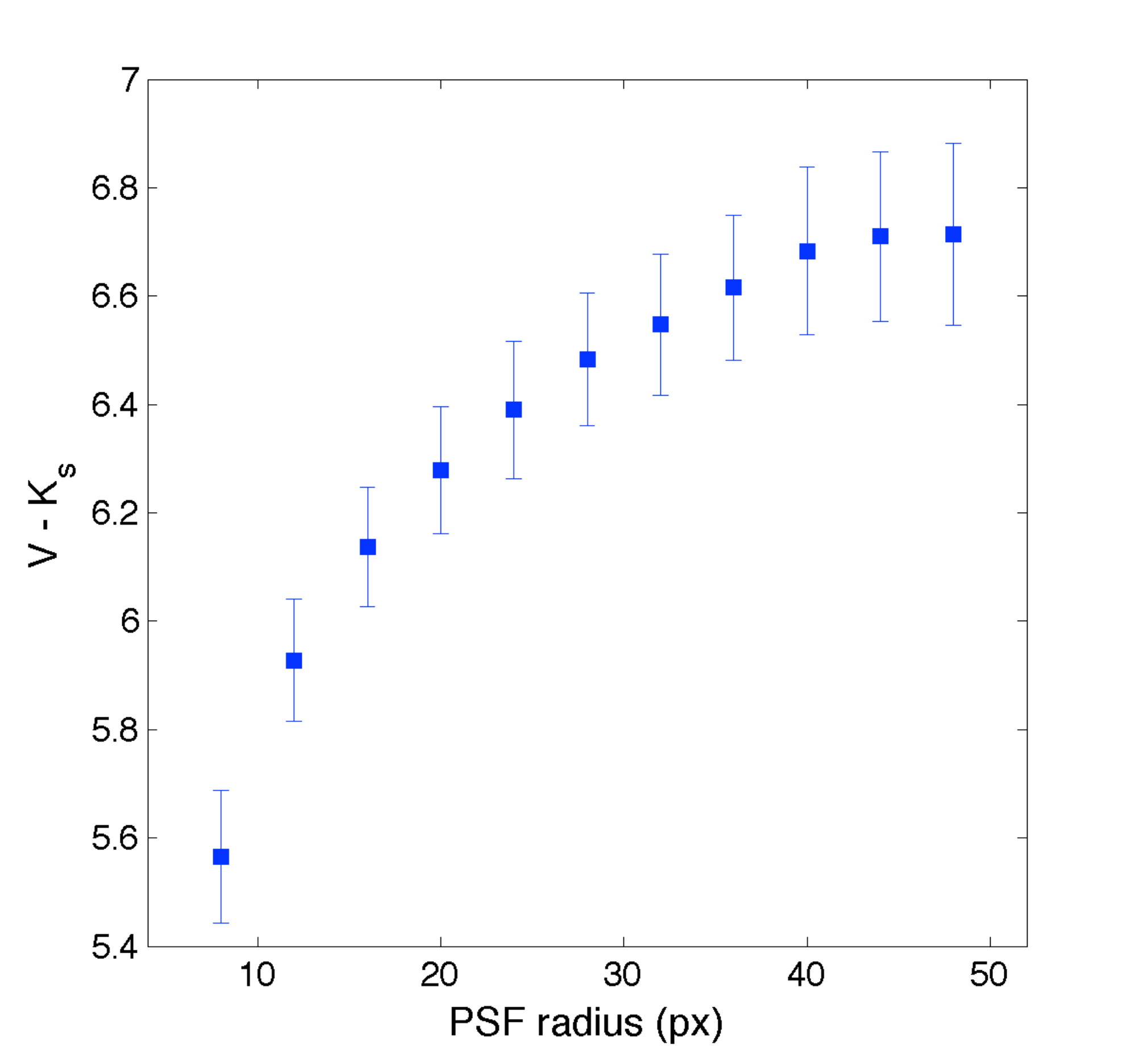}
\caption{Uncalibrated color of the MSTO as a function of the PSF radius in a K$_\mathrm{s}$ image. The width of the sequence is represented by the error bars.\label{fig:psfradstat}}
\end{figure}

The histograms in Figure~\ref{fig:psfradcmd} demonstrate that the width of the MSTO in the CMD  is minimized for PSF radii of $\sim\hbox{16--24}\, px$ (a PSF radius of 24 pixels also appears to produce a narrower red giant branch). We find that this result holds for all exposures, independent of the quality of the AO correction applied. It is clear, therefore, that the optimal PSF model radius for our analysis is close to the value of the control radius $r_{c}$, which itself is considerably smaller than the ``overall'' size of the PSF. Figure~\ref{fig:psfradstat} plots the uncalibrated color of the MSTO as a function of the PSF radius, and it is clear that the color does not change significantly (i.e., the instrumental K$_\mathrm{s}$ magnitude is constant) for a PSF radius between 40 and 50 pixels; this value is also consistent with our discussion relating to Figure~\ref{fig:rad}.

\begin{figure}
\gridline{\fig{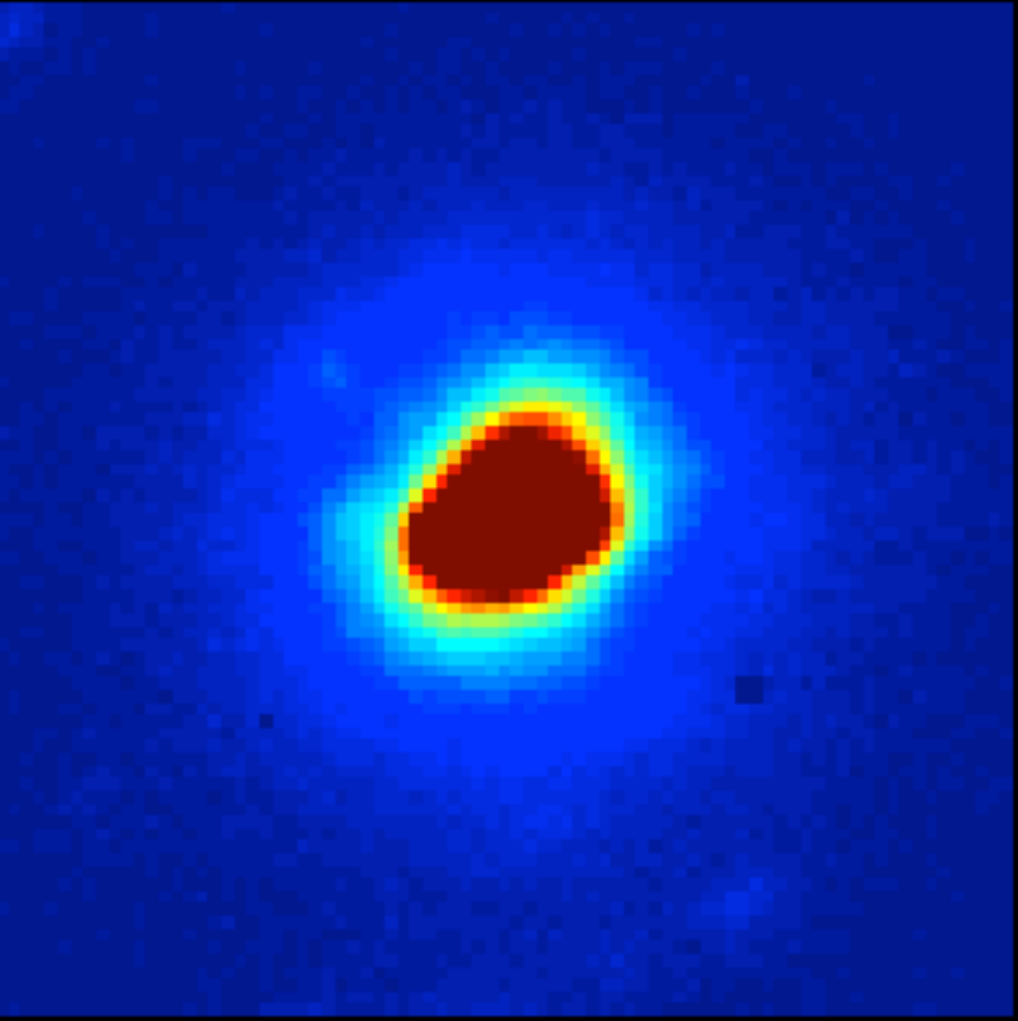}{0.3\textwidth}{} \fig{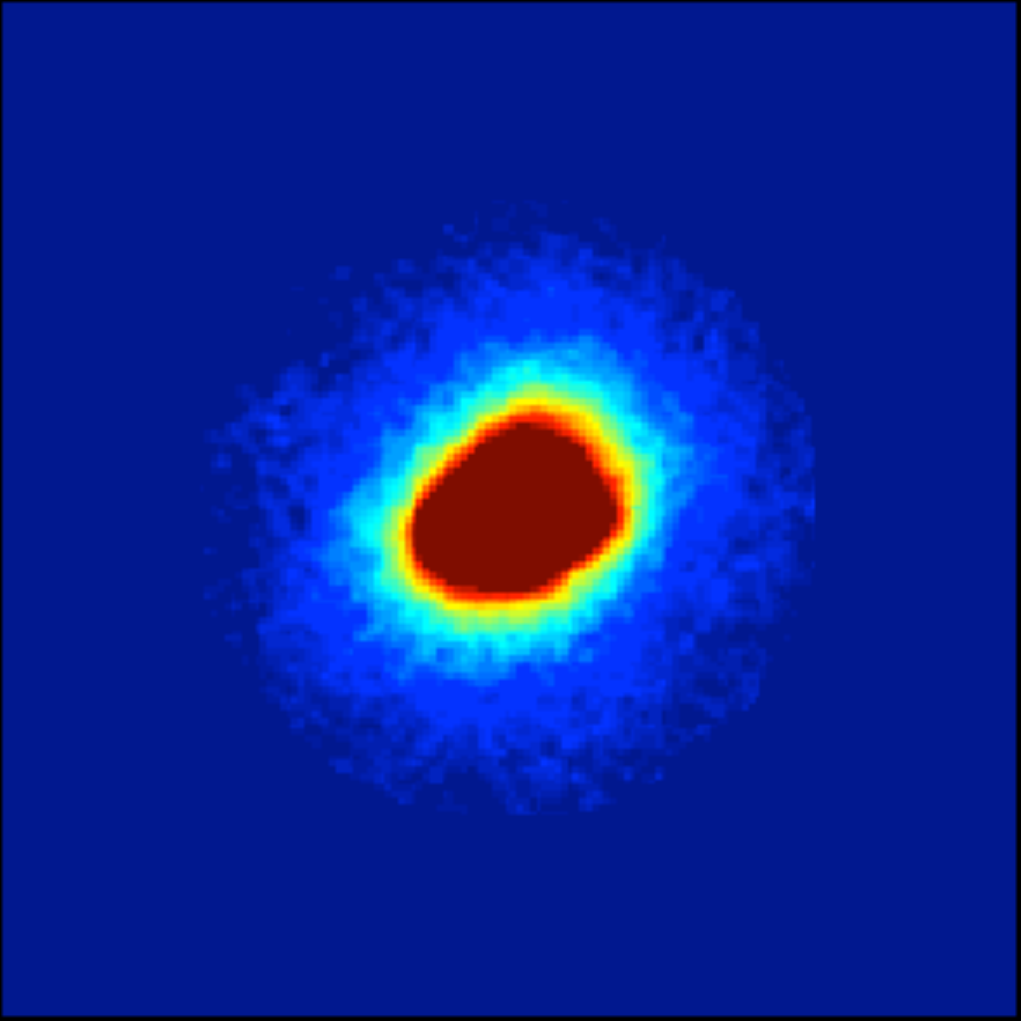}{0.3\textwidth}{} \fig{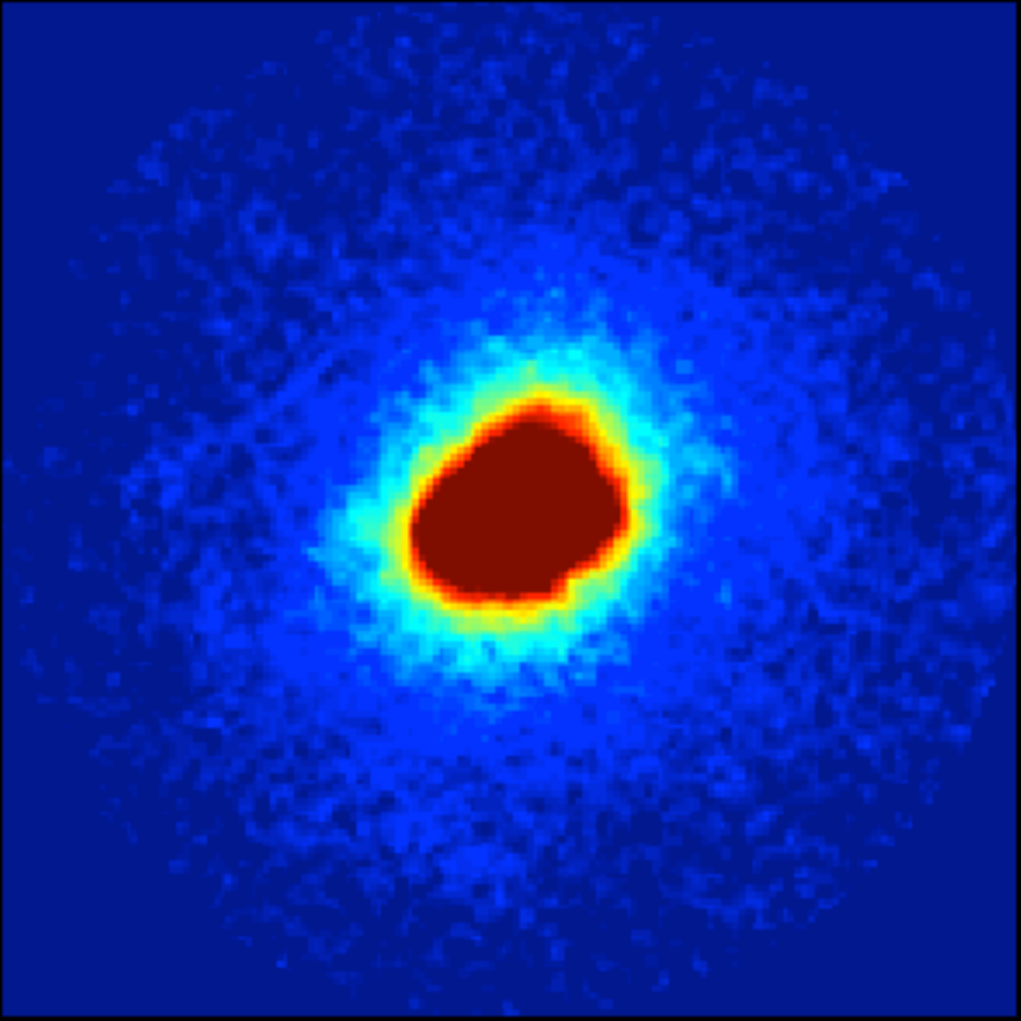}{0.3\textwidth}{}}
\caption{\emph{Left panel}: a bright star in a K$_\mathrm{s}$ exposure. \emph{Second and third panels}: the PSF model of the star adopting a PSF radius of 24 and 40 pixels, respectively. The color scale is linear with a cutoff at 10\% of the peak.\label{fig:psfs}}
\end{figure}

The flux outside the control radius is the same fraction for every star since all the variability in the shape of the PSF is interior to this value. Moreover, at large radii, where the intrinsic flux from the star is low, contamination due to noise, cosmic rays, and nearby stars can be significant (for example, note the presence of several structures at large radii from the center of the bright star shown in Figure~\ref{fig:rad}). Adoption of a large PSF radius means that these spurious structures are considered part of the PSF model, leading to an overall reduction in the precision of the photometry. This is illustrated in Figure \ref{fig:psfs}, where the first panel shows the PSF of a star observed in the  K$_\mathrm{s}$ filter. The second and third panels are the two PSF models for this star, where the former adopts a PSF radius of 24 pixels and the latter a radius of 40 pixels. The latter model has a large amount of small scale structure at large radii that are not intrinsic to the halo, expected to be smooth for these long exposure times. Thus, a PSF radius close to the control radius is large enough to include all the variability of the PSF but small enough to exclude the majority of spurious structure, leading to more precise photometry.

\begin{figure}
\centering
\includegraphics[width=0.3\textwidth]{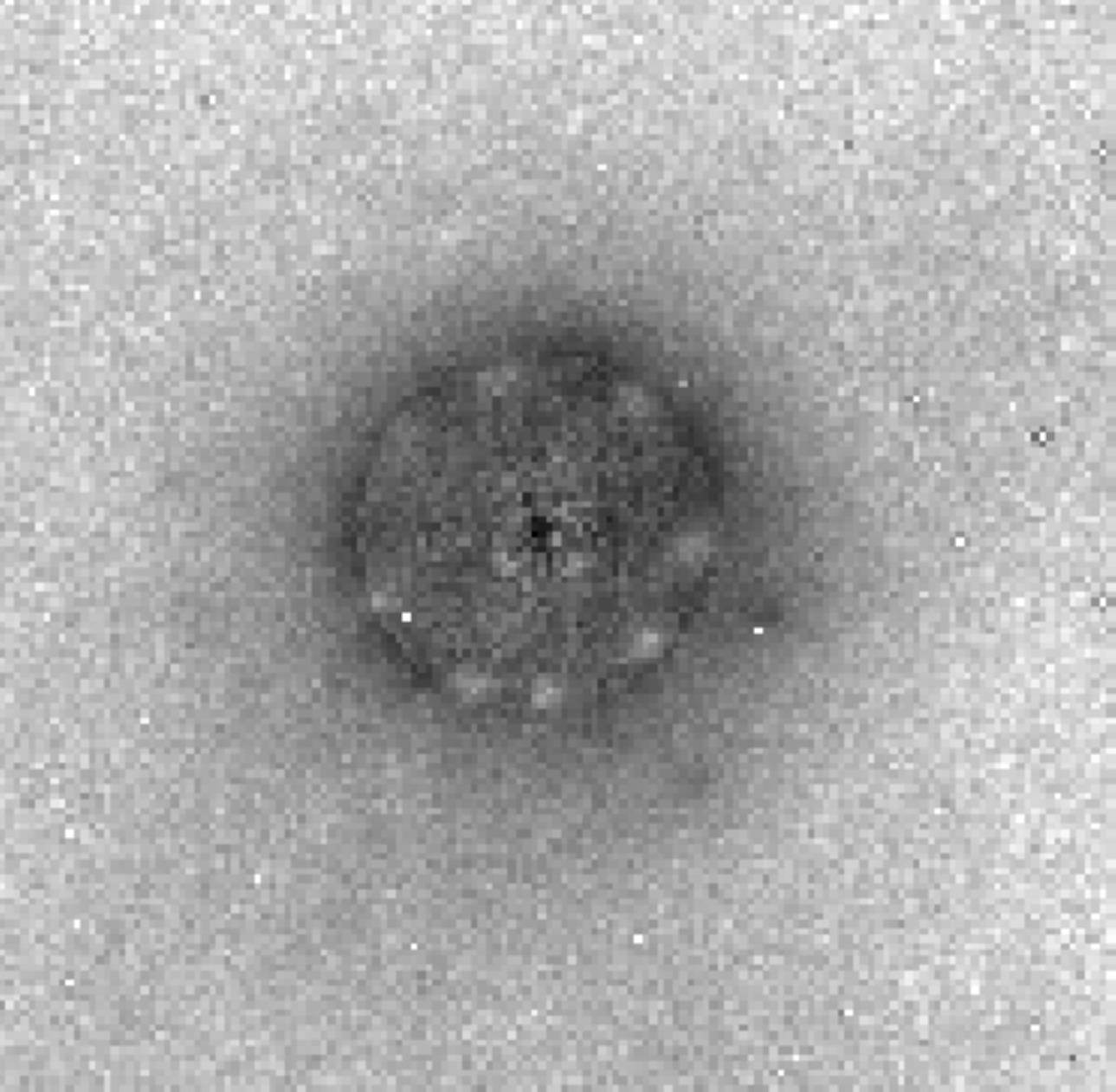}
\caption{A region of a J band long exposure where the stars have been subtracted after the profile fitting. The central bright star has left a clear halo since its PSF model used is smaller than the real PSF. The white spots along the edge of the profile are spurious detections that are automatically fitted using the PSF model.\label{fig:halo}}
\end{figure}

An unwanted side effect of using a smaller value of the PSF radius is that when a star is subtracted from the image during the profile fitting, it leaves an ``orphan'' halo with a central hole the size of the PSF diameter. The inner edge of the halo of a bright star looks like a doughnut, as shown in Figure~\ref{fig:halo}. This discontinuity in brightness at the PSF radius can lead to spurious detections, as is visible in Figure~\ref{fig:halo}. However, they are easily identified and removed from the final catalog during the quality control phase (Section~\ref{sec:quality}).

Following the considerations summarized in this section, we have chosen a PSF radius of 24 pixels for the images in both bands.

\subsubsection{Variability of the PSF}\label{sec:psfvar}

In seeing-limited and space-based images, the PSF is generally uniform across the field. Any variability that does exist is usually a result of static and quasi-static distortions introduced by the optics, with no significant variation between observations. In contrast, in MCAO data the PSF can present a complex spatial and temporal variability in terms of size and overall shape, as demonstrated in Section~\ref{sec:performance}, especially Figure~\ref{fig:elliptmaps}. Such distortions are much less over large fields than using classic AO but they still necessitate careful modeling to obtain accurate photometry.

DAOPHOT uses for the PSF a constant analytical function for all stars in combination with a two dimensional lookup table for the empirical corrections to create the model PSF. Each pixel in this grid can be set to have either the same values everywhere in the field of view or to vary spatially. In the spatially varying case, each element of the grid is associated with a set of coefficients that together describe a bivariate polynomial in x and y, the coordinates of the star in the field. The polynomial can be up to the third order, making the PSF vary linearly, quadratically or cubically. We also note that each grid element varies independently of every other one.

\begin{figure}
\fig{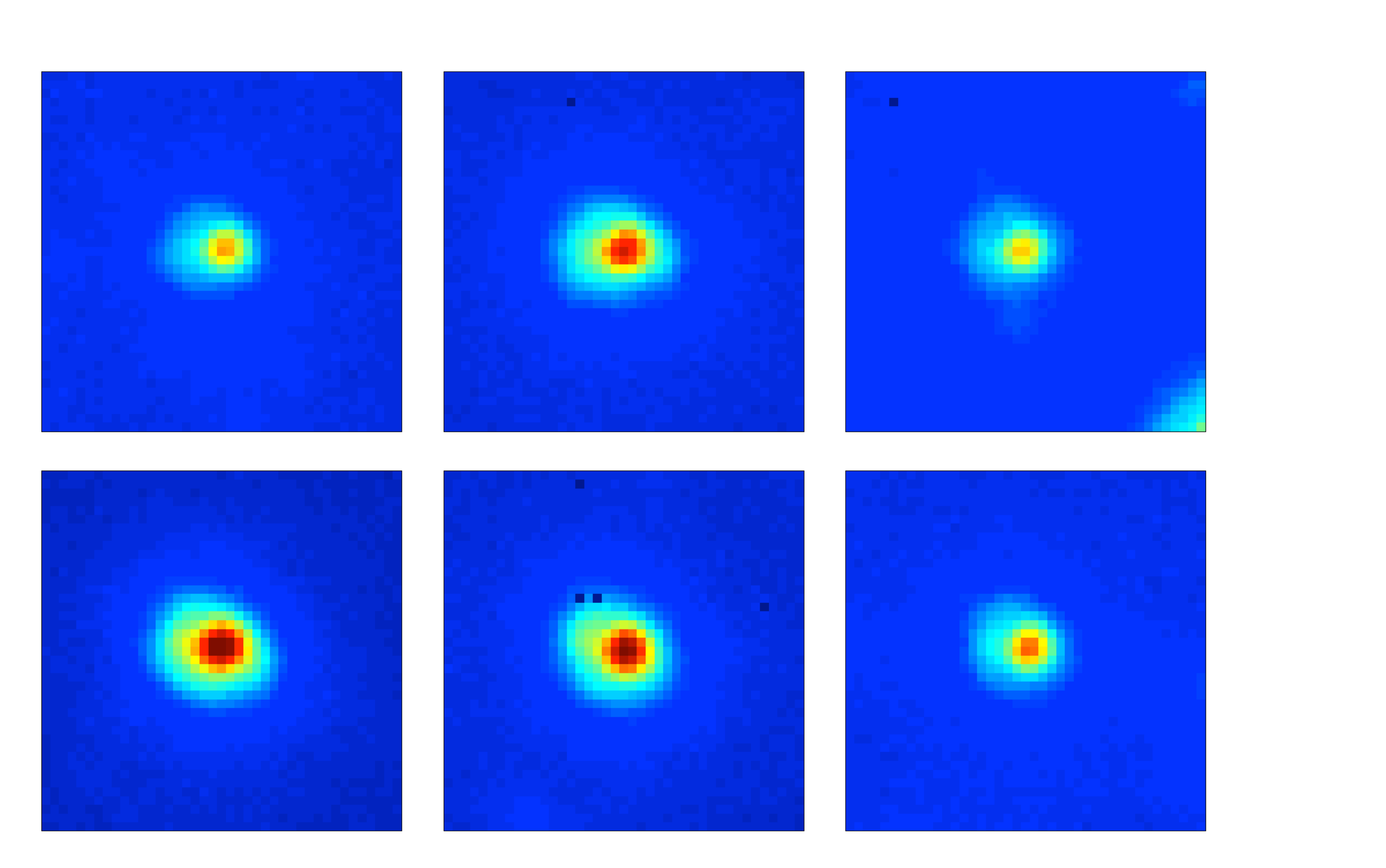}{0.45\textwidth}{Test stars}
\gridline{\fig{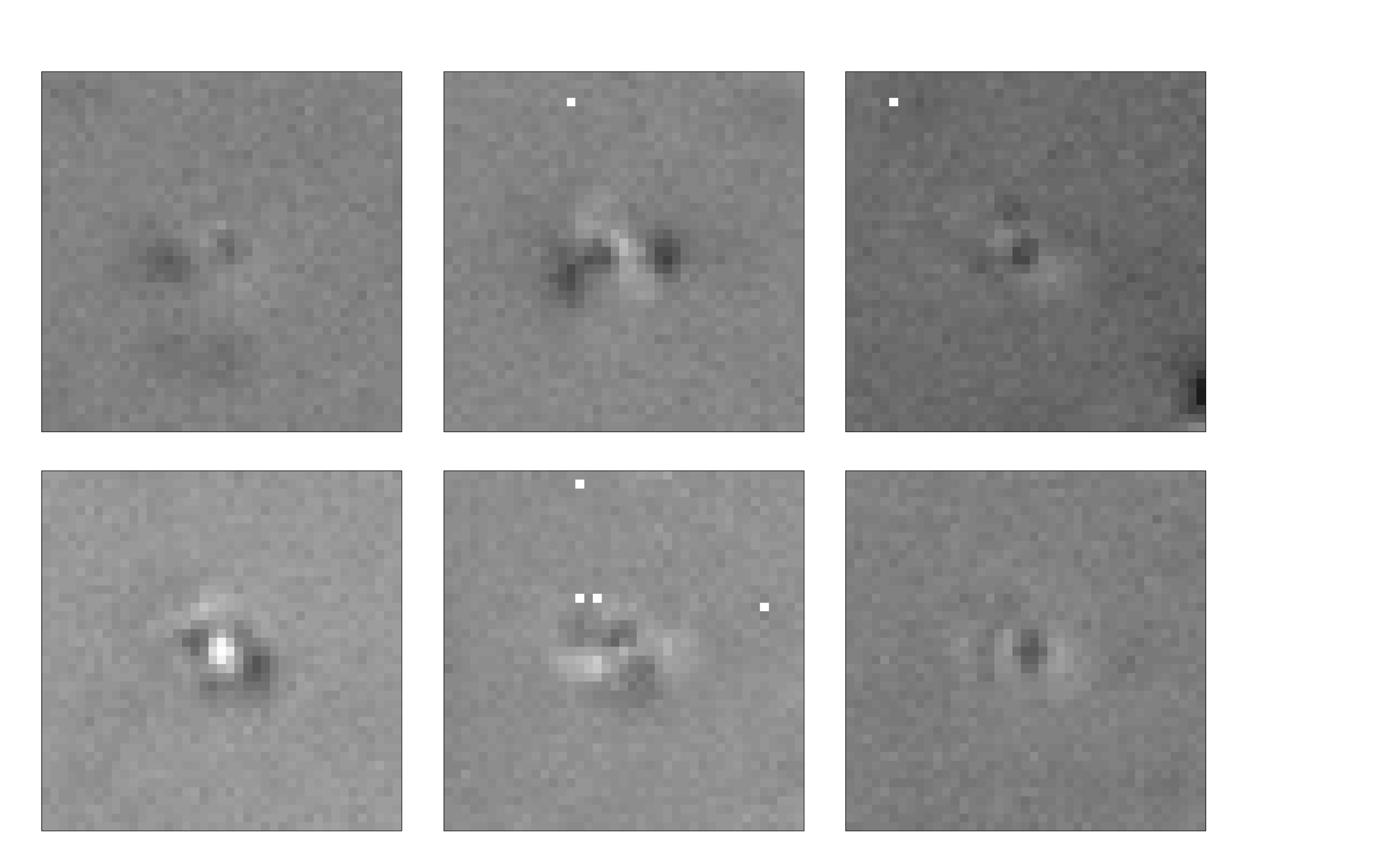}{0.4\textwidth}{Constant} \fig{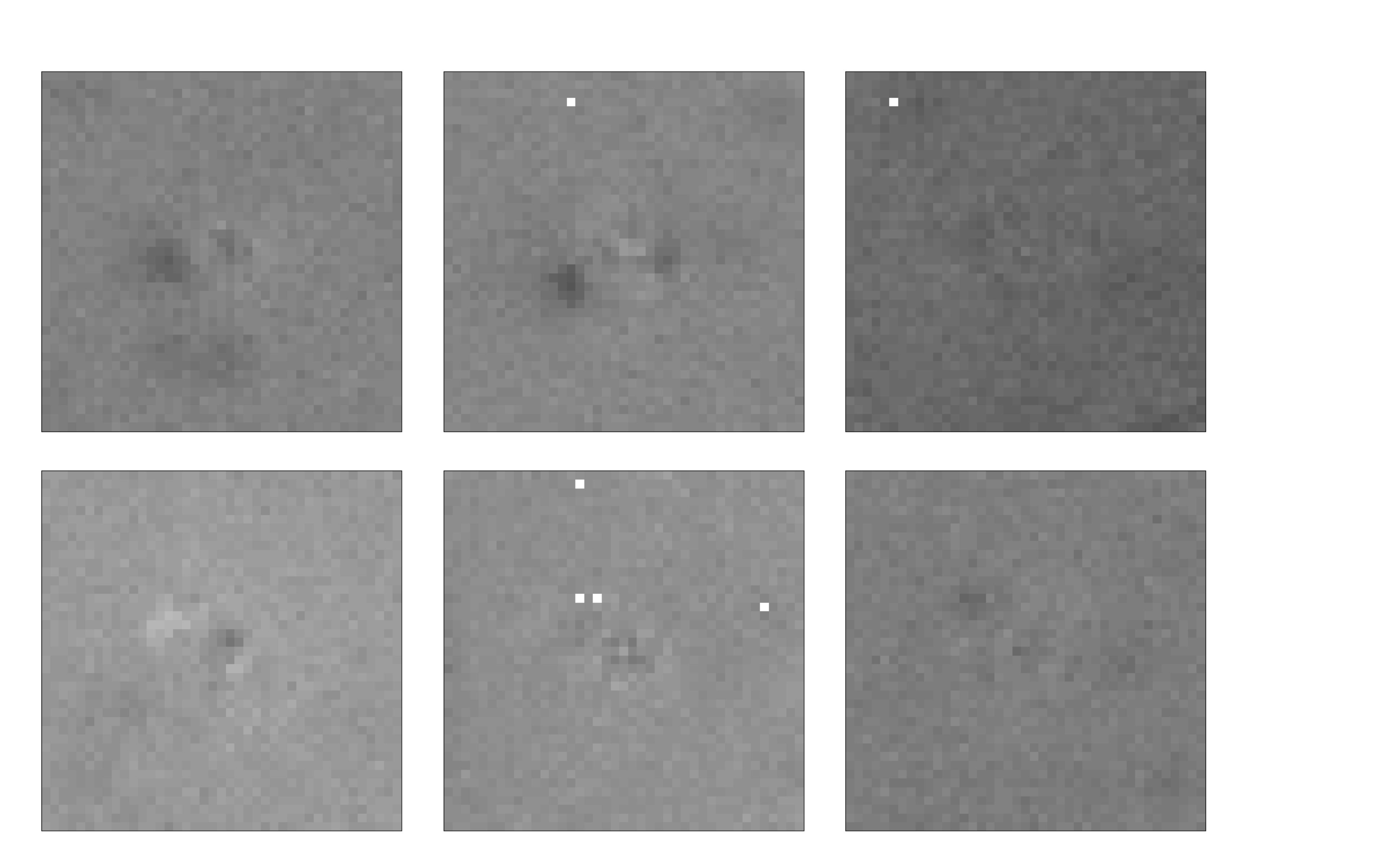}{0.4\textwidth}{Linear}}
\gridline{\fig{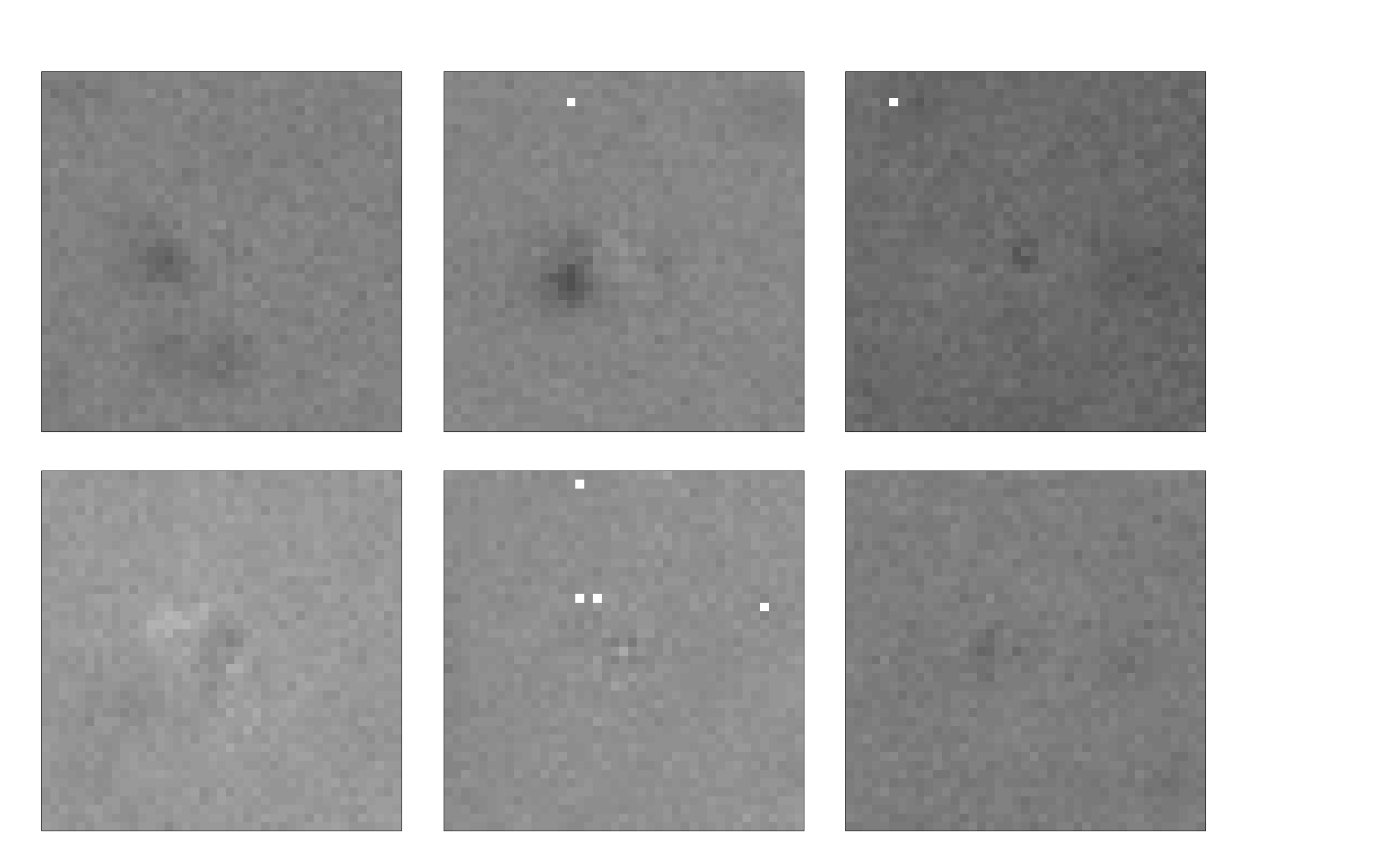}{0.4\textwidth}{Quadratic} \fig{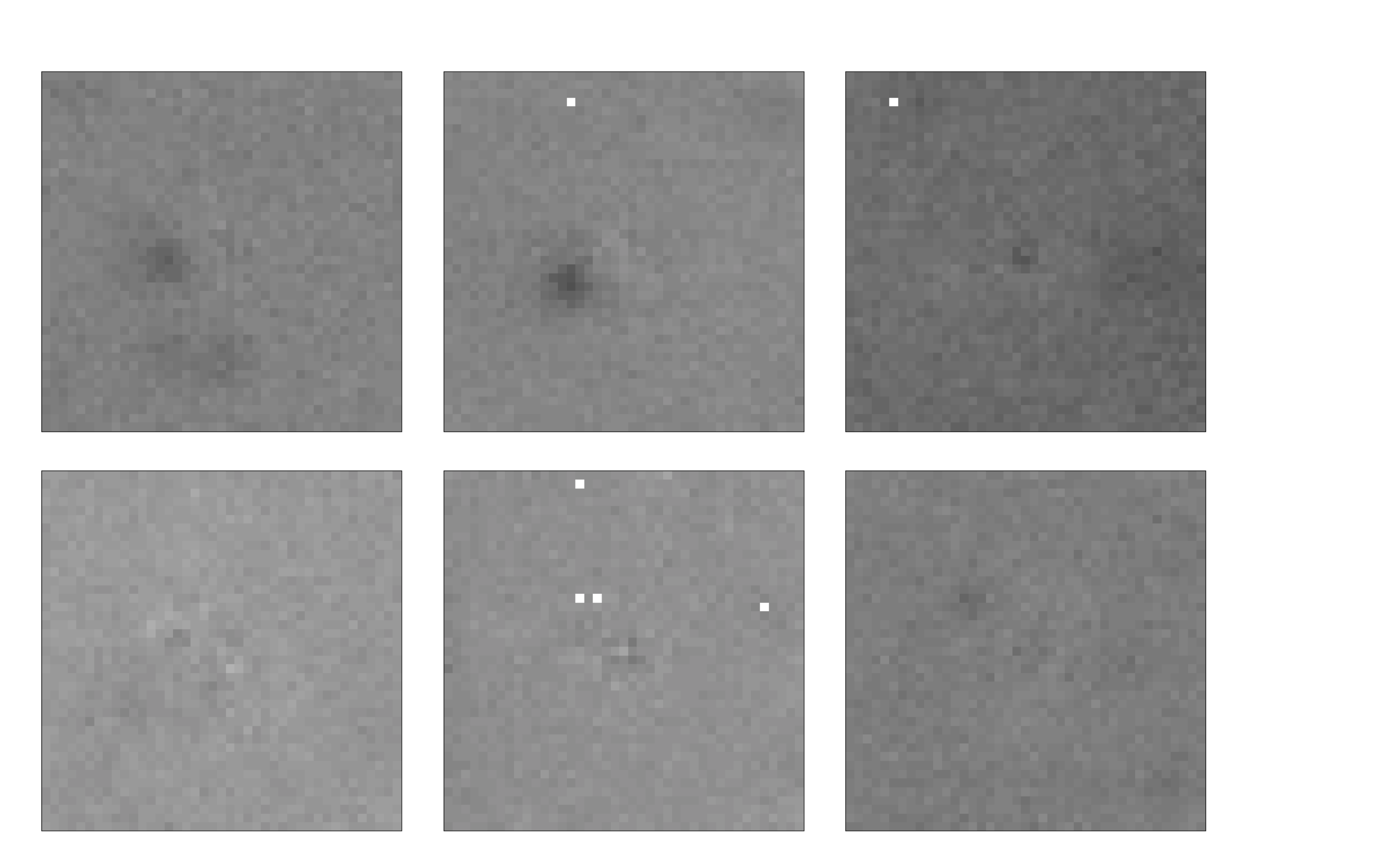}{0.4\textwidth}{Cubic}}
\caption{Profile fitting residuals of six test stars in a K$_\mathrm{s}$ image using different PSF variabilities. Most of the residuals visible in the cubic version are actually faint companion stars.\label{fig:psfvarsub}}
\end{figure}

\begin{figure}
\gridline{\fig{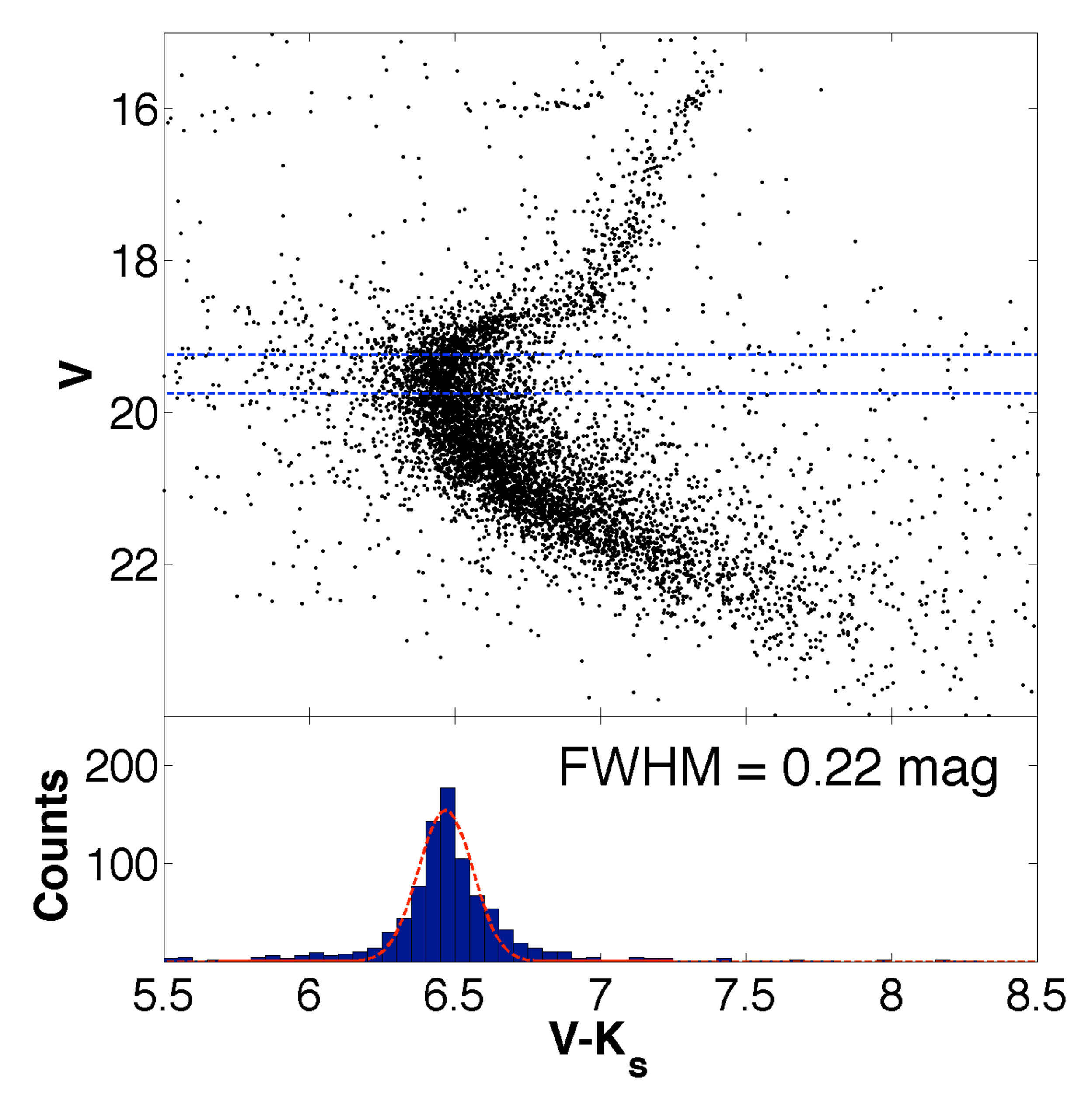}{0.3\textwidth}{Constant} \fig{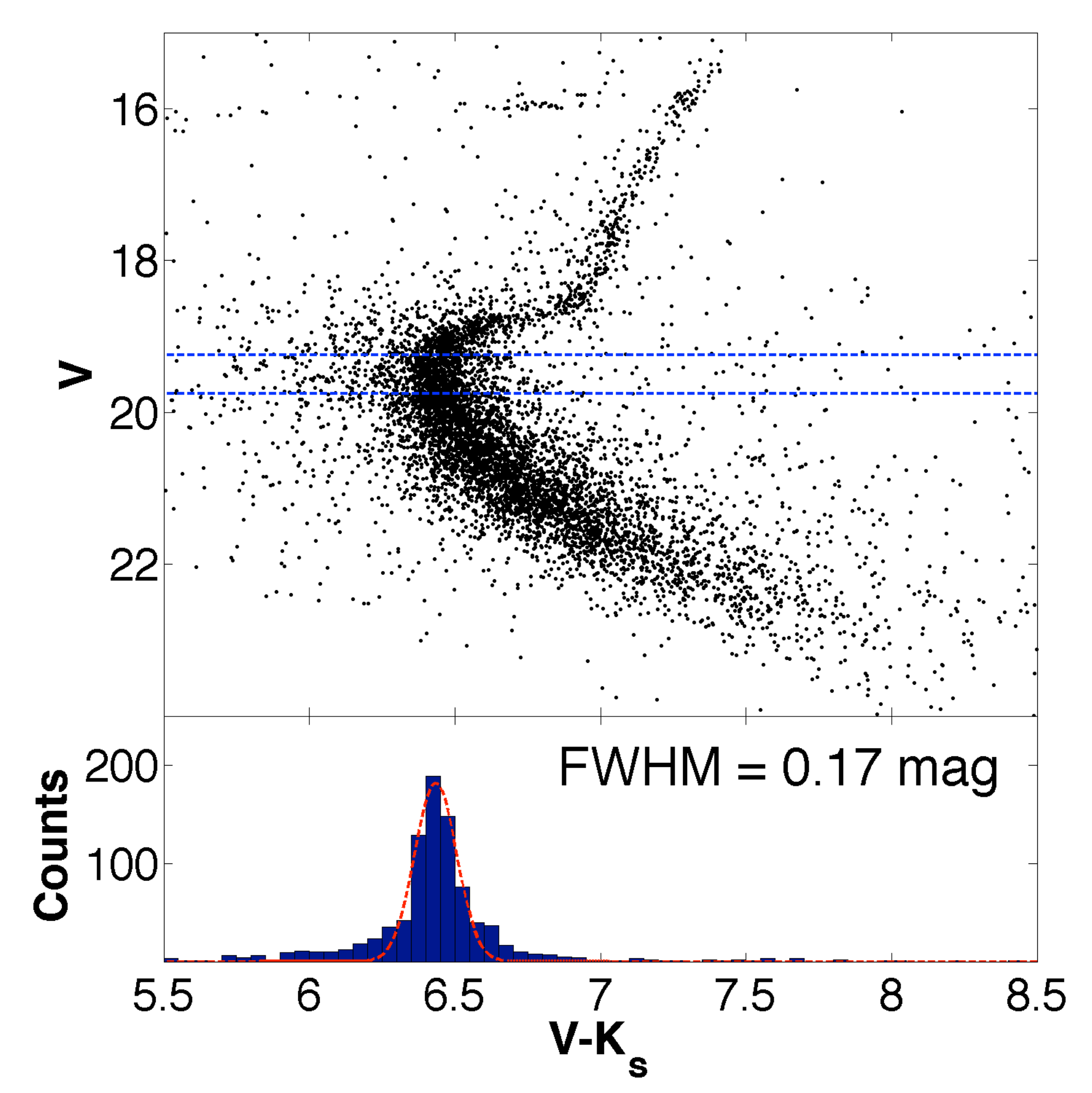}{0.3\textwidth}{Linear}}
\gridline{\fig{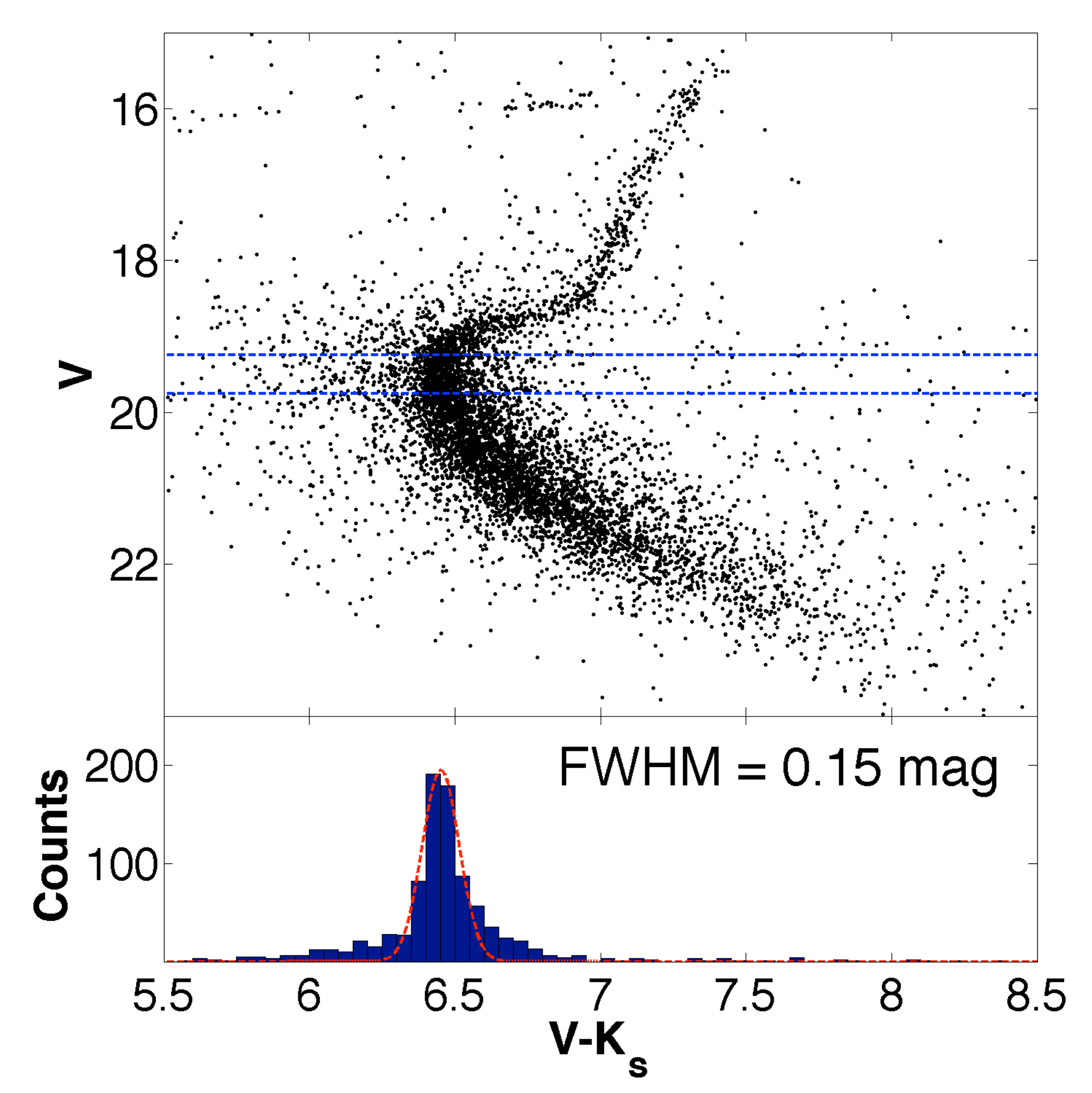}{0.3\textwidth}{Quadratic} \fig{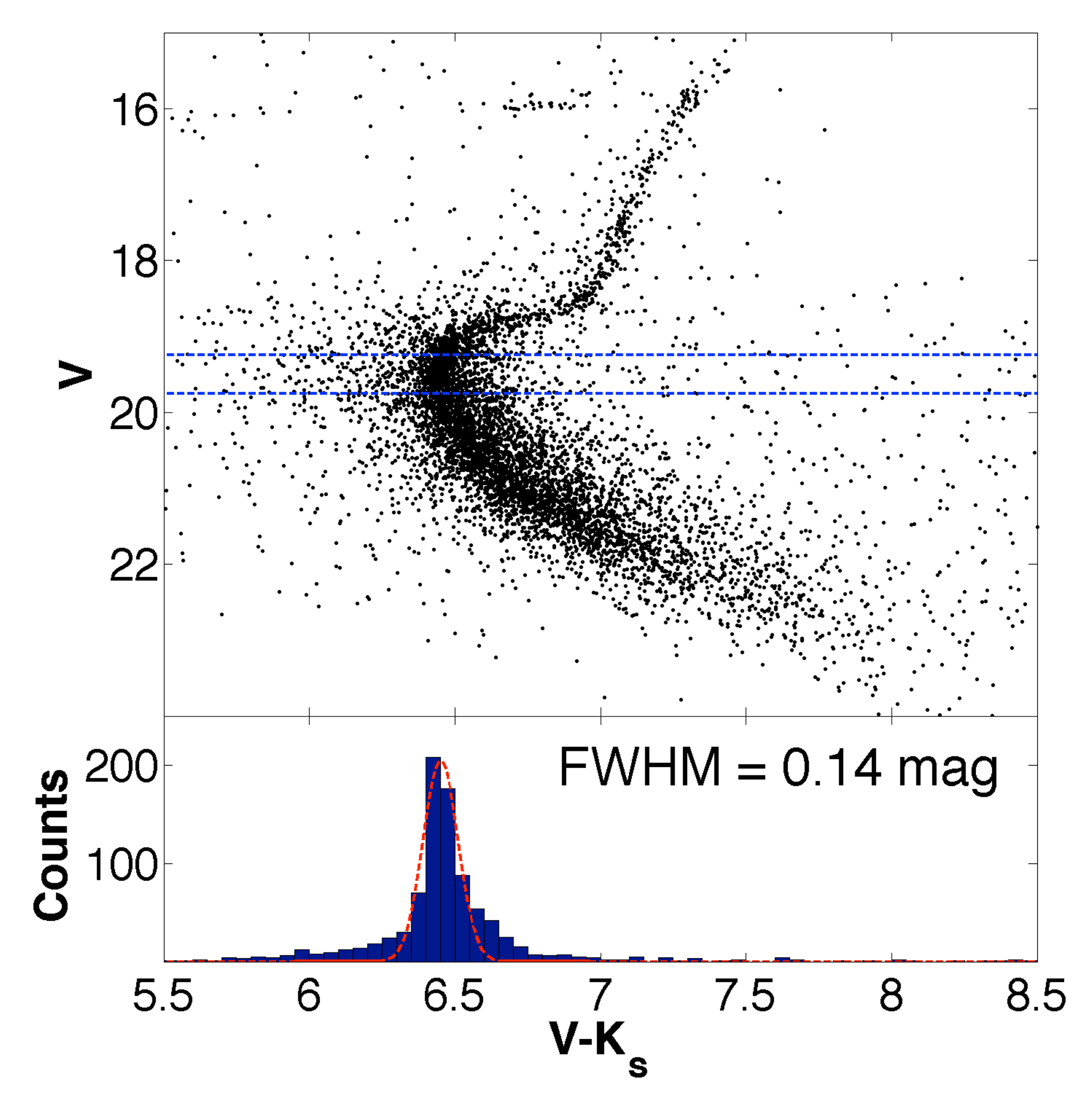}{0.3\textwidth}{Cubic}}
\caption{As Figure~\ref{fig:psfradcmd}, but for different degrees of spatial variability in the PSF model.\label{fig:psfvarcmd}}
\end{figure}

The top panel of Figure~\ref{fig:psfvarsub} shows examples of six PSF stars in a K$_\mathrm{s}$ image from different parts of a frame. 
The other panels show the same regions after profiles with different variability have been fitted and subtracted. The constant and linear models present large residuals, while the quadratic and cubic seem to fit similarly well the profiles. To determine which produces the most accurate photometry, we have created uncalibrated CMDs (Figure~\ref{fig:psfvarcmd}) as in Section~\ref{sec:psfrad} to analyze the width of their MSTOs. On all the tested images we find that the cubic variability has similar or slightly better performance than the quadratic, which is unsurprising given the large distortions of the PSFs that we observed in Section~\ref{sec:performance}. We therefore recommend the use of a highly variable PSF, like the cubic one, provided that the observations include a large enough number of bright PSF stars to measure it.

\subsubsection{Selection of the sky annuli}\label{sec:psfsky}

When testing the PSF radius we have also modified two related sets of radii. The first is the annulus of pixels that we have used to measure and remove the sky value of the PSF stars, measured beyond the PSF radius (see Section~\ref{sec:photometry}). The second set is the annulus to calculate the sky during the profile fitting that is computed at every iteration after the profile subtraction and which is measured interior to the PSF radius. The inner radius of this annulus is set at two pixels from the center to avoid the large residuals that could be near the peak.

We have first tested the influence of the PSF model's sky annulus on the photometry by varying its distance from the edge of the PSF radius and its width. We have tried three annuli with inner and outer radii of [25, 36], [41, 49], and [25, 49], each of which include at least 2000 pixels to have a good signal. The first and third of these options measure the sky relatively close to where the PSF is measured (but with a larger area used in the latter option), whereas the second option measures the sky away from the PSF to reduce any contribution to the flux from the PSF star. Using the same CMD dispersion criterion of Section~\ref{sec:psfrad}, we found no measurable change in the photometric precision for any of the three photometric catalogs. We therefore adopted the first of the three choices.

\begin{figure}
\gridline{\fig{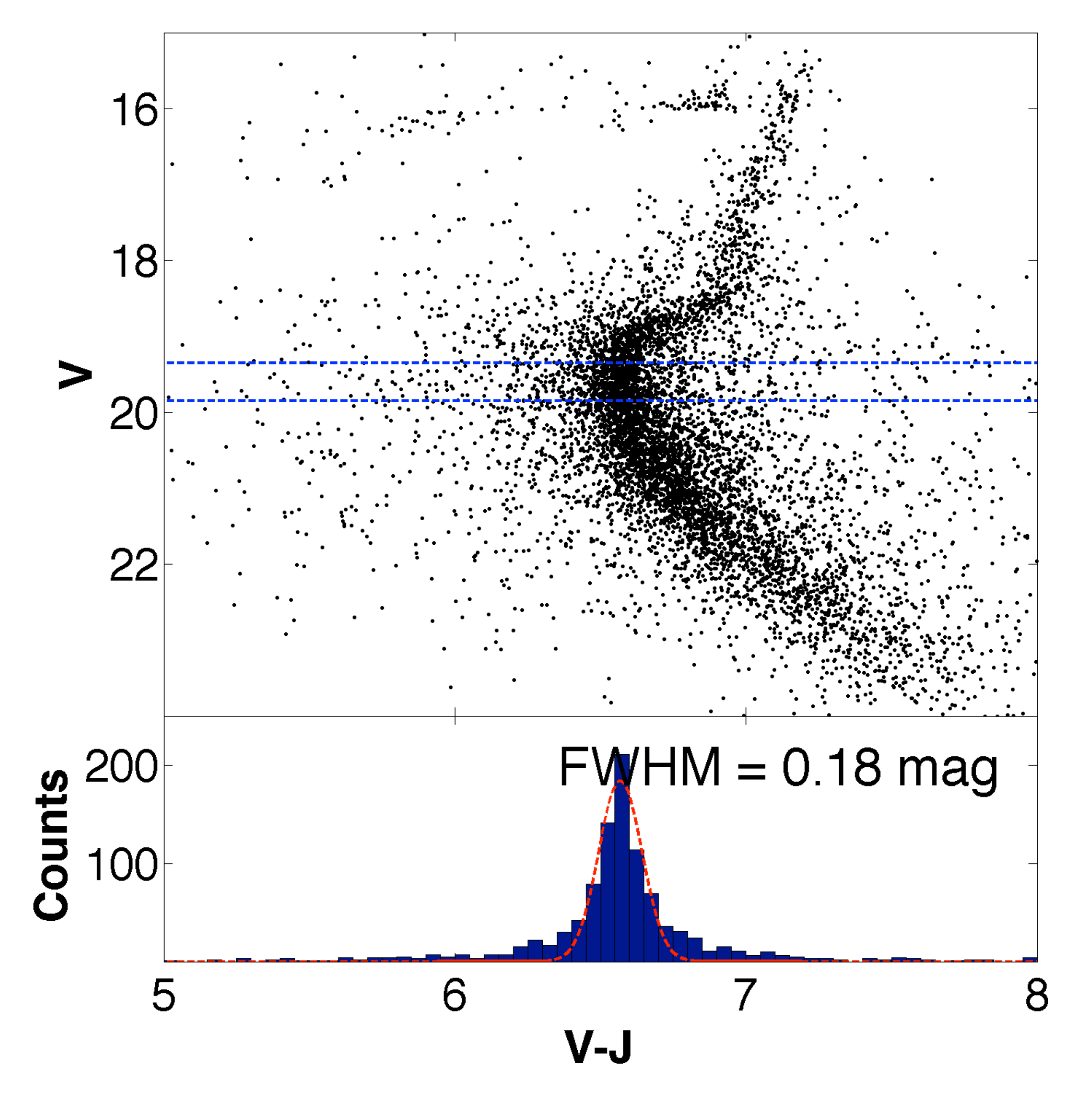}{0.3\textwidth}{Outer sky radius: 16 px} \fig{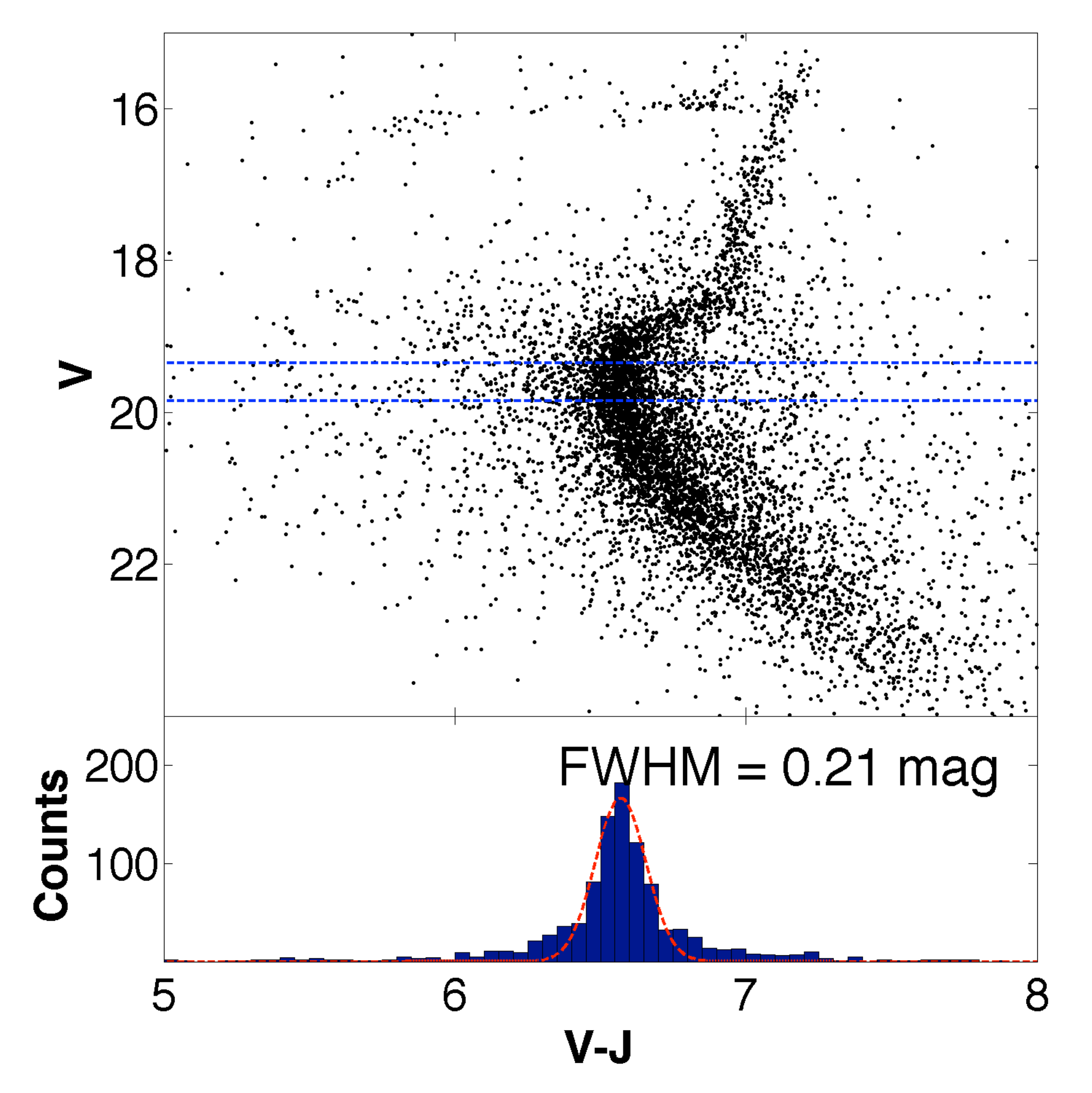}{0.3\textwidth}{Outer sky radius: 20 px} \fig{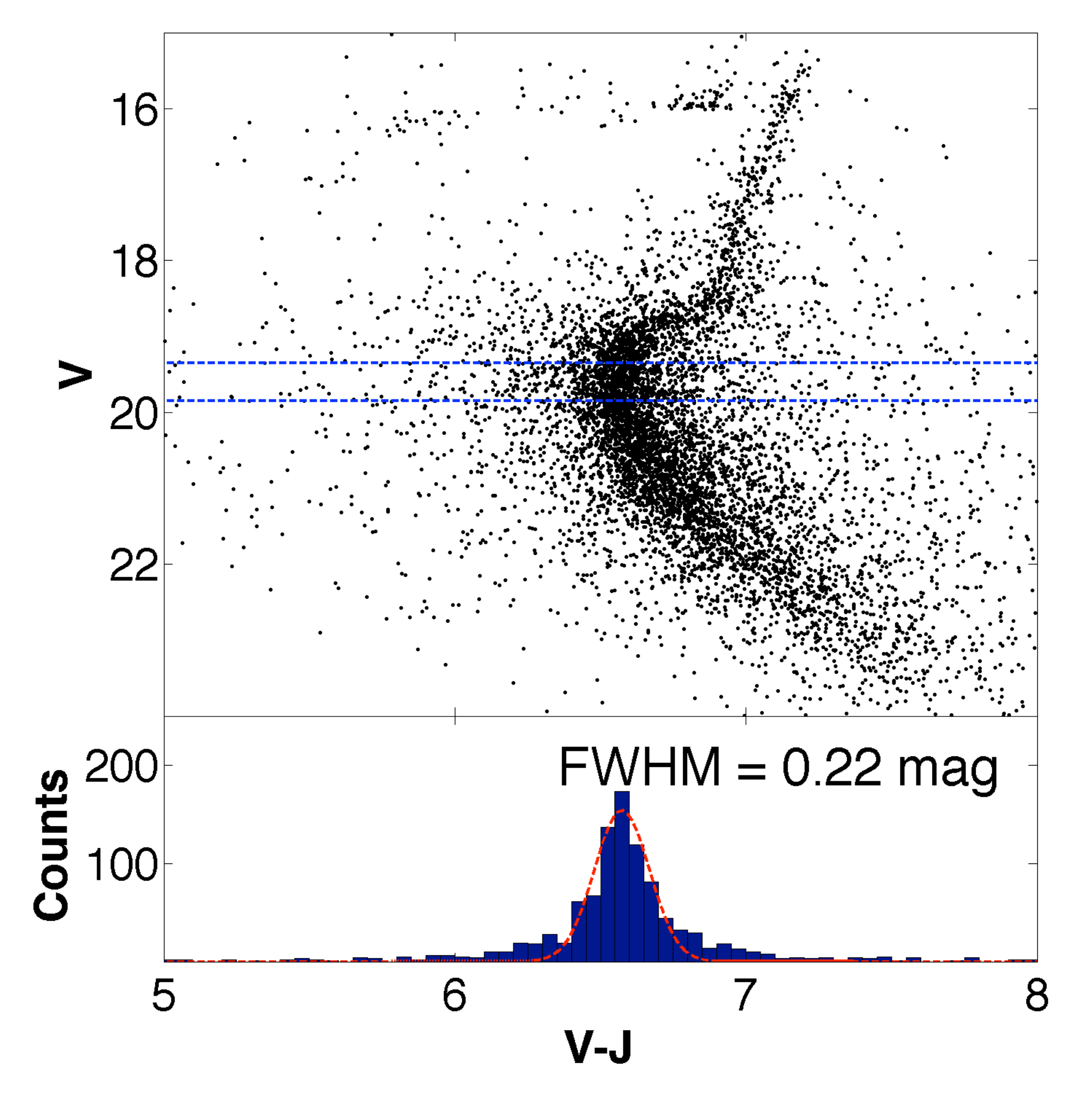}{0.3\textwidth}{Outer sky radius: 24 px}}
\caption{As Figure~\ref{fig:psfradcmd}, but for J band images that adopt different outer radii for the sky annulus.\label{fig:psfskycmd}}
\end{figure}

For the sky annulus used during profile fitting, we have tested an outer radius of 16, 20, and 24 pixels. For most of the images in K$_\mathrm{s}$ filter, the photometric precision does not measurably change. However, for the K$_\mathrm{s}$ exposures with poorer MCAO correction and for all the J band exposures we tested, we find that a small outer radius is preferable, as is shown in Figure~\ref{fig:psfskycmd}. We have therefore adopted a sky annulus with an inner radius of 2 pixels and an outer radius of 16 pixels. We postulate that a small outer radius is preferred because, when the Strehl ratio is lower, more faint stars are left undetected. A larger outer sky radius then increases the chances of these faint stars contributing to the ``sky'' measurement, reducing the accuracy of this measurement.

\subsubsection{Minimizing contamination light in the PSF stars}\label{sec:psfclean}

\begin{figure}
\gridline{\fig{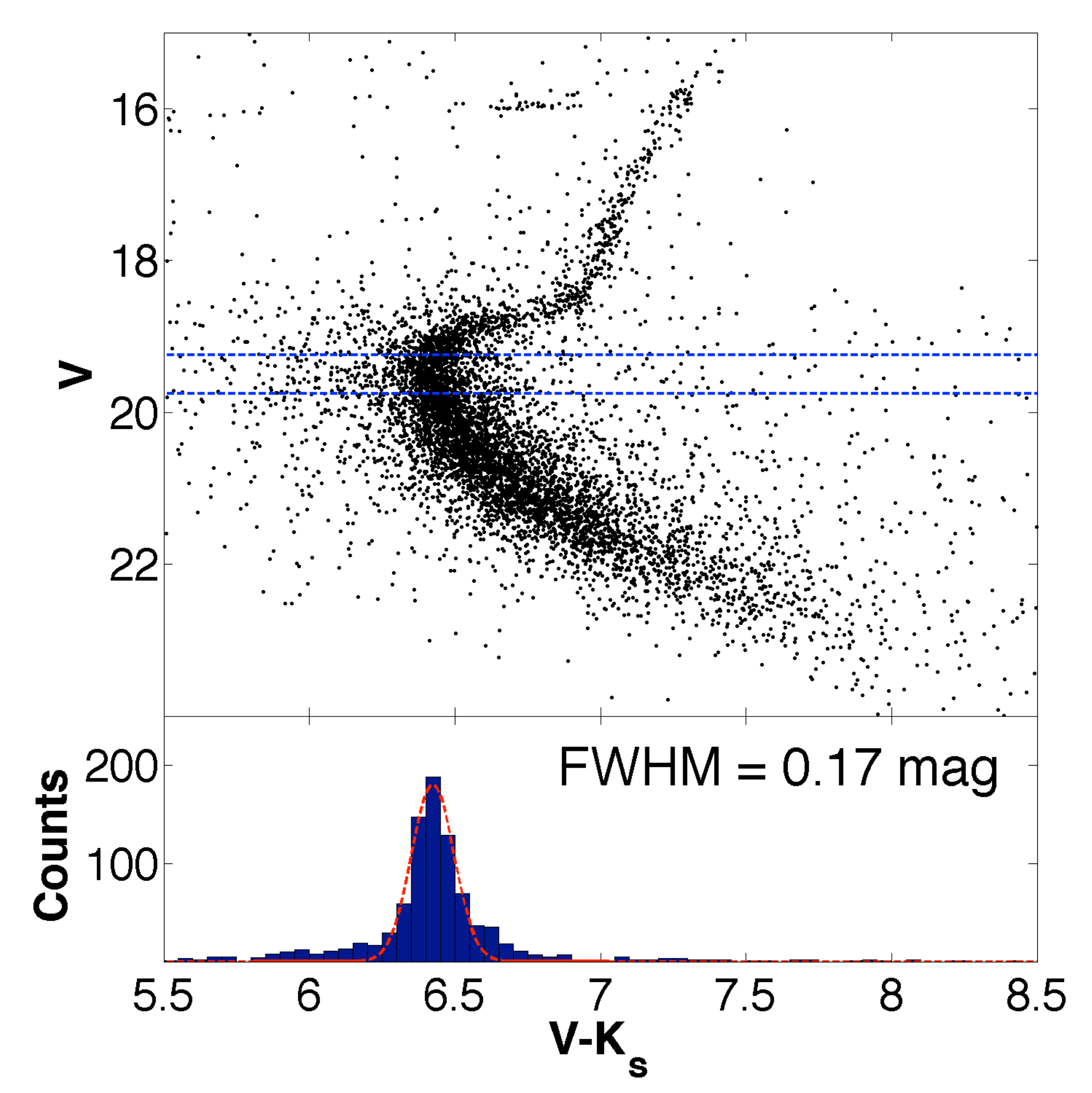}{0.3\textwidth}{First iteration} \fig{fig47.pdf}{0.3\textwidth}{Second iteration} \fig{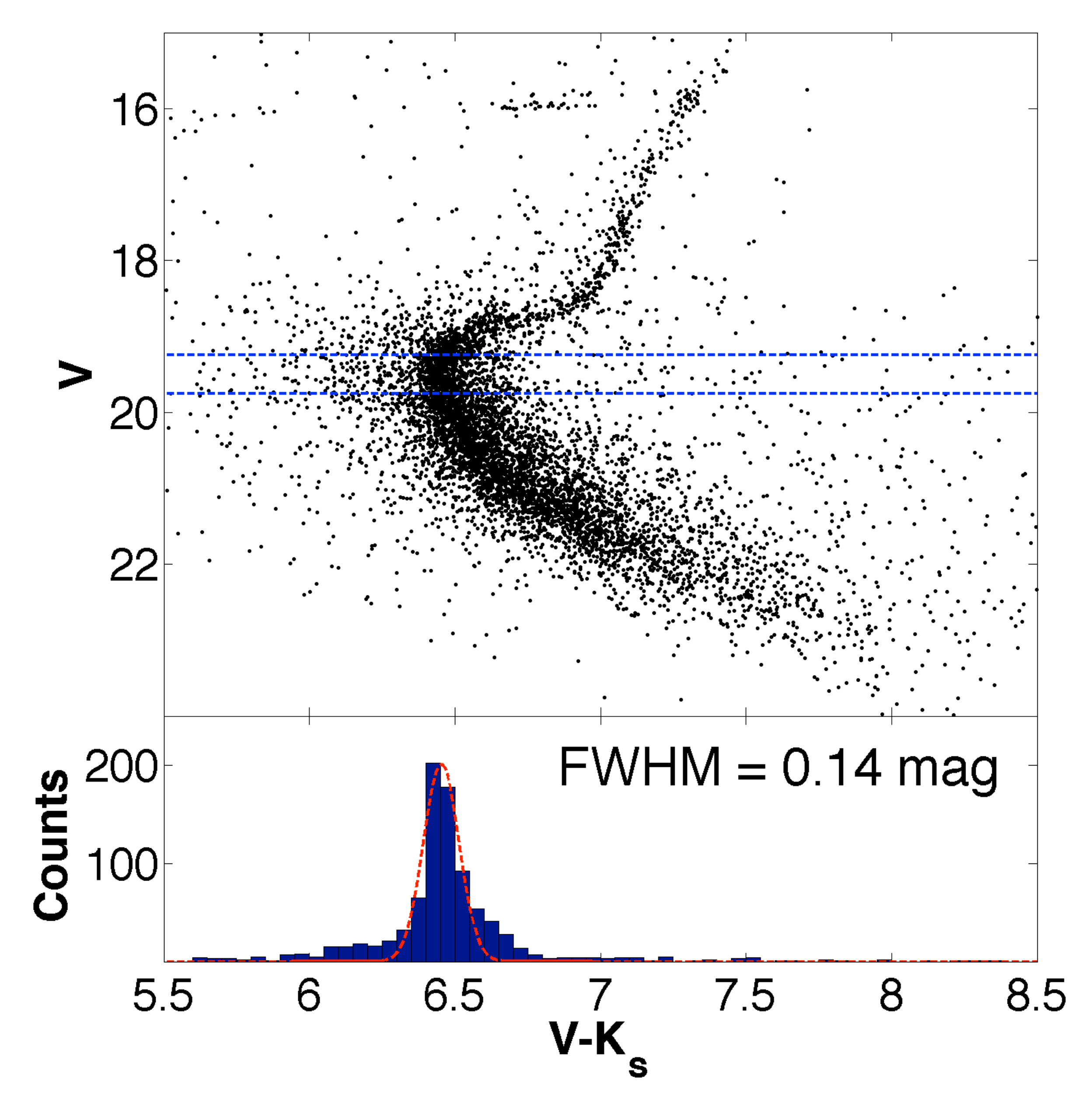}{0.3\textwidth}{Third iteration}}
\caption{As Figure~\ref{fig:psfradcmd}, but for uncalibrated CMDs of a K$_\mathrm{s}$ image using different iterations of PSF cleaning.\label{fig:psfitercmd}}
\end{figure}

In a crowded image, the PSF is inevitably going to be measured using stars that have their profiles contaminated by nearby sources, causing inaccuracies in its estimation (for example, see the left panel of Figure~\ref{fig:rad}). To improve the resulting model, we have performed an iterative estimation of the PSF. Namely, we measure a first estimate of the PSF on the crowded image (first iteration) and with it we subtract all the stars from the image except for the PSF stars. We then use this ``cleaned'' image to redetermine a better PSF (second iteration). Figure~\ref{fig:psf_stars} shows an example of an original reduced frame and a ``cleaned'' one, where the PSF stars are marked. By measuring the width of the MSTO in the usual way, we determine that there is a gain in photometric precision when using the PSF estimated in the second iteration compared to the first (Figure~\ref{fig:psfitercmd}). No benefit has been observed from increasing the number of iterations that we use beyond two, an effect of the use of uncrowded PSF stars since the first iteration.

\subsubsection{Independent profile fitting}\label{sec:psfindep}

The ALLFRAME task in the DAOPHOT suite can perform the profile fitting photometry on multiple images at the same time. At each iteration, this program uses the geometric transformations of each frame on the master catalog to determine the position of the stars on the images. Then the PSF models are centered on these positions and are fitted to the image by subtracting the nearby stars determined in the previous iteration. The estimated residuals in the positions of the stars are minimized in the following iterations by adjusting both the stars' positions in the master catalog and the coefficients of the geometric transformation. By fixing the positions of the stars in the reference frame for all the images, the combined photometry is potentially more precise by reducing the uncertainty caused by the noise in positioning the fitted PSF.

\begin{figure}
\gridline{\fig{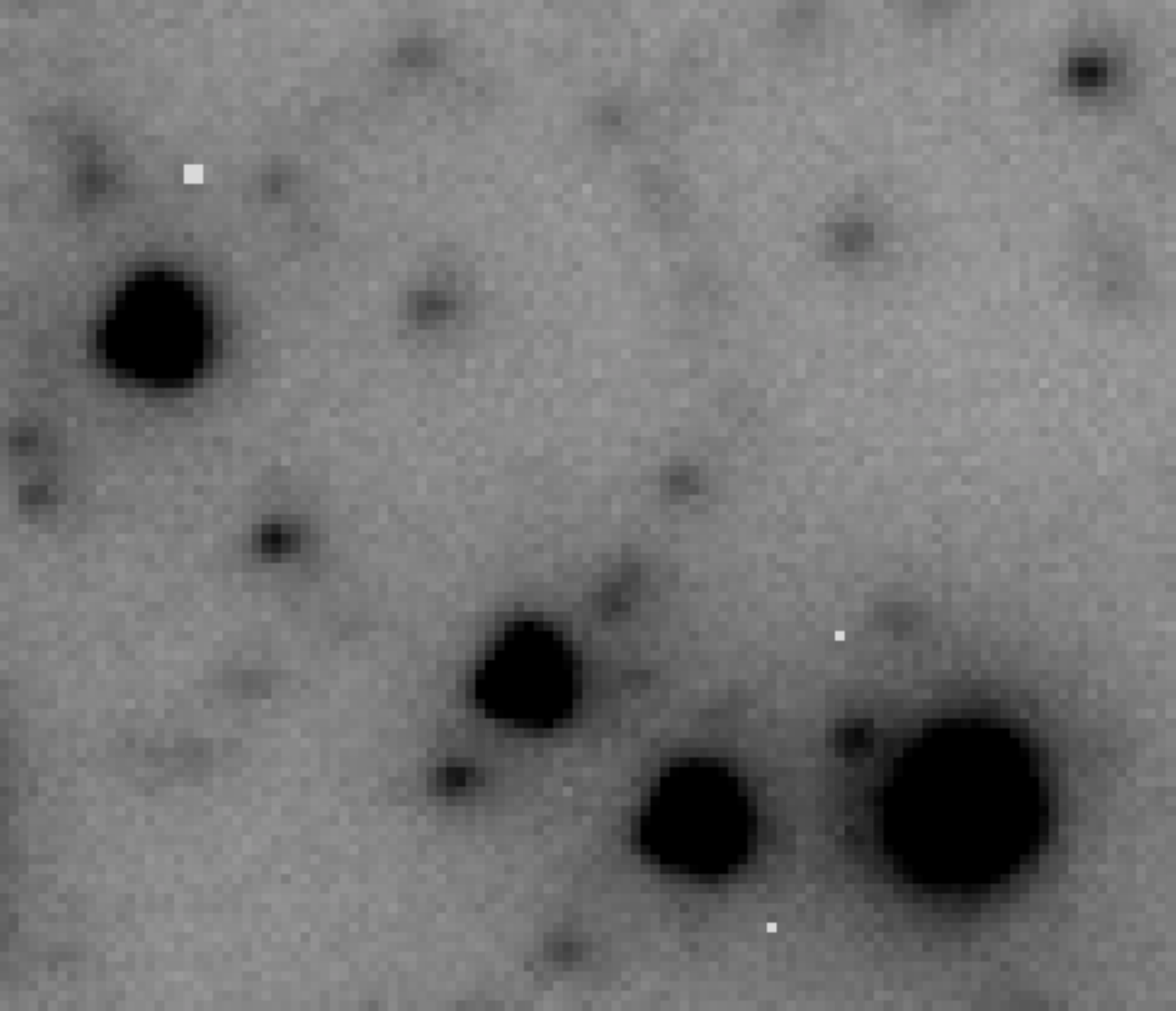}{0.3\textwidth}{} \fig{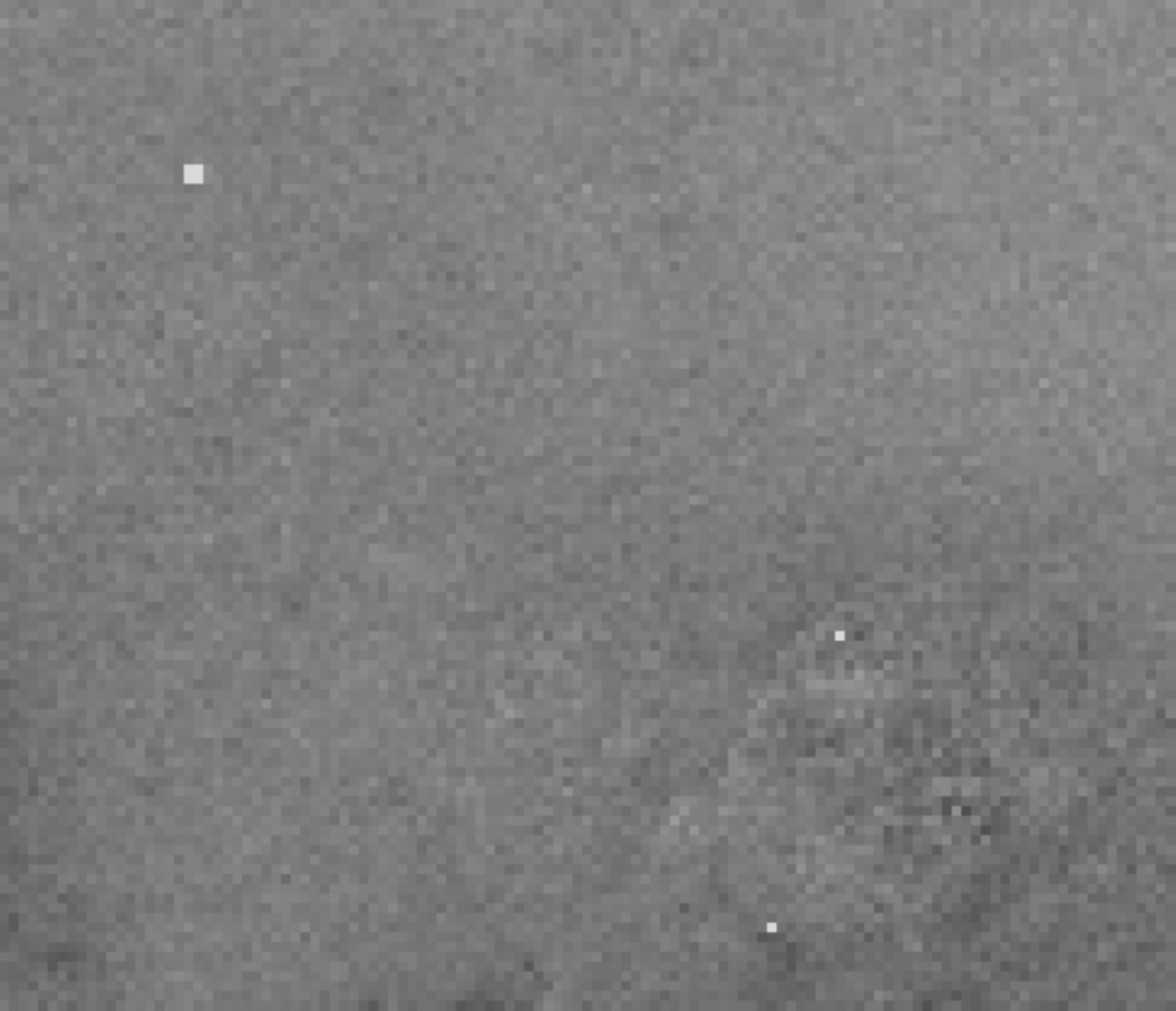}{0.3\textwidth}{} \fig{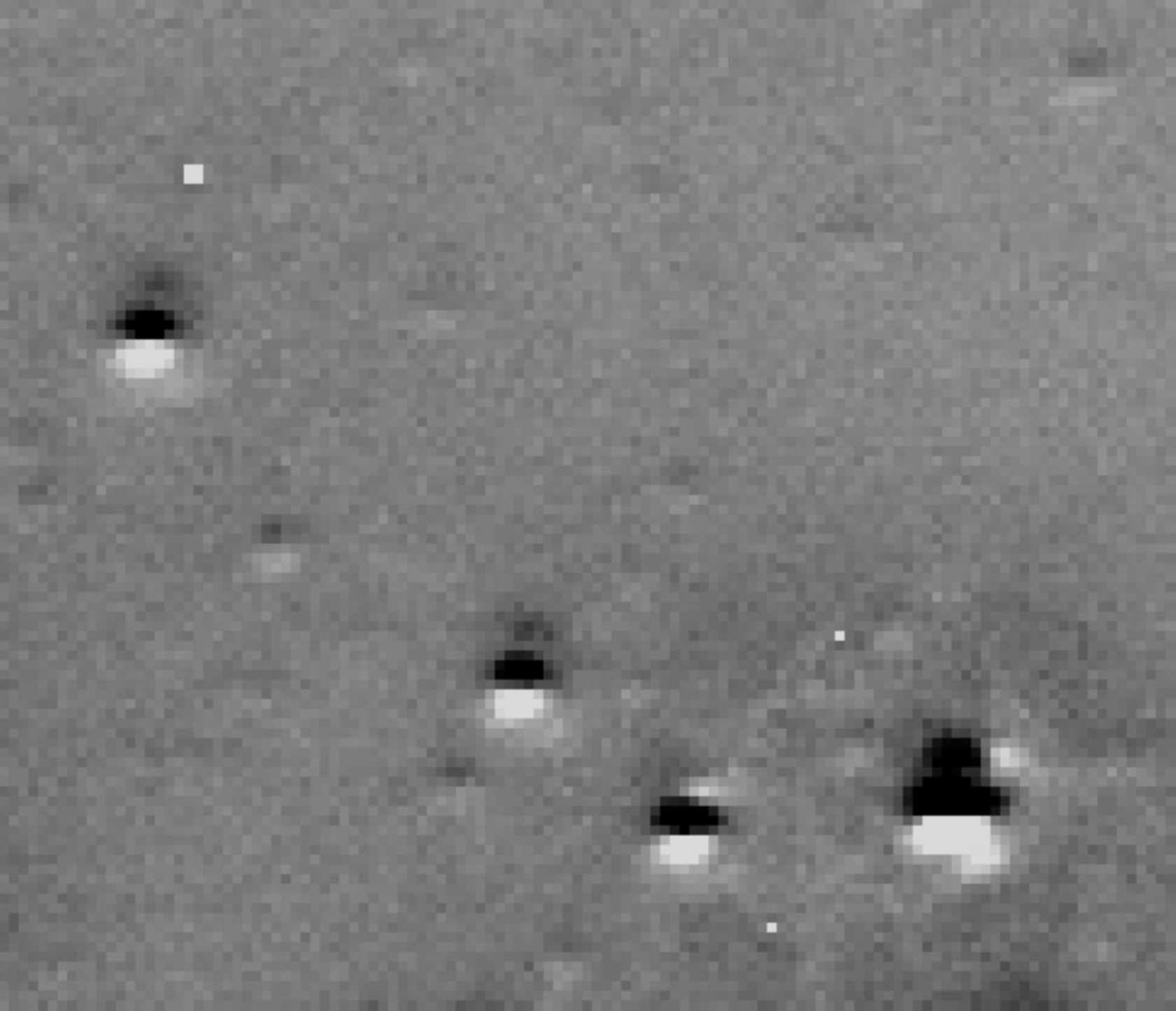}{0.3\textwidth}{}}
\caption{Example of the difference in using single- and multi-frame photometry on the edge of a K$_\mathrm{s}$ image. (\emph{Left:}) a region of a $K_s$ band image near the edge of the field. \emph{Center:} the subtracted image once all the PSFs have been fitted without placing constraints on their positions.\emph{Right:} the subtracted image where the positions of the stars are fixed using the ALLFRAME transformation.\label{fig:alsalf}}
\end{figure}

\begin{figure}
\centering
\includegraphics[width=0.4\textwidth]{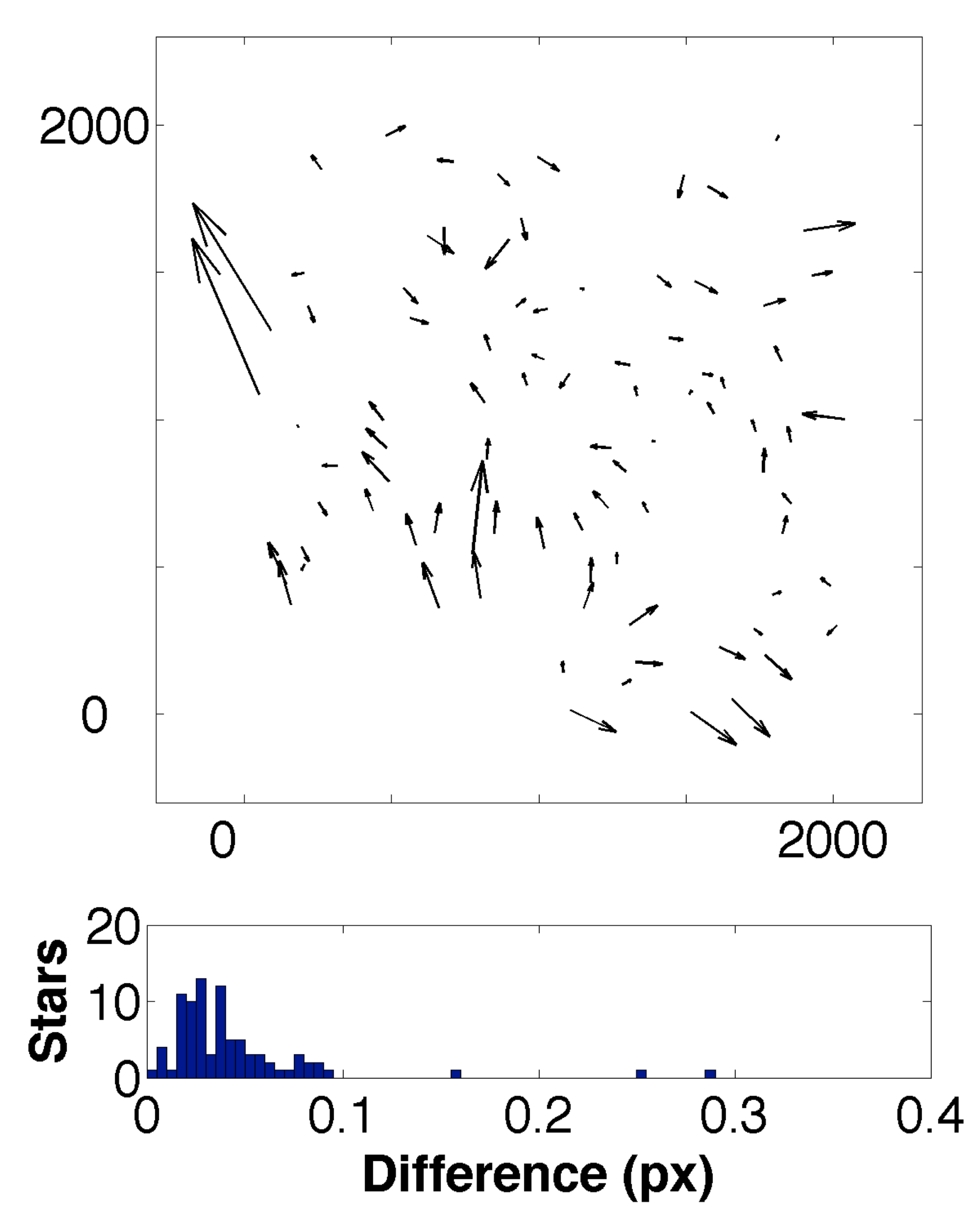}
\caption{\emph{Top}: difference of the PSF stars' positions determined by single- and multi-frame profile fitting on a K$_\mathrm{s}$ image. The arrows' lengths are magnified by 2000. \emph{Bottom}: histogram of the differences in position between the two methods.\label{fig:alsalfpos}}
\end{figure}

In Figure~\ref{fig:alsalf}, the first panel shows a region of a K$_\mathrm{s}$ band image near the edge of the field. The middle panel shows the subtracted image once all the PSFs have been fitted without placing constraints on their positions from the master catalog. The right panel shows the subtracted image where the positions of the stars are fixed using the ALLFRAME transformation. Clearly, the latter image displays strong residuals caused by a similar misplacement of the PSF centers. We note that the region selected for this example is one where the effect is particularly dramatic and that in most of the images it is more subtle. While not a focus of this paper, the resulting astrometry is also affected: Figure~\ref{fig:alsalfpos} shows the difference between the positions of PSF stars in a K$_\mathrm{s}$ band image determined with the two methods. Arrows are magnified by a factor of 2000 for clarity and the histogram of the offset of the positions is also shown.

Clearly, Figure~\ref{fig:alsalfpos} indicates a problem using DAOPHOT in the multi-frame mode for GeMS images. We note that ALLFRAME uses a polynomial fit of maximum third order. It has been shown by \cite{bib:massari16c} that GeMS delivers a field of view with geometric distortions well described by a fifth order polynomial with an rms of about 0.01 px respect to the third order. For the exposure in Figure~\ref{fig:alsalfpos}, we find a median difference in positions of 0.03 px between no constrains and the third order. However, the difference between polynomials could be more severe near the edges. By relying on the DAOPHOT transformations, we are potentially producing differences between the real position on the image of the stars and where the PSF is fitted by ALLFRAME.

\begin{figure}
\gridline{\fig{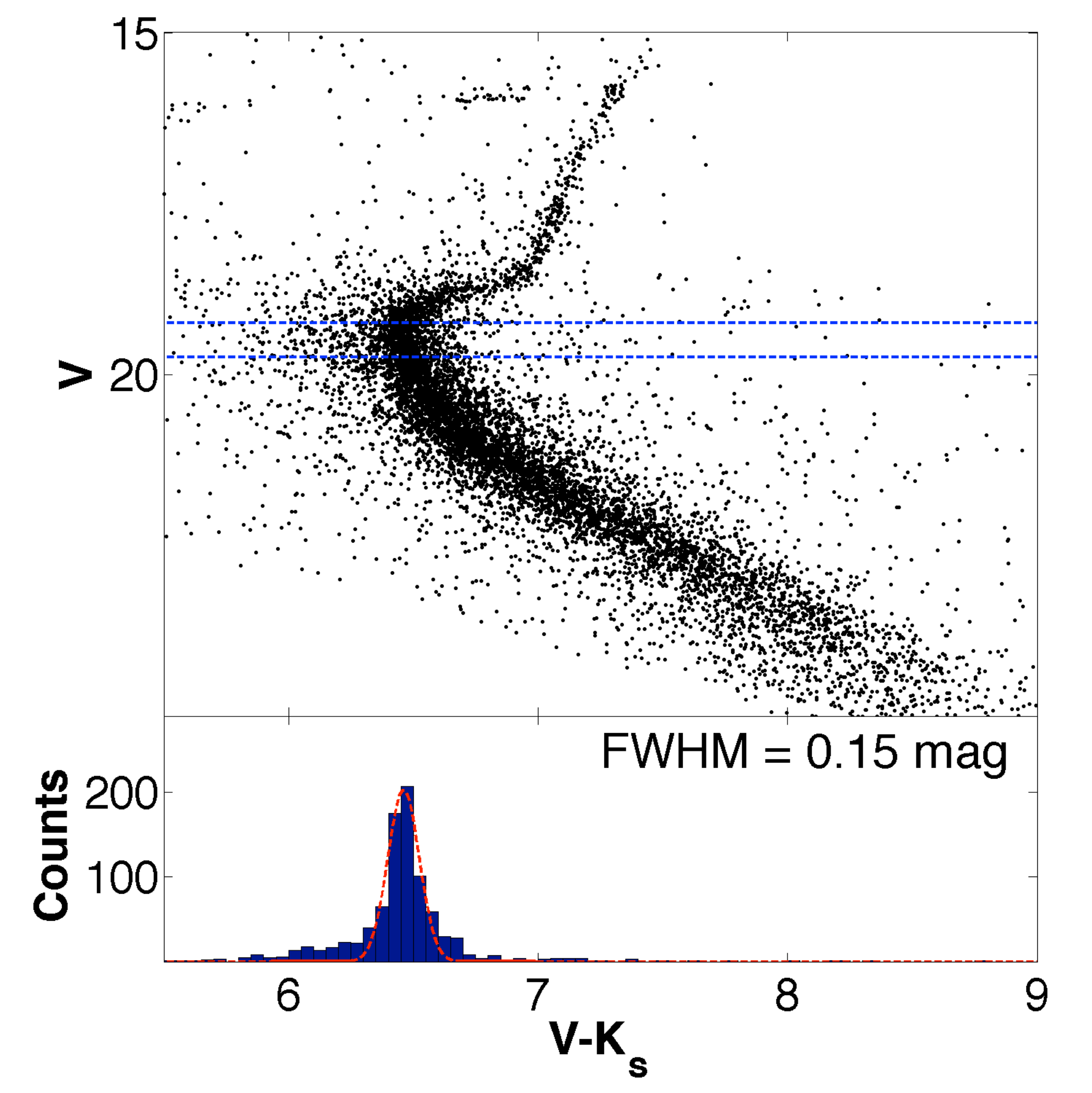}{0.45\textwidth}{Single-frame profile fitting} \fig{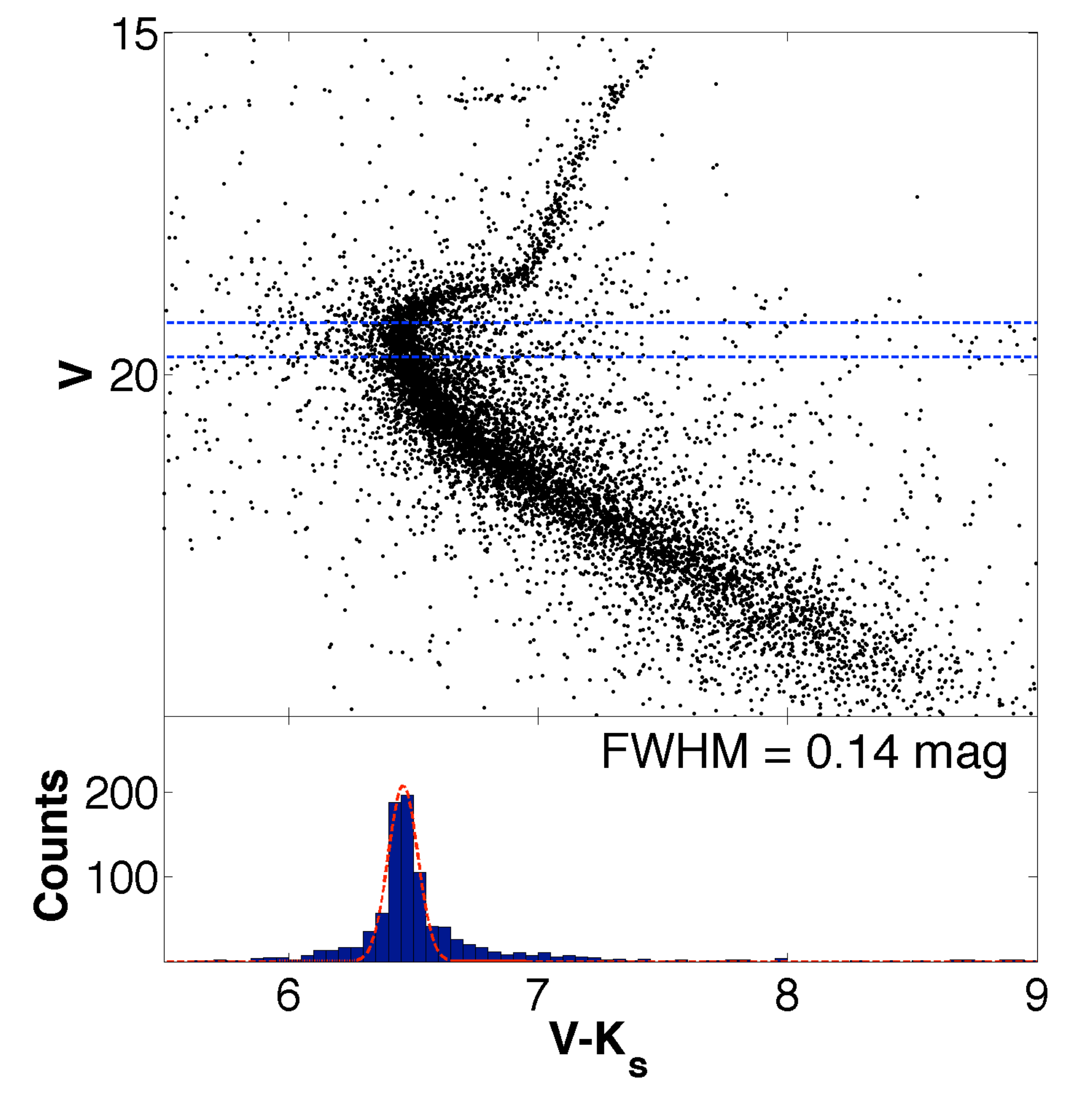}{0.45\textwidth}{Multi-frame profile fitting}}
\caption{As Figure~\ref{fig:psfradcmd}, for uncalibrated CMDs of a K$_\mathrm{s}$ image using single- \emph{(left)} and multi-frame \emph{(right)} profile fitting. This photometry is deeper than the uncalibrated CMDs in the previous sections because the input catalog includes the faint stars detected in the stacked images. In this particular exposure, the same used for testing the previous PSF parameters, the multi-frame profile fitting performs marginally better than the single-frame.\label{fig:alsalfcmd}}
\end{figure}

By analyzing the test CMDs for the two cases in the usual way (Figure~\ref{fig:alsalfcmd}), we generally do not find substantial differences around the MSTO between the two methods. However, to reduce astrometric systematic errors and eliminate residuals such as those in Figure~\ref{fig:alsalf}, we use the single-frame method in the profile fitting.

\subsection{Optimization of the photometric calibration}\label{sec:calibration}

\subsubsection{Photometric zeropoints}

\begin{figure}
\gridline{\fig{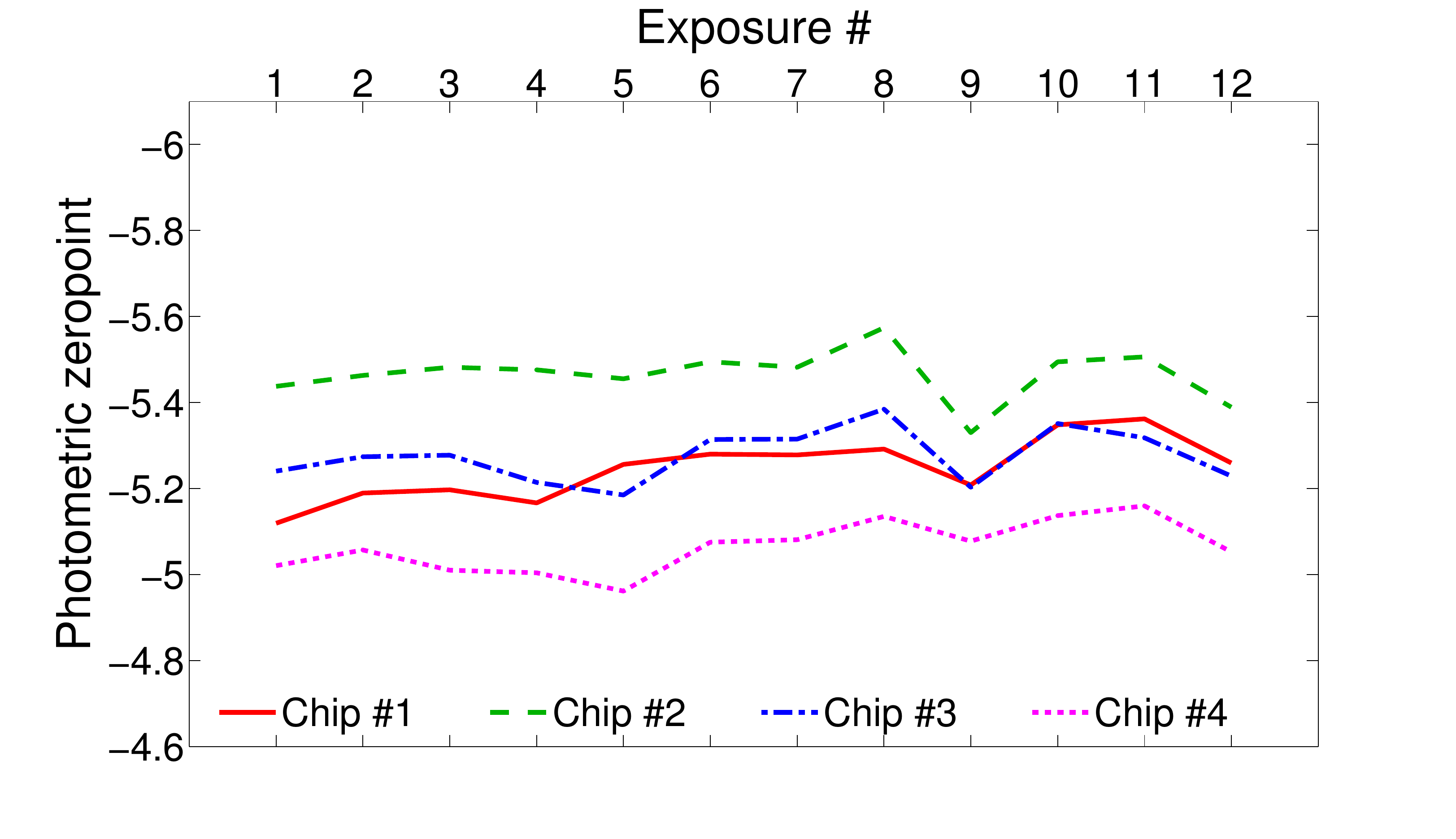}{0.45\textwidth}{} \fig{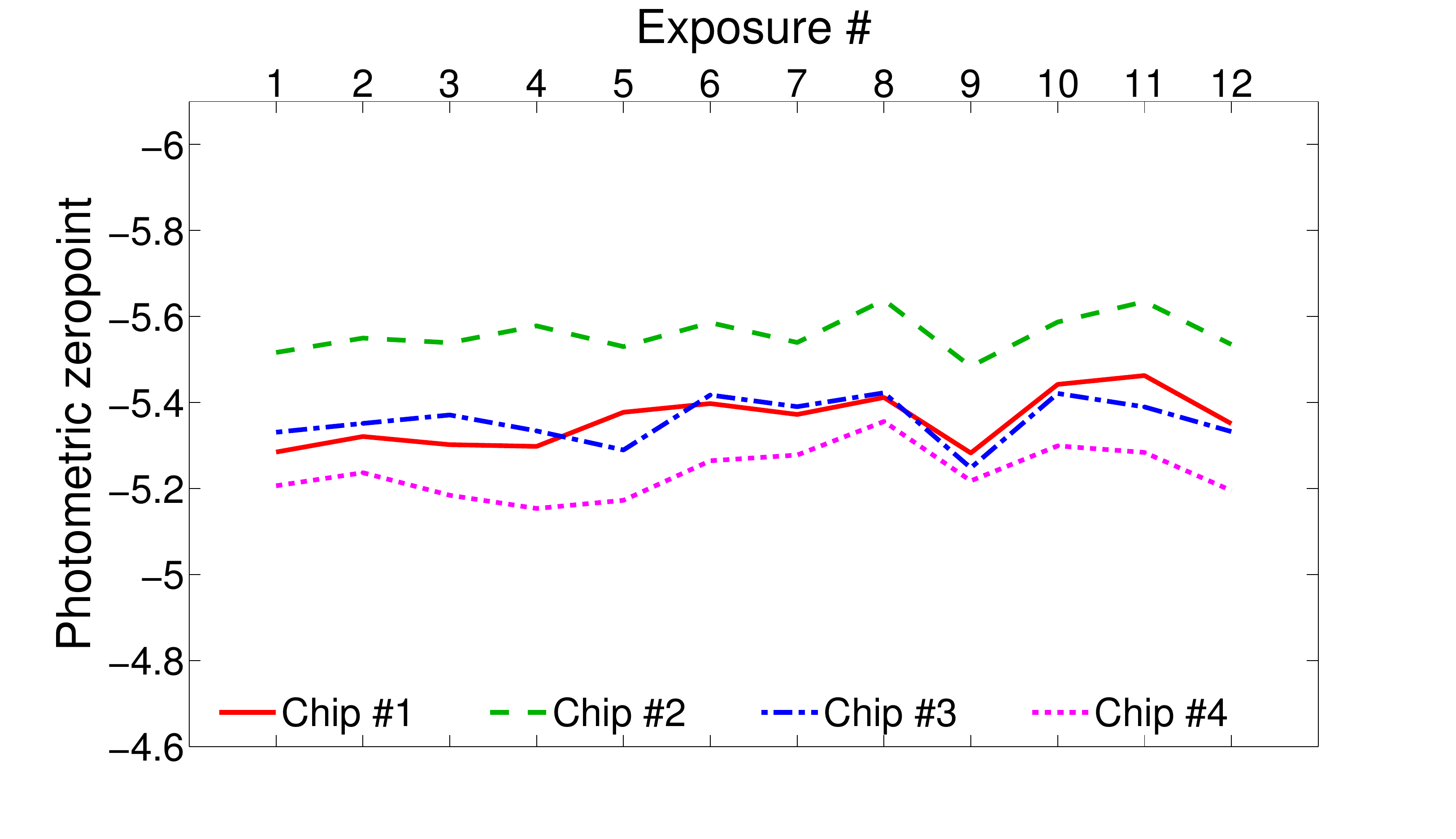}{0.45\textwidth}{}}
\caption{K$_\mathrm{s}$ photometric zeropoints without (\emph{left}) and with (\emph{right}) the correction for the seeing in the reference catalog.\label{fig:zerop}}
\end{figure}

The instrumental magnitudes of stars measured in our data are calibrated relatively to our reference catalog described in Section~\ref{sec:calibrationover} by applying a zeropoint ($Z$) and a color correction ($C$). To determine these coefficients, we have matched our PSF stars to the reference catalog obtaining more than 50 matches in every exposure/chip. We begin with an initial estimation of the zeropoint $Z$ defined as the median of the difference between the instrumental and reference magnitude for all stars in our exposures that are matched to the reference catalog. Zeropoints are worked out for each chip in each exposure, independently. The resulting values are shown in the left panel of Figure~\ref{fig:zerop}.

\begin{figure}
\centering
\gridline{\fig{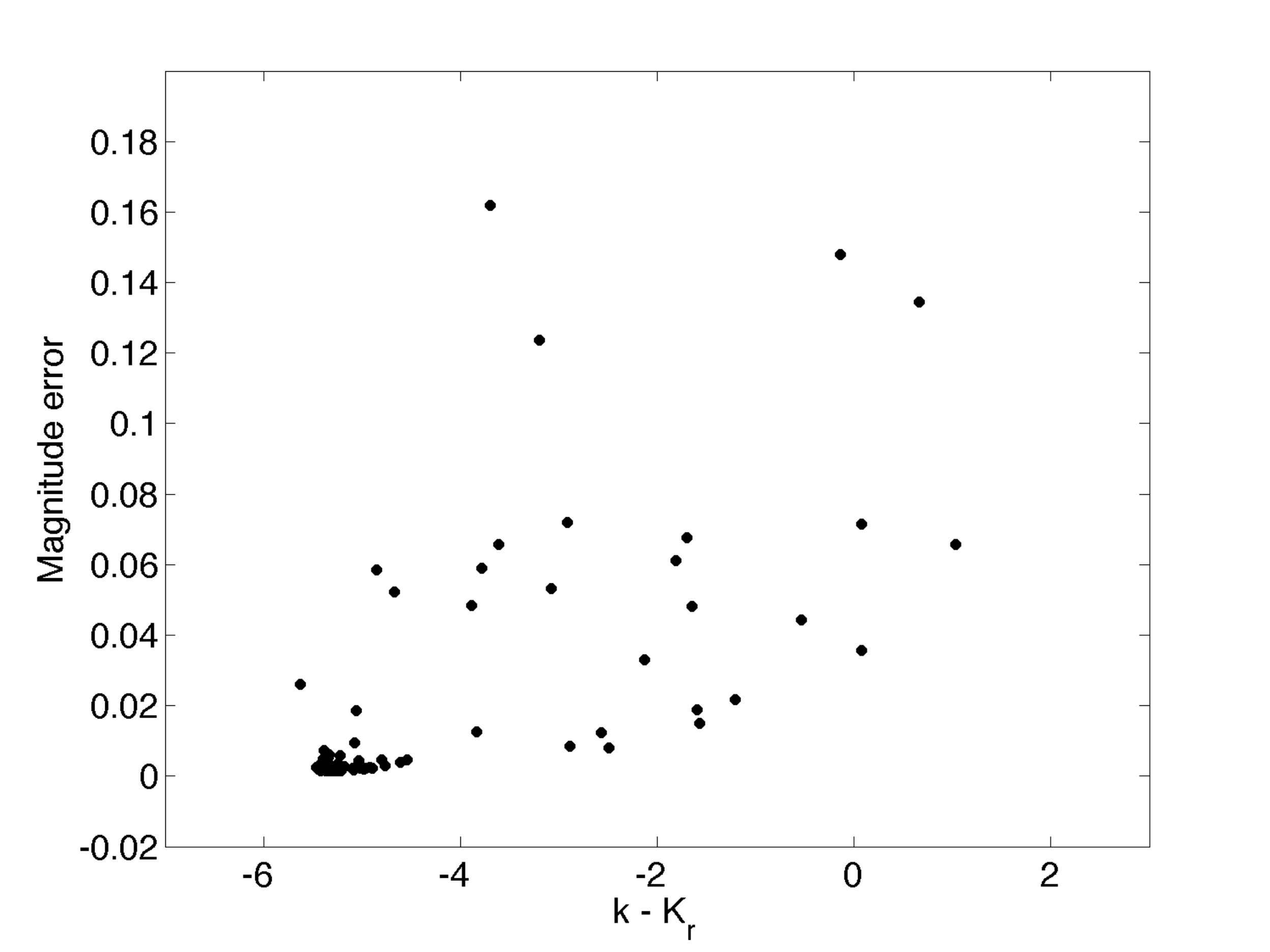}{0.45\textwidth}{} \fig{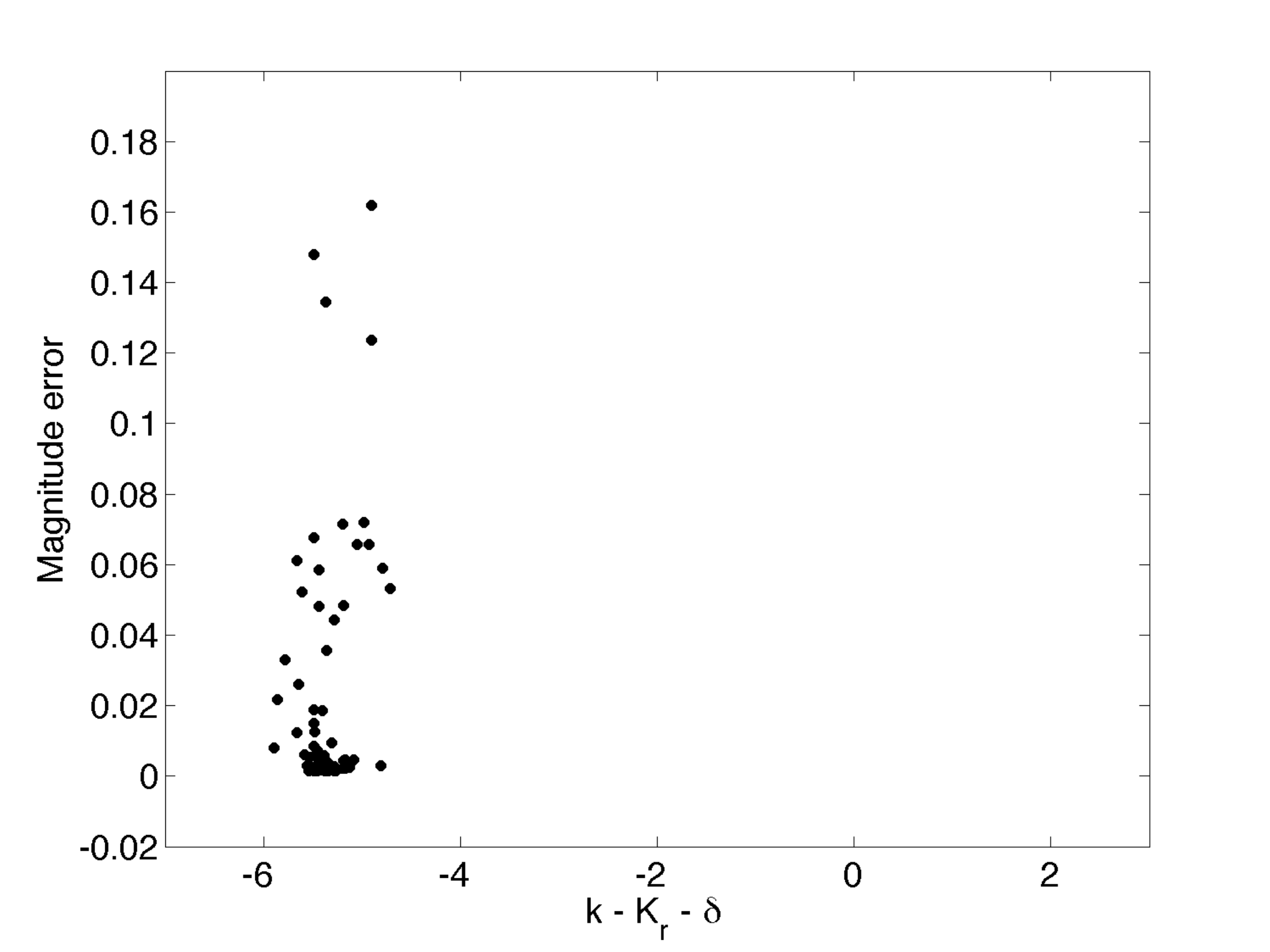}{0.45\textwidth}{}}
\caption{Magnitude error of the PSF stars estimated by DAOPHOT for one of the K$_\mathrm{s}$ images. \emph{Left:} as a function of the difference between instrumental and calibrated magnitude. \emph{Right:} as a function of the instrumental minus calibrated magnitude minus the $\delta$ correction for the effect of nearby neighbors. In both plots, no clear correlation is observed.\label{fig:seeingmagerr}}
\end{figure}

\begin{figure}
\gridline{\fig{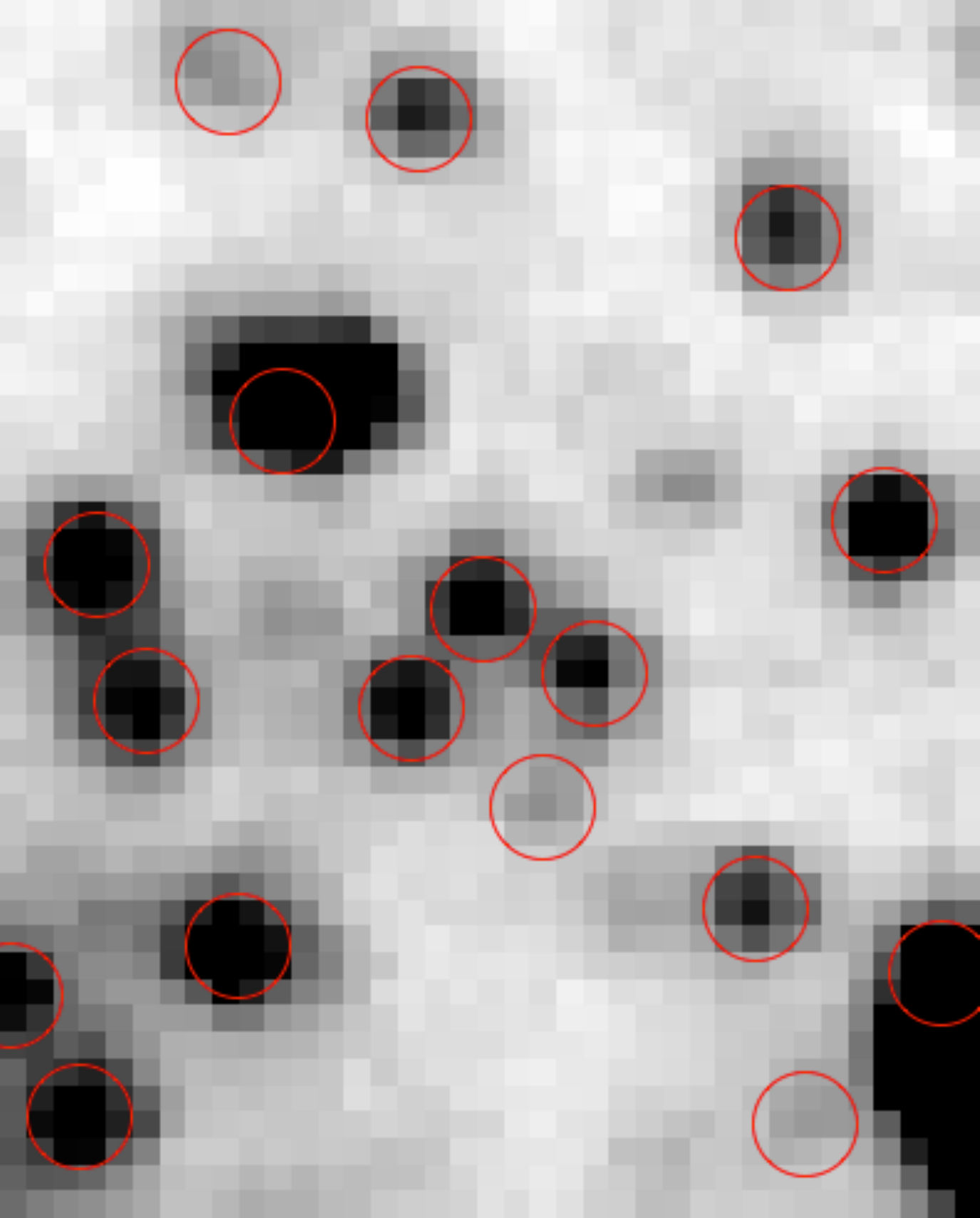}{0.35\textwidth}{} \fig{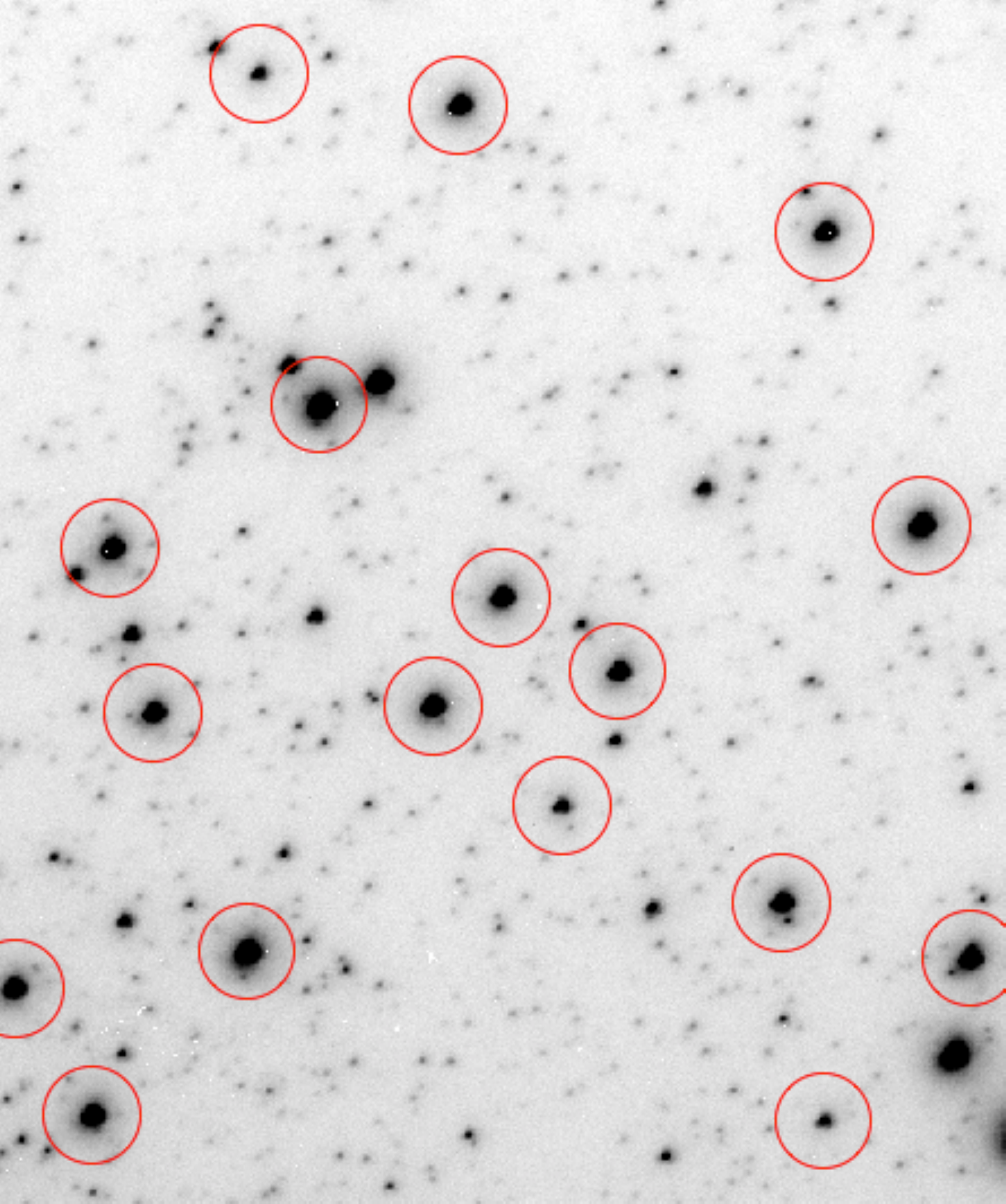}{0.35\textwidth}{}}
\caption{\emph{Left:} detail of the the composite frame of NGC 1851 created using the seeing-limited images from NEWFIRM. \emph{Right:} same region in a single GeMS K$_\mathrm{s}$ exposure. The red circles are centered on the stars used for the photometric calibration and have a diameter of 1.2\arcsec .\label{fig:seeingcalib}}
\end{figure}

It is worthwhile to examine the distribution of values of the difference between the instrumental and reference magnitudes for matched stars in one of the chips/exposures, on which this initial estimate of the zeropoint is based. These values are shown on the left panel of Figure~\ref{fig:seeingmagerr} and are typical of all the chips/exposures. The range of values is very large, of the order of several magnitudes. While the median operation reduces the effect of outliers, the number of them is so large that it implies that there is an issue with the matched magnitudes. The origin of this issue is that we are comparing two catalogs with very different angular resolutions: our reference catalog is seeing-limited while our observations with GeMS are diffraction-limited. Therefore, our reference stars include multiple stars that are resolved in our images but not in the reference catalog. This can be clearly seen in Figure~\ref{fig:seeingcalib}, where the left image is from our reference dataset and the right image is the corresponding part of the image from our data.

\begin{figure}
\centering
\includegraphics[width=0.6\textwidth]{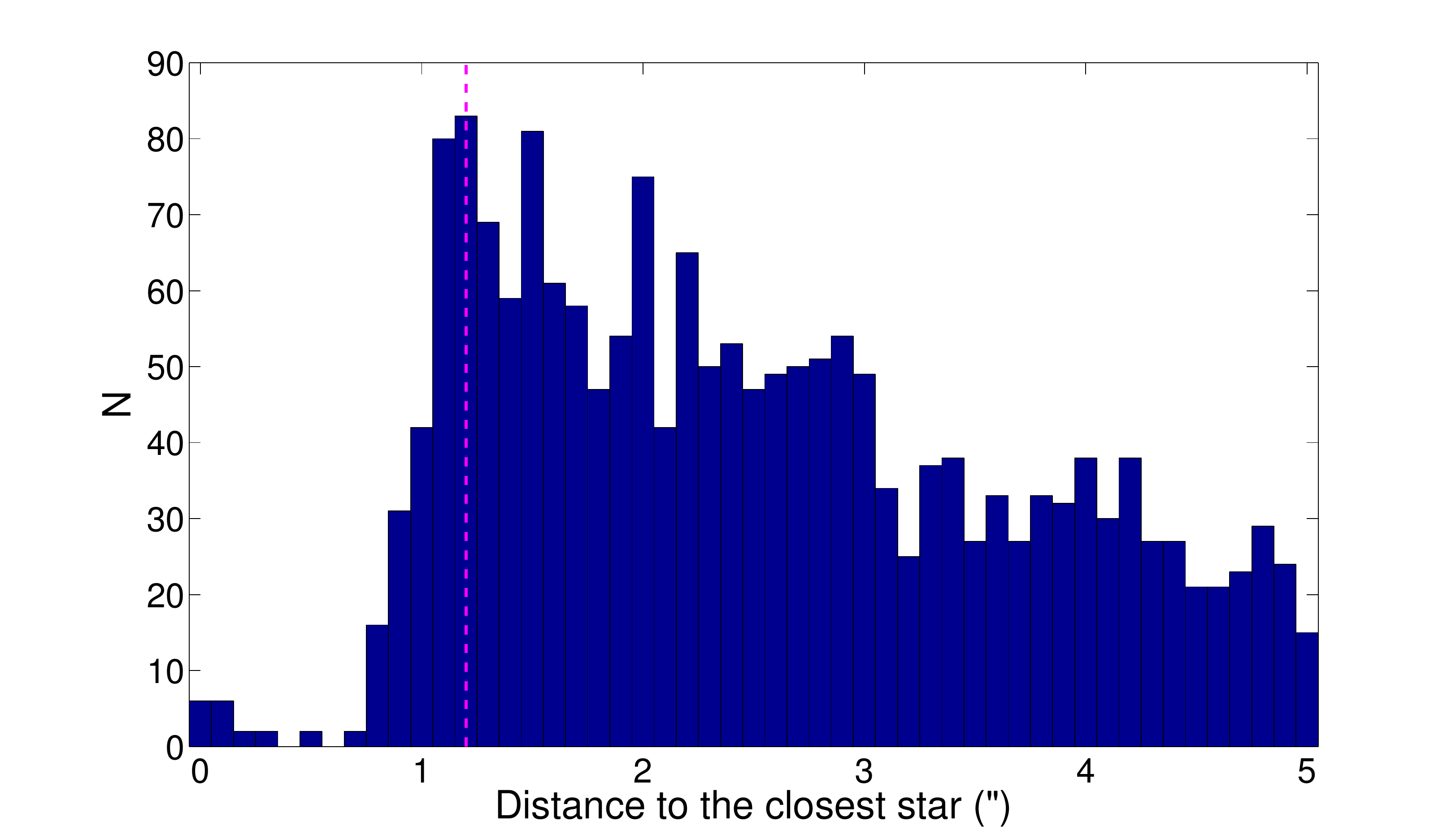}
\caption{Histogram of the distance of every star in the reference catalogue to their closest neighbor. The peak at 1.2\arcsec\, is indicated by the dashed line.\label{fig:seeingrad}}
\end{figure}

It is therefore necessary to reformulate our estimate of $Z$ for each chip/exposure. For a star in our reference catalog, its true magnitude $K_{t}$ is related to its reported magnitude $K_{r}$ via $K_{t}=K_{r}+\delta$, where $\delta$ is a correction that accounts for unresolved companions located within a distance equal to the spatial resolution of our reference catalog. Normally, this would equal the seeing of the observation. However, given our reference catalog is based on a composite dataset, we determine its spatial resolution by making a histogram of the distances of every star to their closest neighbor in the catalog, shown in Figure~\ref{fig:seeingrad}. The peak of this distribution is at 1.2\arcsec (close to the median seeing value of 1.32\arcsec\, of the images that make this composite dataset).

For a star $n$ with true magnitude $K_{t}^{n}$ that is present both in the reference and the GeMS catalog, we can identify all neighbors, that are found within a radius of 1.2\arcsec\, in our Gemini data. Further, we know the instrumental magnitudes $k^{m}$ of each of these neighboring stars. Therefore, 
\begin{displaymath}
\delta =2.5\,\log_{10}{\left( 1+\sum_{m}{\left( 10^{\left( K_{t}^{n}-K_{t}^{m}\right) /2.5}\right)}\right)}\text{.}
\end{displaymath}
Assuming small color corrections, $K_{t}^{n}-K_{t}^{m}=k^{n}-k^{m}$, where the lowercase $k$ denotes instrumental magnitudes.

\begin{figure}
\centering
\includegraphics[width=0.5\textwidth]{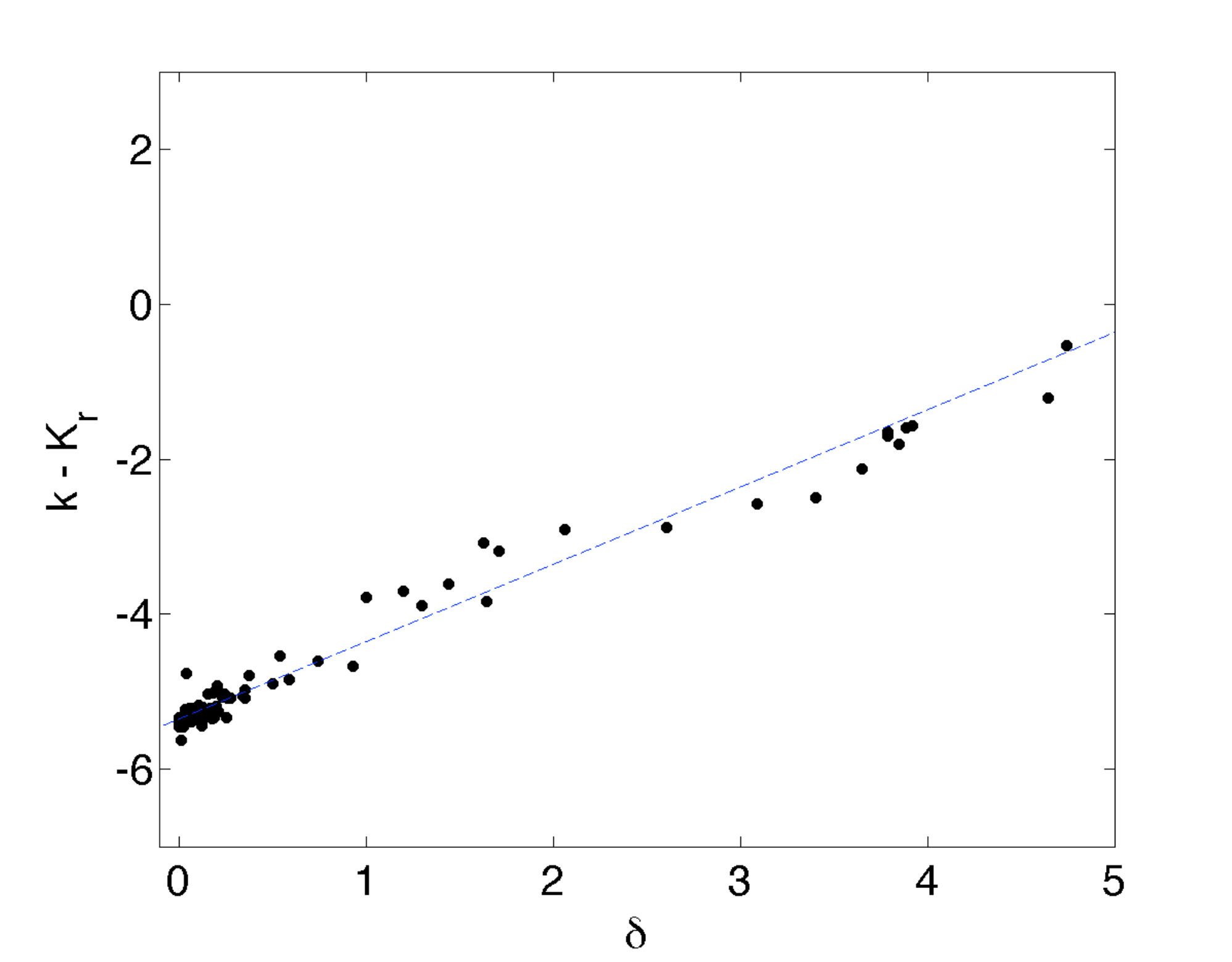}
\caption{Relation between the correction for the effect of nearby neighbors $\delta$ and the difference between the instrumental magnitudes $k$ in our data and the corresponding reference magnitudes $K_{r}$ in the seeing limited catalog (see text for details). The best fit linear relationship with a slope of 1 is shown by the dashed line.\label{fig:seeingmag}}
\end{figure}

Figure~\ref{fig:seeingmag} plots $\delta$ against $k-K_{r}$ for all matched stars in a single chip/exposure. There is a clearly defined linear relationship between the reference magnitude and the instrumental magnitude corrected for the nearby neighbors that are unresolved in the reference catalog. The large spread in the difference between instrumental and reference magnitudes that we observed in the left panel of Figure~\ref{fig:seeingmagerr} can be explained by the correction for crowding, as demonstrated by the reduction of the spread in the right panel of the same figure. The appropriate value for the zeropoint is then the y-intercept of the best fit linear model with a slope equal to 1, shown by the dashed line in Figure~\ref{fig:seeingmag}. We note that to determine the zeropoint to high accuracy as is done here requires us to eliminate all spurious detections in the vicinity of bright stars, especially those relating to the ``orphan'' halos discussed previously in Section~\ref{sec:psfrad}.

\begin{figure}
\centering
\includegraphics[width=0.5\textwidth]{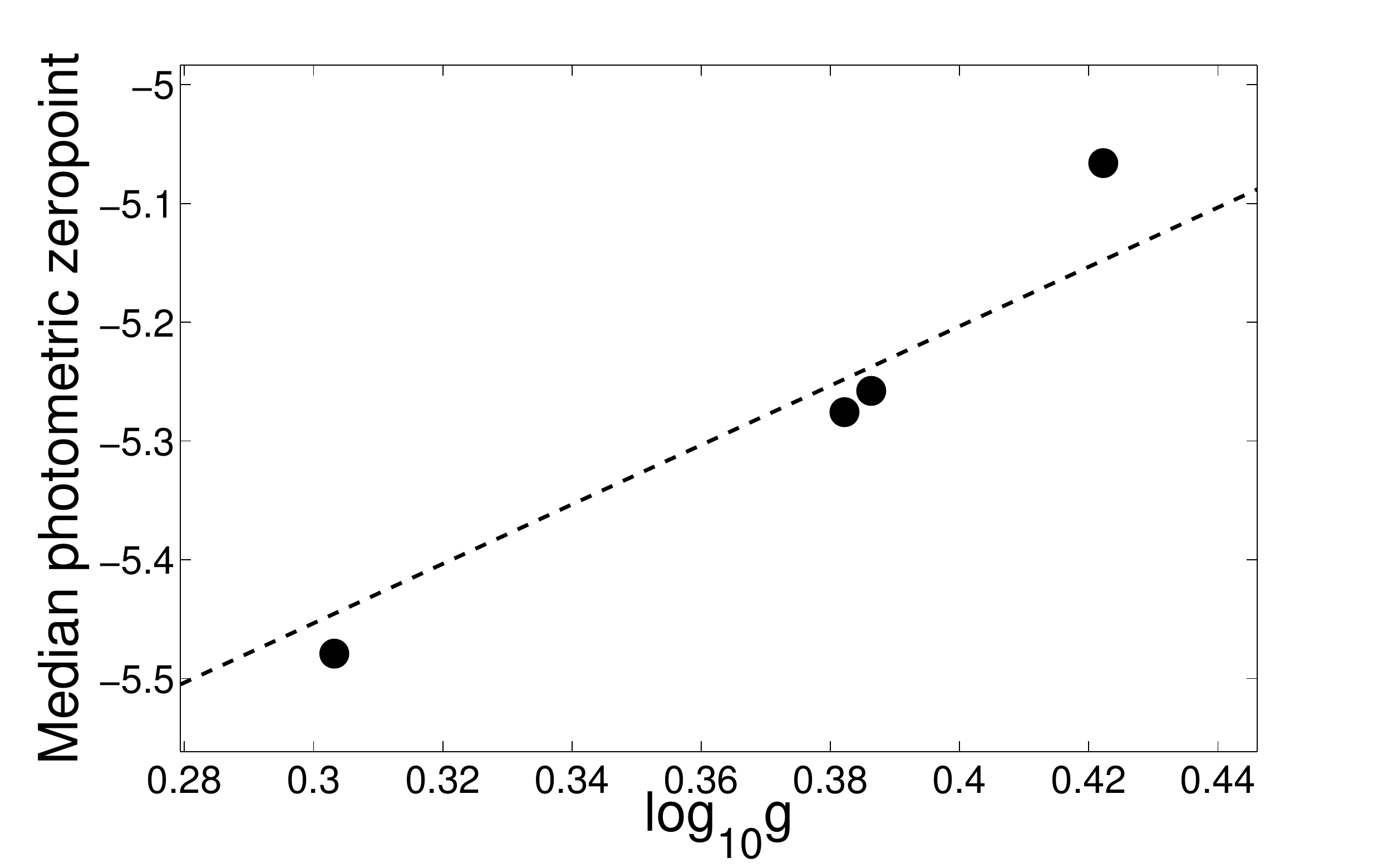}
\caption{Median photometric zeropoints in K$_\mathrm{s}$ of the four chips, compared to the contribution $\log_{10}{g}$ to the instrumental magnitudes from their gain. The dashed line has a slope of 2.5.\label{fig:gain}}
\end{figure}

The K$_\mathrm{s}$ zeropoints calculated with the correction for the crowding are shown in the right panel of Figure~\ref{fig:zerop}. They still show a temporal variation but the high correlation between the chips indicates that it is caused by a global effect like a change in atmospheric transparency, seeing or quality of the MCAO correction. We notice that the systematic difference of the zeropoints among the four chips is consistent with their gain values, as shown in Figure~\ref{fig:gain}.

\begin{figure}
\gridline{\fig{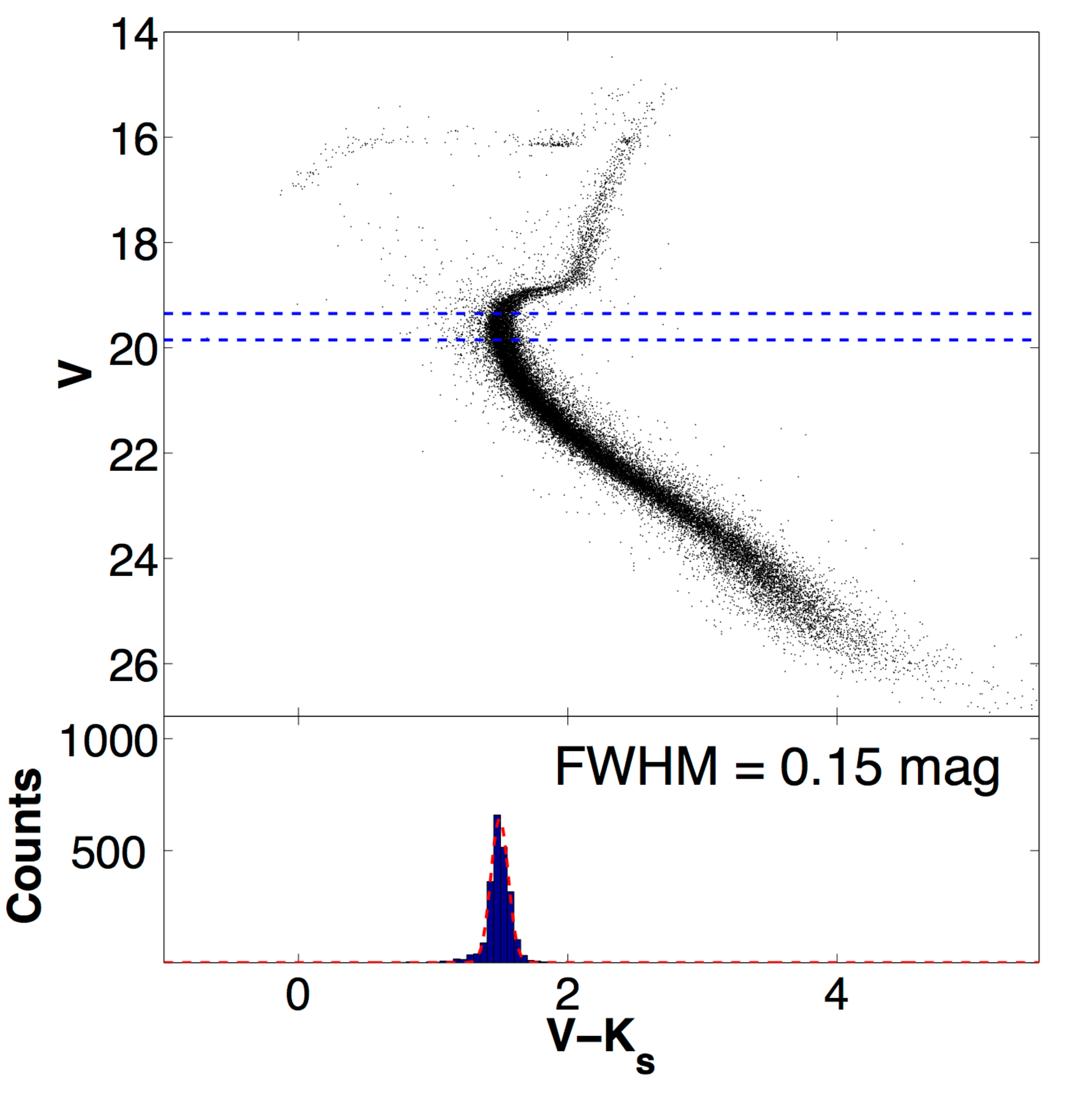}{0.3\textwidth}{} \fig{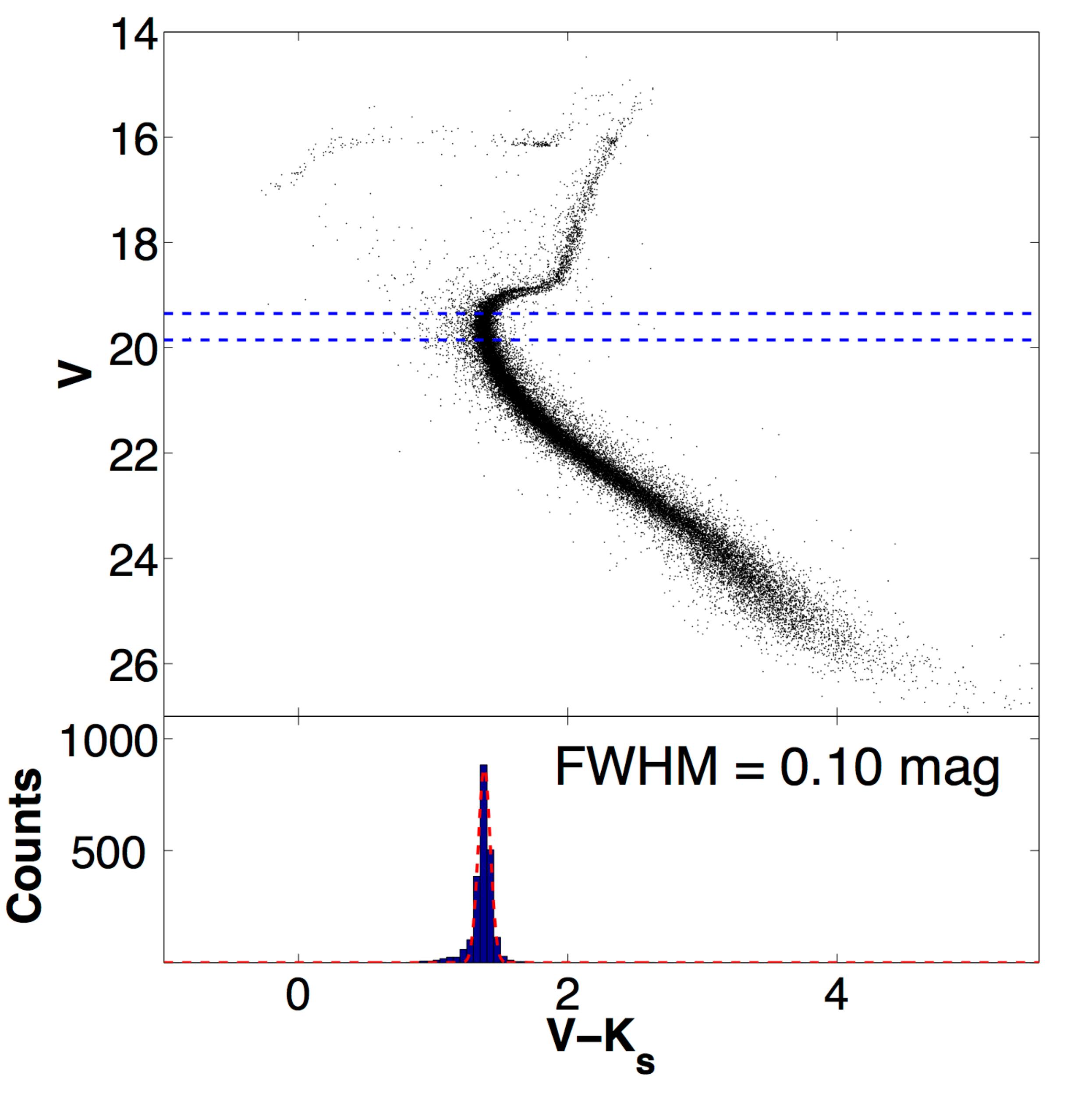}{0.3\textwidth}{} \fig{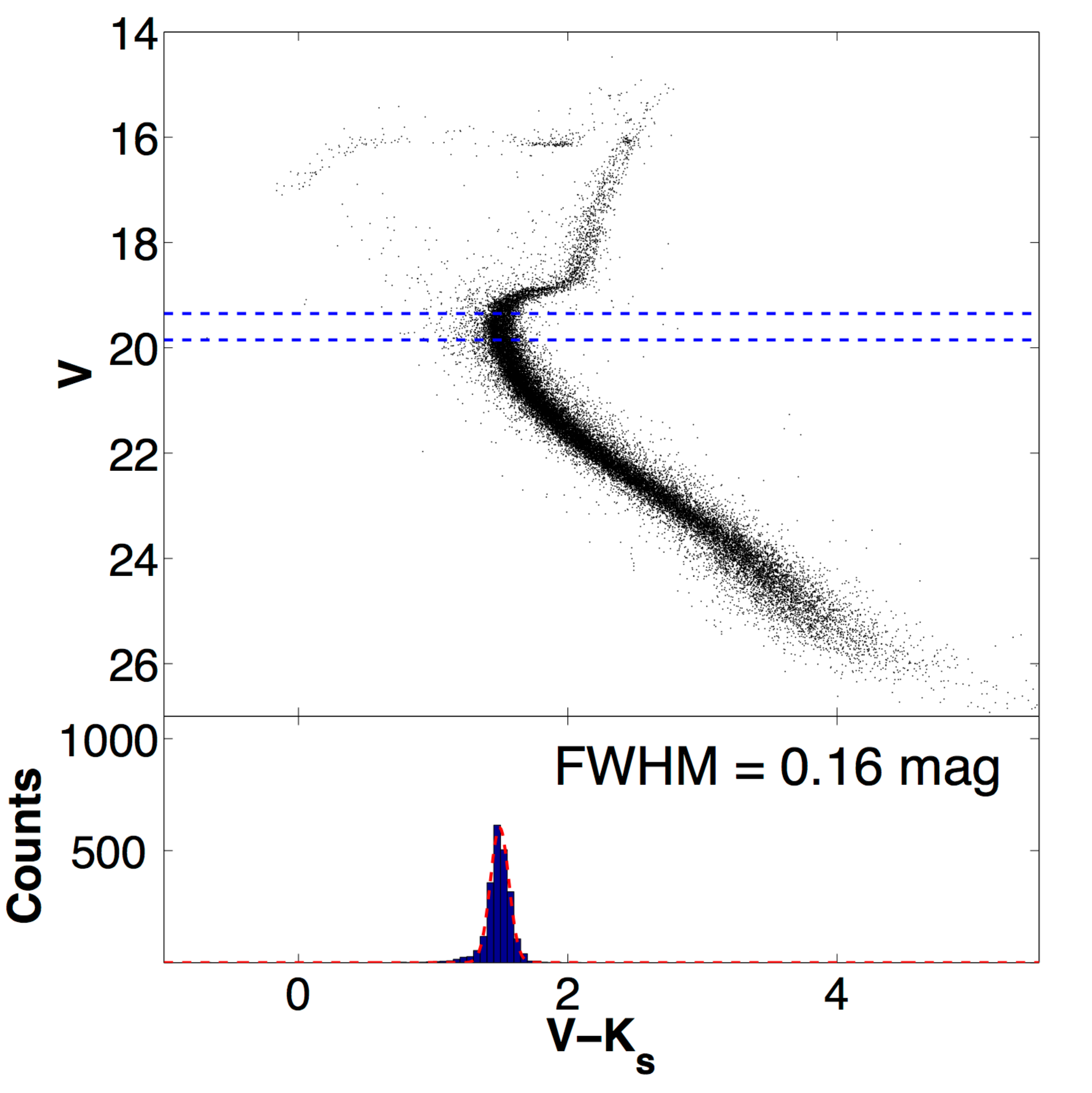}{0.3\textwidth}{}}
\caption{K$_\mathrm{s}$ CMDs calibrated using the reference catalog, without (\emph{left}) and with (\emph{center}) the correction for the seeing. The calibration for the CMD \emph{on the right} is done by averaging the photometric zeropoints.\label{fig:seeingcmd}}
\end{figure}

The improvement in the photometric calibration can be quantified by comparing the left and middle panels in Figure~\ref{fig:seeingcmd}, where we have used the same metric of the width of the MSTO as used earlier. Indeed, the improvement is also visible by eye in comparing the widths of the red giant branch. We do not find a significant spatial variation of the zeropoint in any of the images.

We note that it is a common photometric calibration strategy to observe a standard star or stars near in time and space to the science observation. However, the variation in the zeropoint of each chip between subsequent exposures, that is visible in Figure~\ref{fig:zerop}, warns against this strategy for MCAO observations. Adopting a zeropoint for a given image based on observations in a different image, even one taken immediately before in the same direction, could be a source of error for precision MCAO photometric observations. There are many ways to simulate this; we demonstrate this here by averaging the zeropoints in one band of each quadrant of GSAOI, and adopting these as the appropriate zeropoint corrections for all exposures. The result is a widening of the sequences in the CMD, as seen in the right panel of Figure~\ref{fig:seeingcmd}.

\subsubsection{Color correction}

\begin{figure}
\gridline{\fig{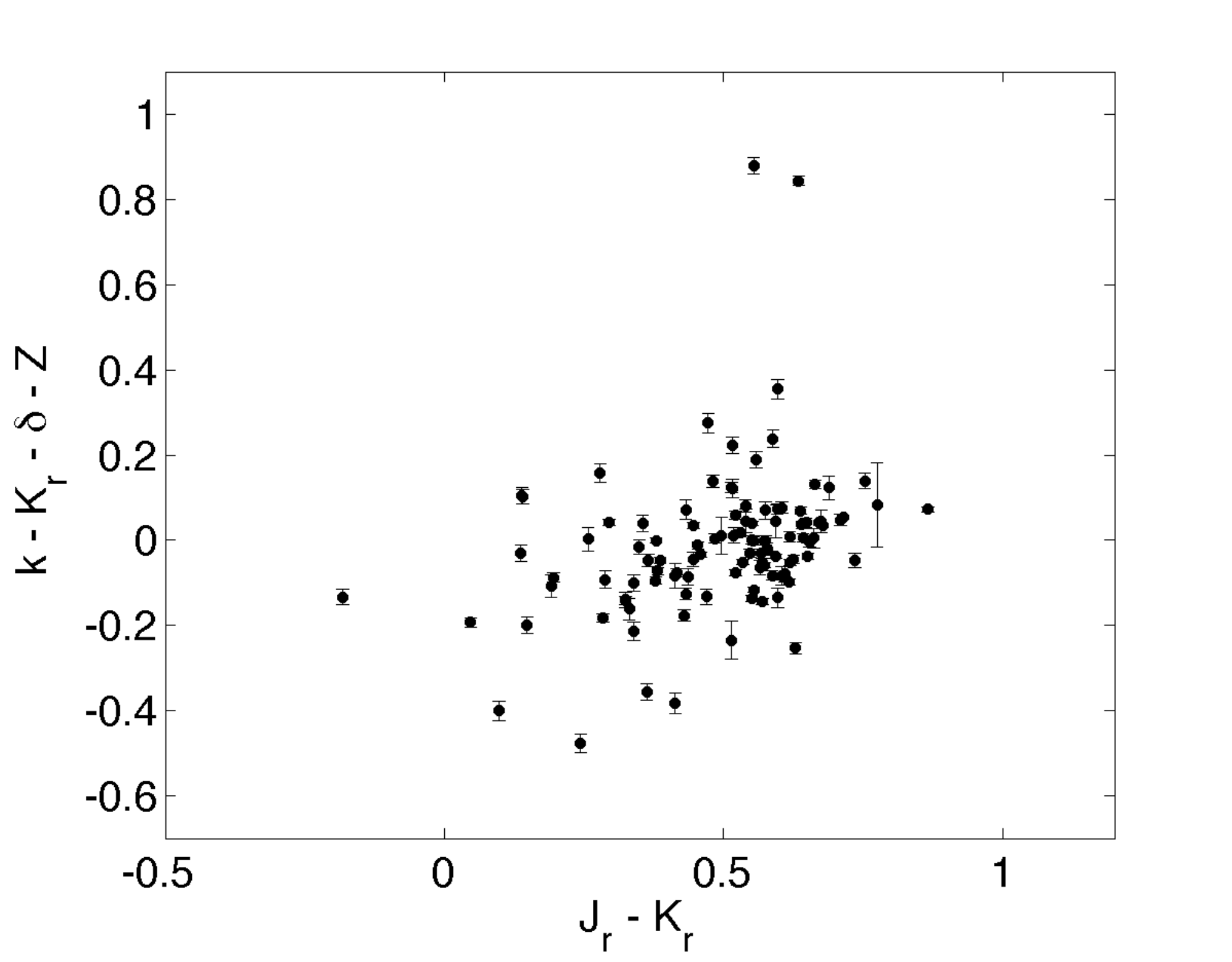}{0.4\textwidth}{Chip \#1} \fig{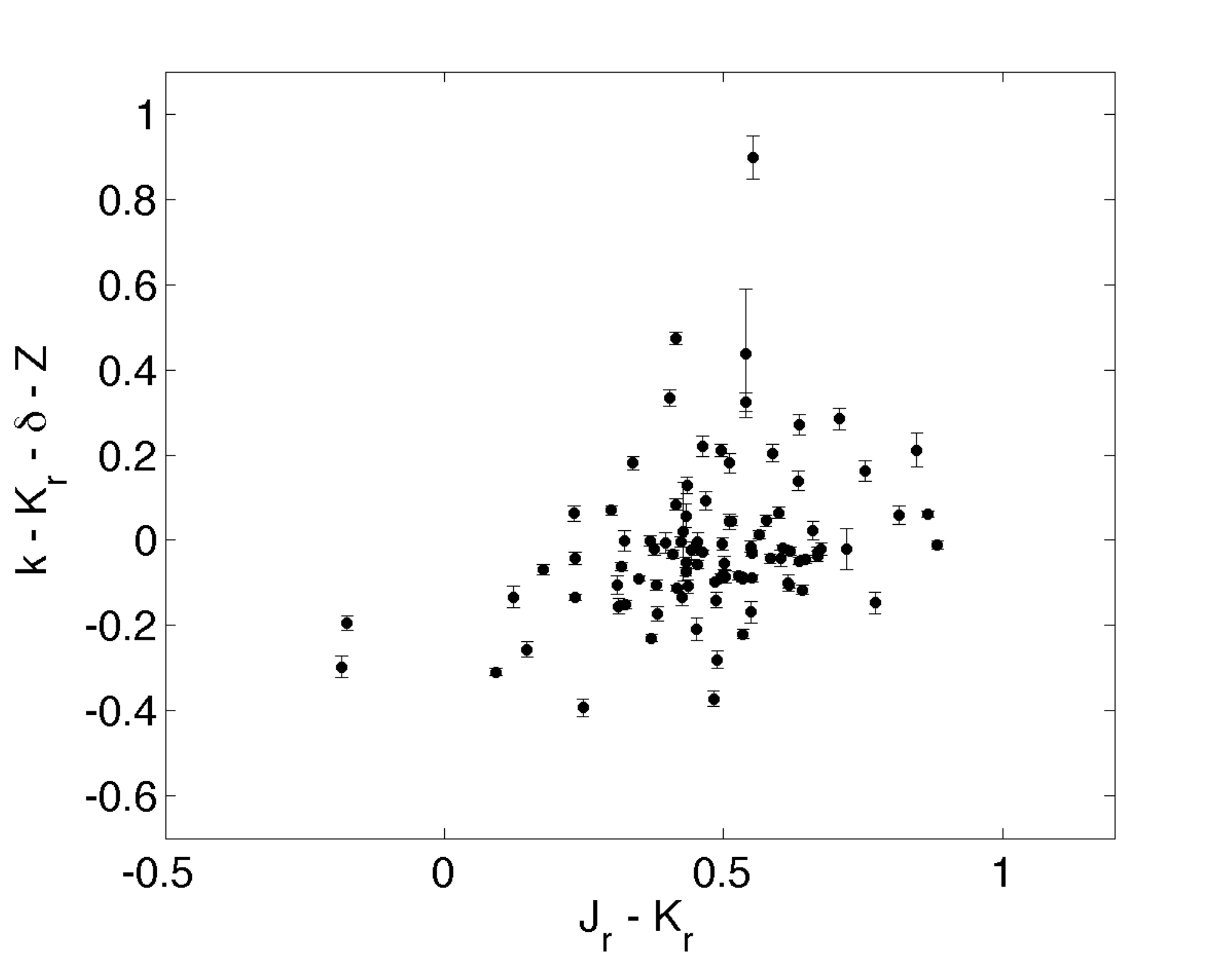}{0.4\textwidth}{Chip \#2}}
\gridline{\fig{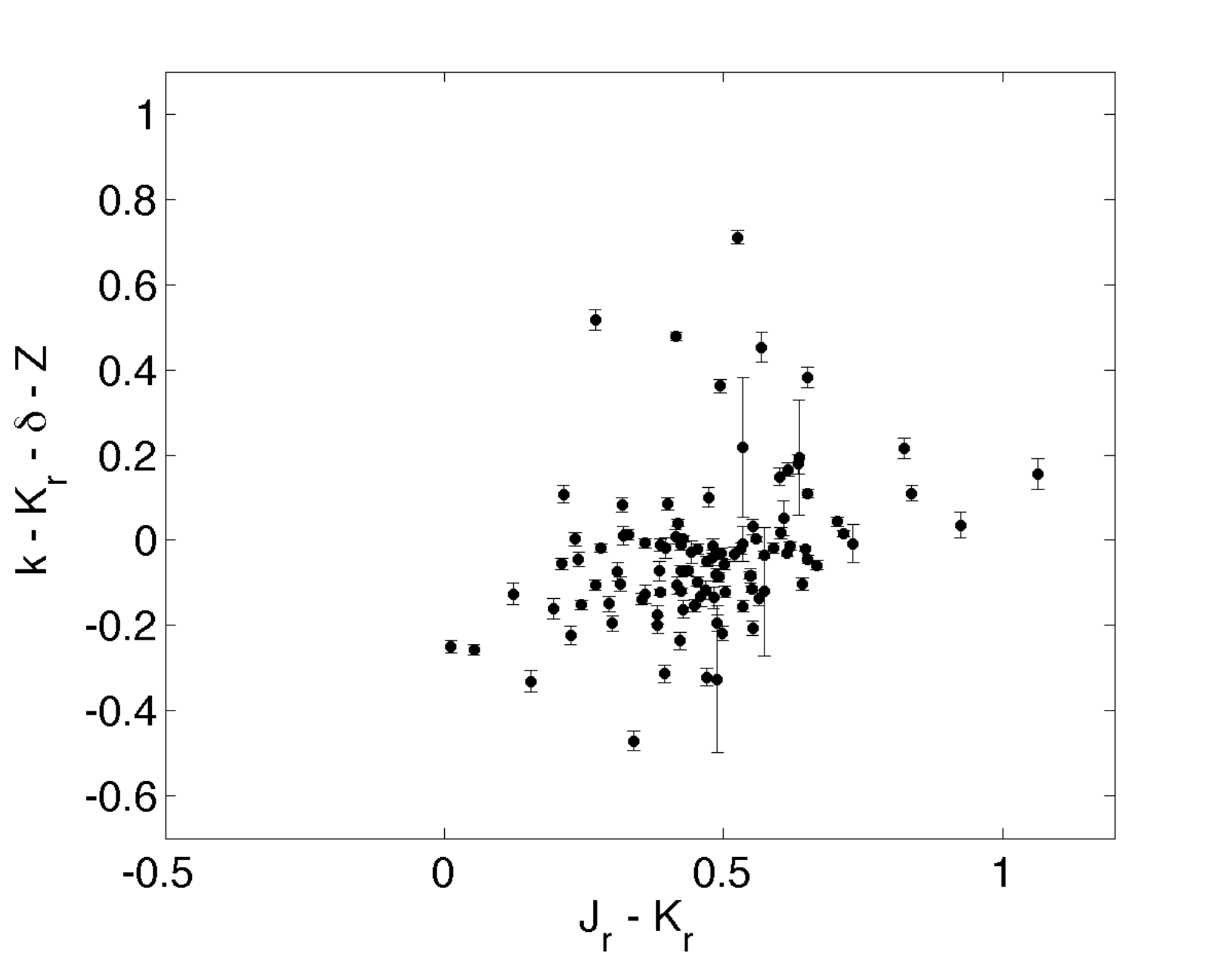}{0.4\textwidth}{Chip \#3} \fig{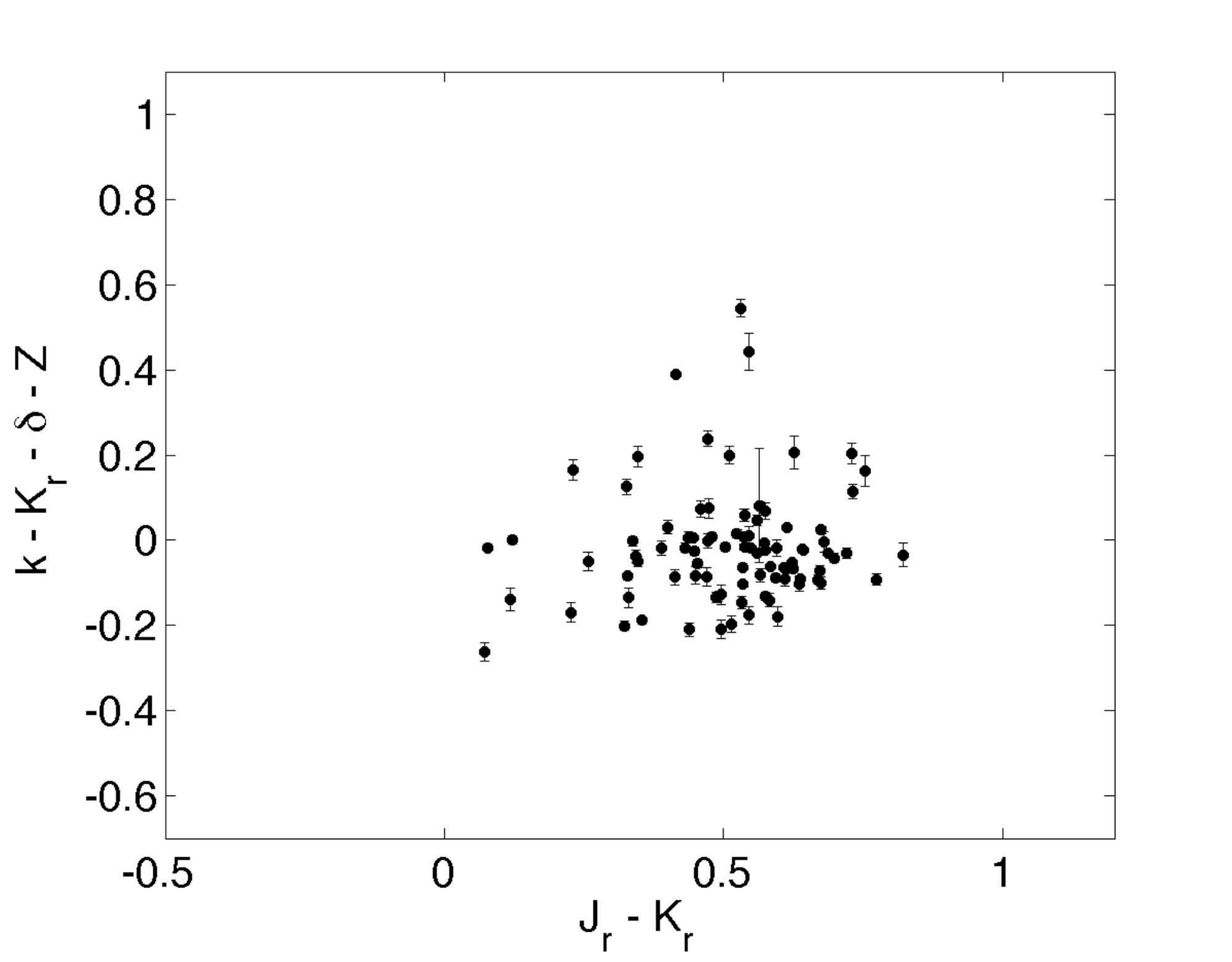}{0.4\textwidth}{Chip \#4}}
\caption{Difference between instrumental and reference magnitudes, corrected for zeropoint and nearby neighbors, as a function of color. The Pearson correlation coefficients show that there are no significant trends with color for any of the chips, indicating that the color coefficients of the chips are zero.\label{fig:colterm}}
\end{figure}

Having estimated the zeropoints, we can now determine the color terms $C$. We assume that these do not change as a function of exposure/time. We attempt to calculate only a single term $C$ for each chip and can therefore use all the exposures. To simplify our analysis, we consider only those matched stars that have little contamination from nearby stars ($\delta <0.5$), thus we can assume that the color reported in the reference catalog is close to the real one. For every of those stars, we calculate $k-K_{r}-\delta -Z$ as a function of the color $J_{r}-K_{r}$. Where the star is measured in multiple exposures, we average these measurement and associate an error bar that is a combination of the reported magnitude error and the standard error in the mean. Figure~\ref{fig:colterm} shows the results for each chip. For each chip, it is clear that the scatter is significant and it is unclear if there is a trend with color. To determine if there is a significant correlation for any of the chips, we perform a Pearson correlation test. The Pearson correlation coefficients of the K$_\mathrm{s}$ band are 0.34, 0.27, 0.23, and 0.11 for the four chips, indicating that there is not a significant correlation with color. Since the color terms have not been measured by the Gemini Observatory and---based on the filters and detectors characteristics---they are not expected to be significant (private communication with Gemini staff), we assume their values to be zero.

All the stars found in each image are used to calculate the geometric transformation, weighted for their magnitude error. They are then matched to create the final catalog and the magnitude of each is determined from all its available calibrated measurements in each of the different exposures. These magnitudes are combined using the method of ``artificial skepticism'' \citep[chap. 3]{bib:stetson89}.

\section{Color magnitude diagrams}\label{sec:cmds}

\begin{figure}
\gridline{\fig{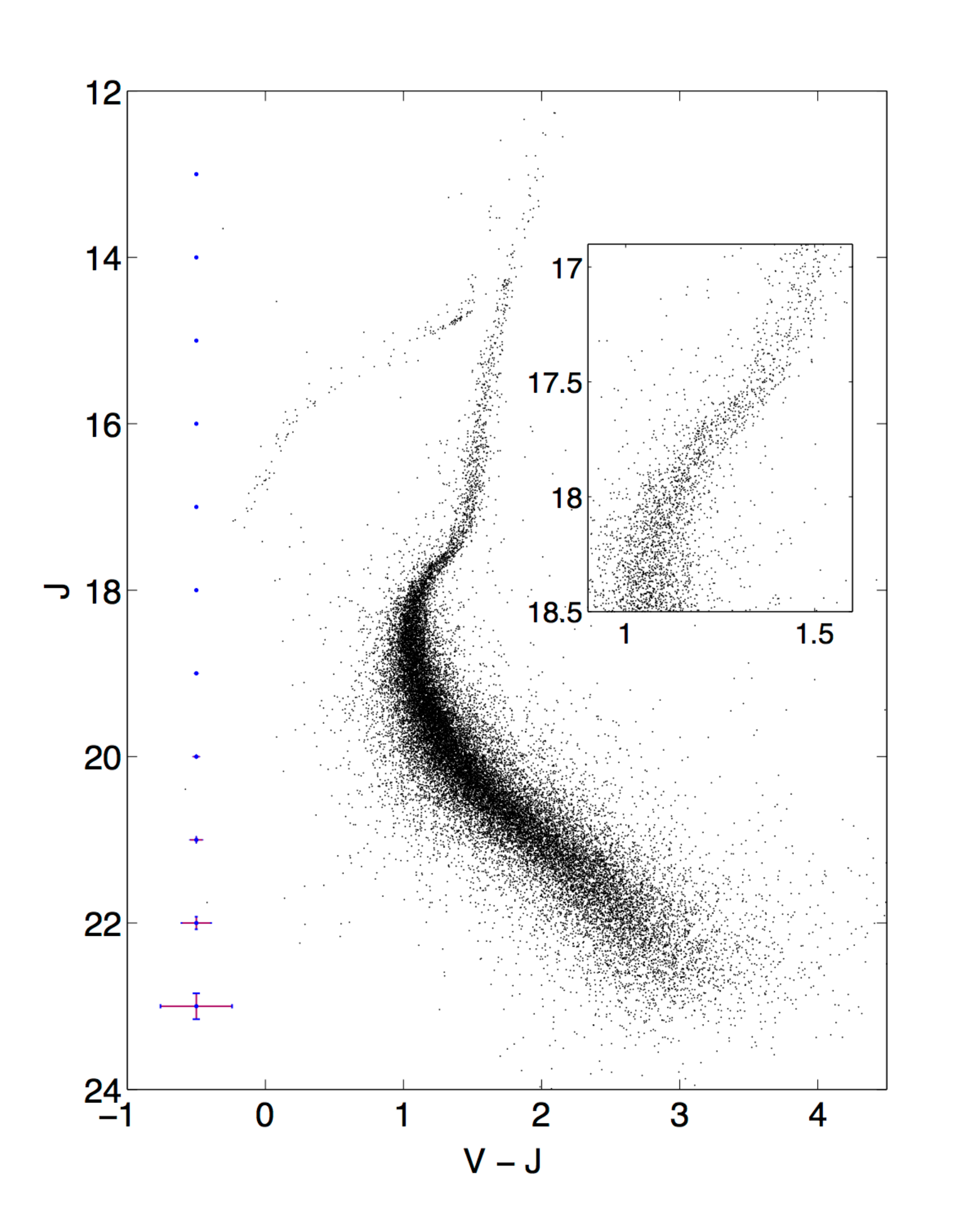}{0.3\textwidth}{} \fig{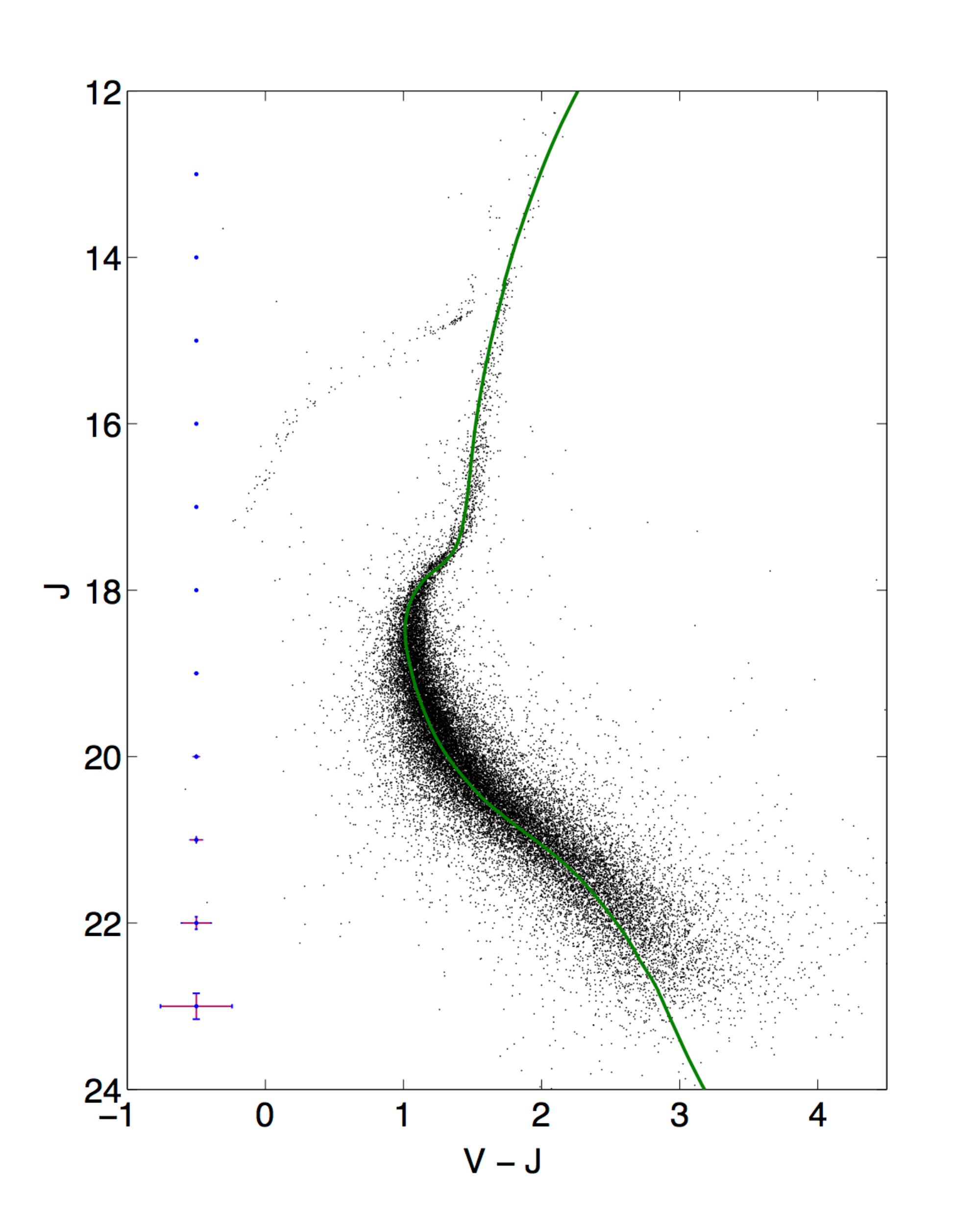}{0.3\textwidth}{}}
\vspace{-10mm}
\gridline{\fig{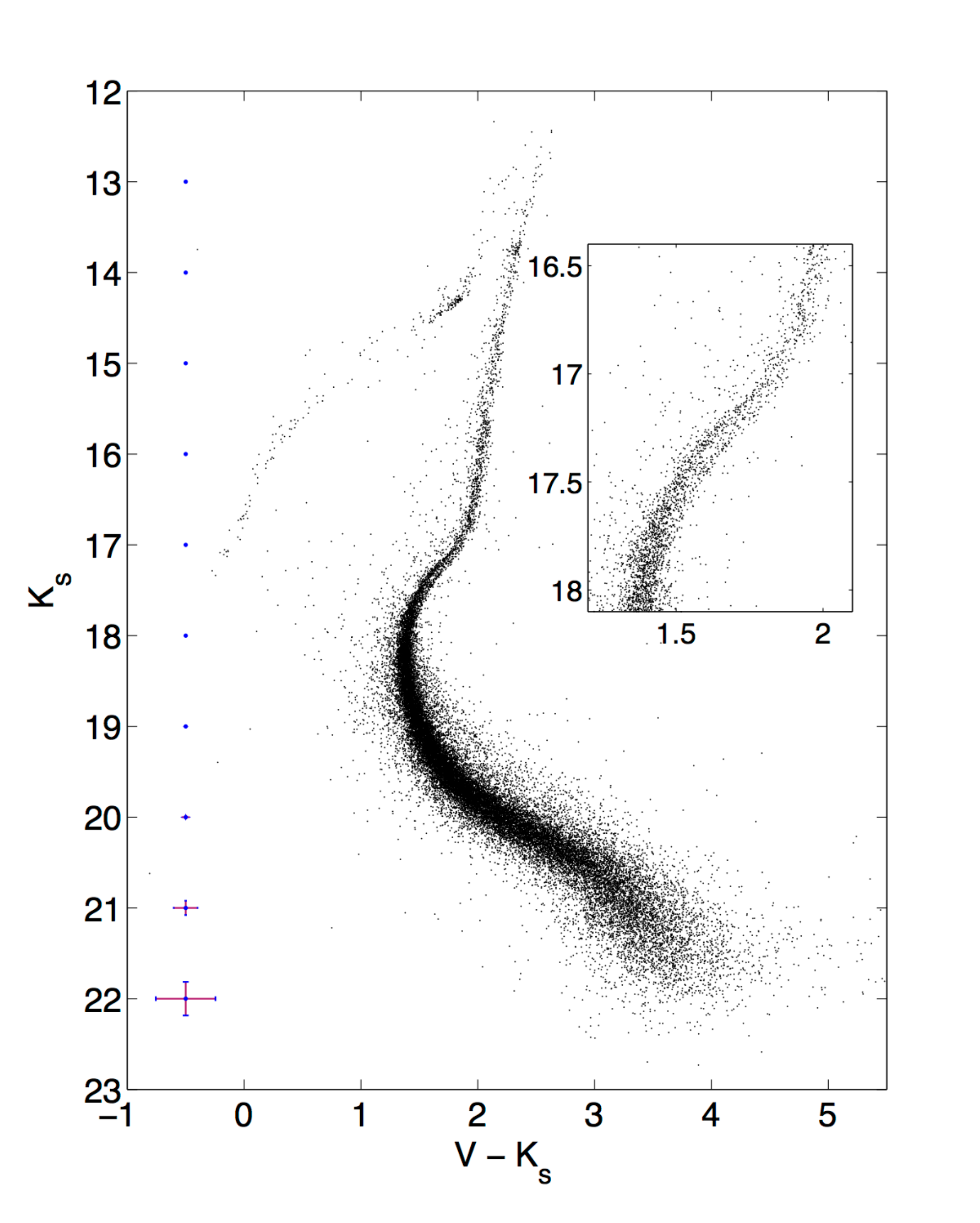}{0.3\textwidth}{} \fig{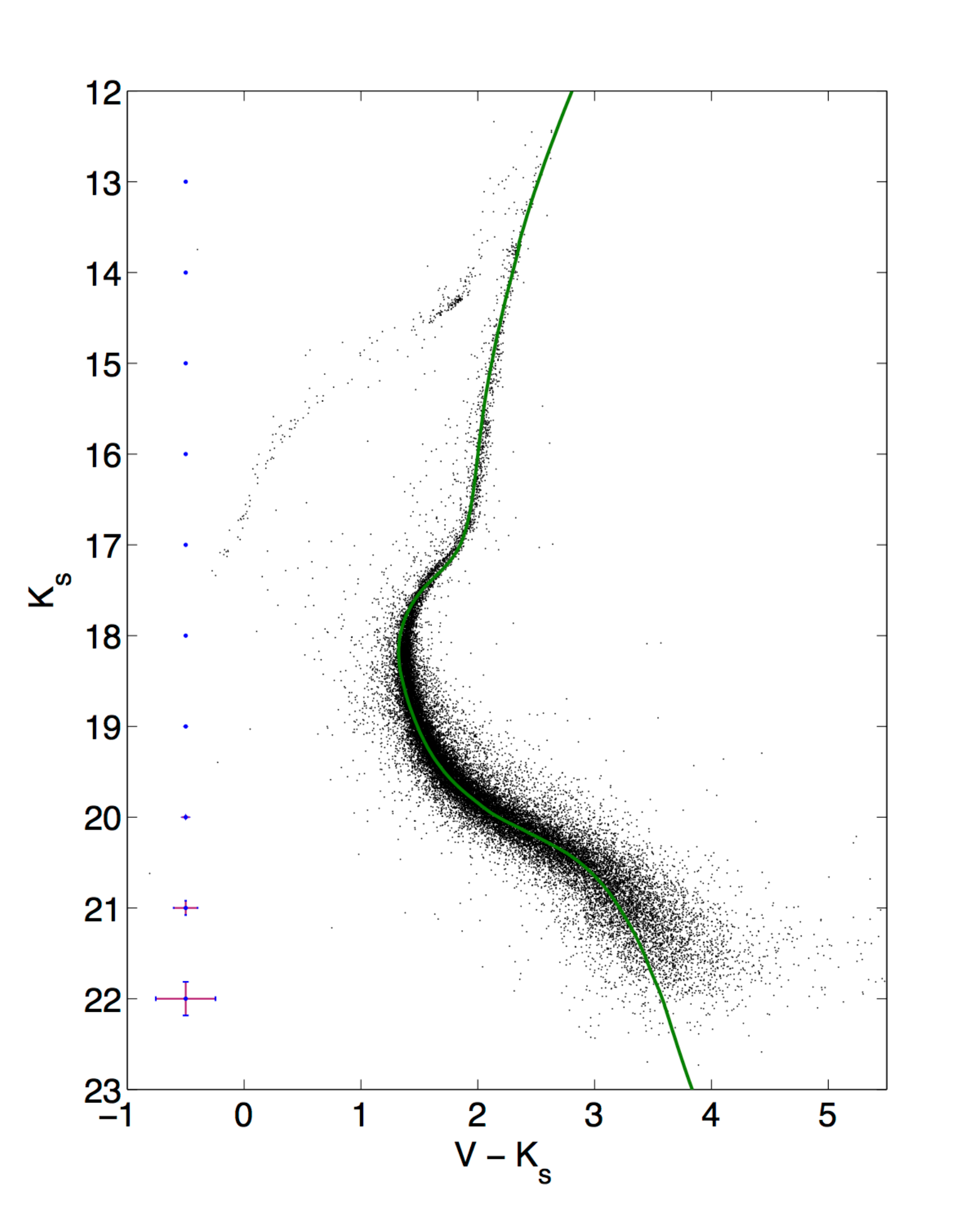}{0.3\textwidth}{}}
\vspace{-10mm}
\gridline{\fig{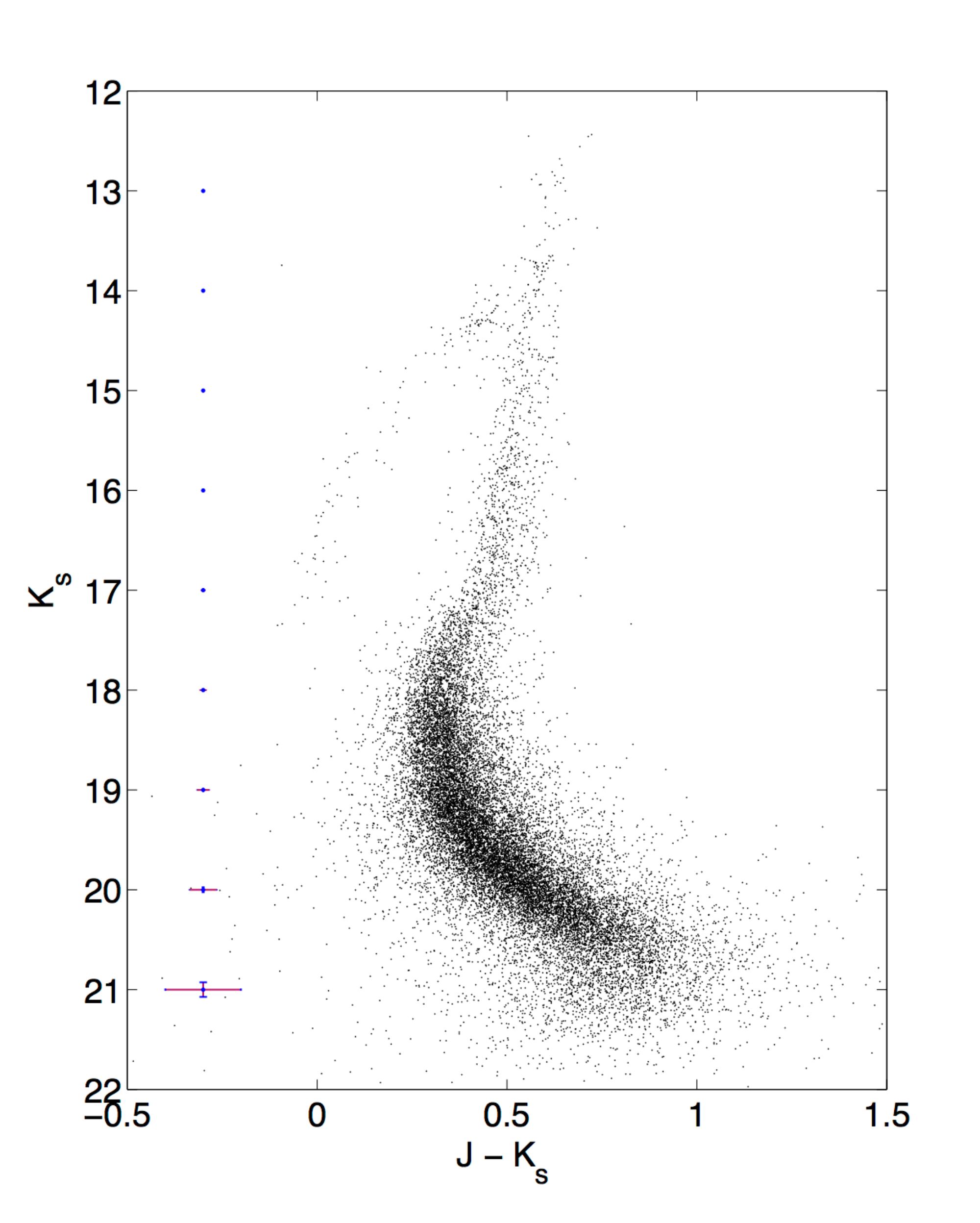}{0.3\textwidth}{} \fig{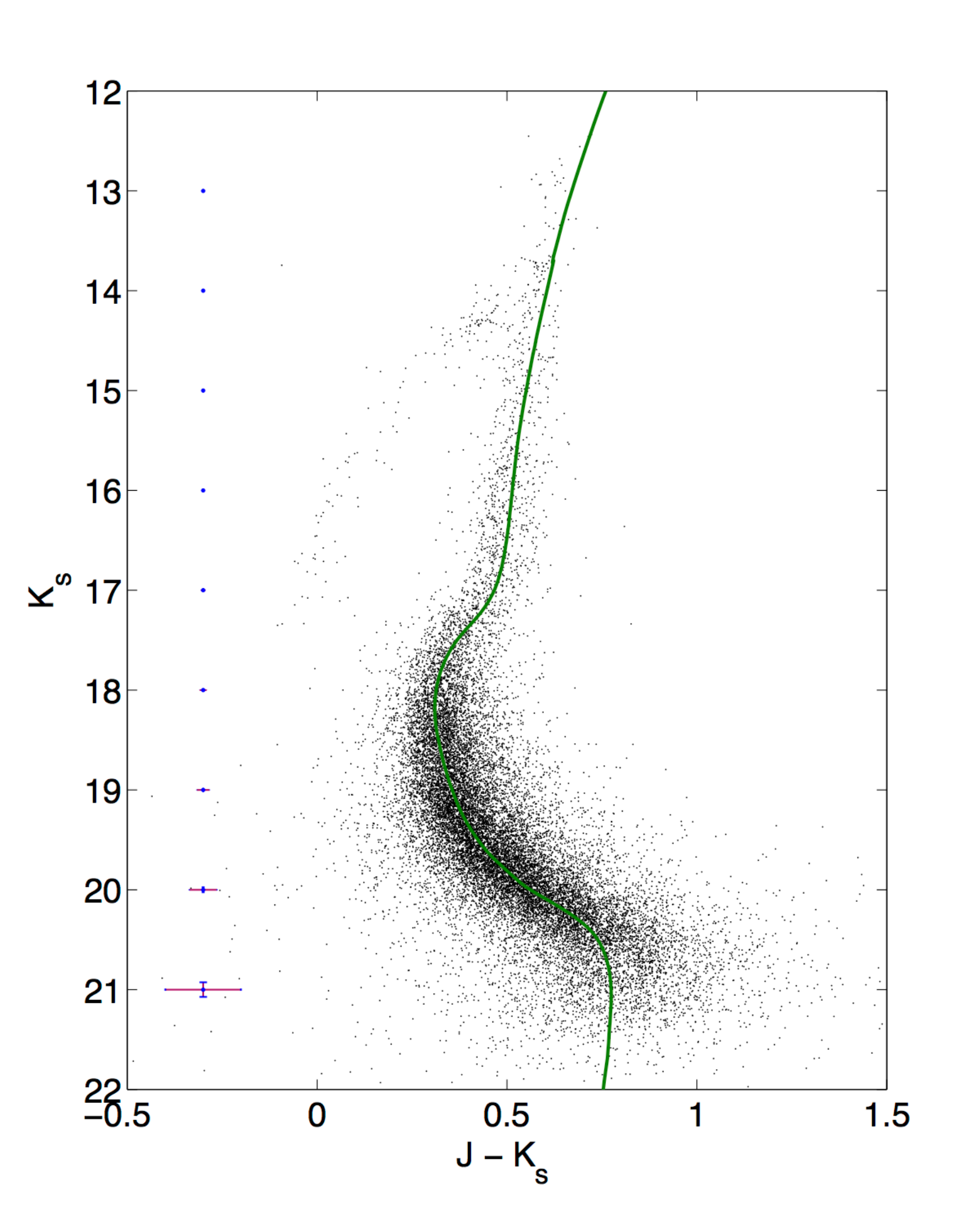}{0.3\textwidth}{}}
\vspace{-5mm}
\caption{Final and calibrated CMDs showing the NIR GeMS/GSAOI photometry of NGC 1851, in combination with HST ACS optical data. The optical-NIR CMDs \emph{on the left} have inset panels showing a zoom-in around the subgiant branch, where the split in stellar populations is clearly visible. \emph{On the right}, the same CMDs have isochrones overlaid (see text for details). The theoretical isochrones match the stellar locus with remarkable accuracy over the entire range in magnitude, from the main sequence knee to the giant branch.\label{fig:cmds}}
\end{figure}

Figure~\ref{fig:cmds} shows the final, fully calibrated and cleaned by photometric parameters and proper motions J vs (V-J), K$_\mathrm{s}$ vs (V-K$_\mathrm{s}$), and K$_\mathrm{s}$ vs (J-K$_\mathrm{s}$) CMDs for NGC 1851. The (V-J) and (V-K$_\mathrm{s}$) CMDs on the left have inset panels that show a zoom-in around the split subgiant branch where this feature is visible. \cite{bib:milone08} suggested the presence of two distinct stellar populations, where the faint SGB is $\sim$1 Gyr older than the bright SGB. \cite{bib:cassisi08} proposed instead two coeval populations of 10 Gyr with different chemical abundances, the bright SGB with normal $\alpha$-enhancement and the faint SGB with a strong correlation among the CNONa abundances. While we recognize the complexity of modeling this cluster, we chose to overlay the CMDs in the right panels of Figure~\ref{fig:cmds} with a single isochrone to demonstrate the broad agreement in color-magnitude space between our data and the general characteristics of NGC 1851 as reported in the literature. The isochrones are generated with the Dartmouth models of stellar evolution \citep{bib:dotter08} using $[Fe/H]=-1.16$, $[\alpha /H]=0.4$ \citep{bib:carretta11} and an age of 11 Gyr \citep{bib:vandenberg13}. These have been corrected for distance and reddening by
\begin{displaymath}
V=V_{0}+(m-M)_{V}
\end{displaymath}
\begin{displaymath}
J=J_{0}+((m-M)_{V}-R_{V}E(B-V))+R_{J}E(B-V)
\end{displaymath}
\begin{displaymath}
K=K_{0}+((m-M)_{V}-R_{V}E(B-V))+R_{K}E(B-V)\text{,}
\end{displaymath}
where the extinction ratios $R_{V}=3.1$, $R_{J}=0.72$, $R_{K}=0.31$ are from \cite{bib:yuan13} and the visual distance modulus $(m-M)_{V}=15.47$ and the color excess $E(B-V)=0.02$ are from \cite{bib:harris96}.

The overall depth of these data is impressive, reaching 23 and 22 magnitudes in J and K$_\mathrm{s}$ band, respectively, with formal uncertainties of 0.16 and 0.26 magnitudes. The knee of the main sequence \citep{bib:bono10} is visible in the NIR-only CMD, some three magnitudes fainter than the MSTO. The extended horizontal branch is divided in two by the RR Lyrae gap and, at around the same magnitude, we can observe the red giant bump on the red giant branch. We can also notice a weak population of blue stragglers. In all three CMDs, the agreement between the theoretical isochrones and the actual stellar locus is quite remarkable considering the isochrones were derived solely on the preexisting relevant values in the literature and no effort has been made to adjust them to our data. In each case, the isochrone follows very nearly the central spine of the stellar locus, from the main sequence knee up to the upper red giant branch, although the section around the MSTO appears slightly bluer than our data. This gives us further confidence that the strategies that we have employed in our analysis, especially regarding the photometric calibration, are valid across the full magnitude and color range of the target stars.

\cite{bib:turri15} presented an earlier version of our K$_\mathrm{s}$ band photometry, in combination with the HST ACS data, that was of sufficient quality to exhibit the double subgiant branch of this globular cluster. Here again the split is present and we also see weak evidence for this feature in the J band data, when combined with the HST ACS data. We do not see any evidence of the split subgiant branch in the NIR-only CMD. The combined uncertainties on the J and K$_\mathrm{s}$ data is too large to clearly see such a structure.

\begin{figure}
\centering
\includegraphics[width=0.5\textwidth]{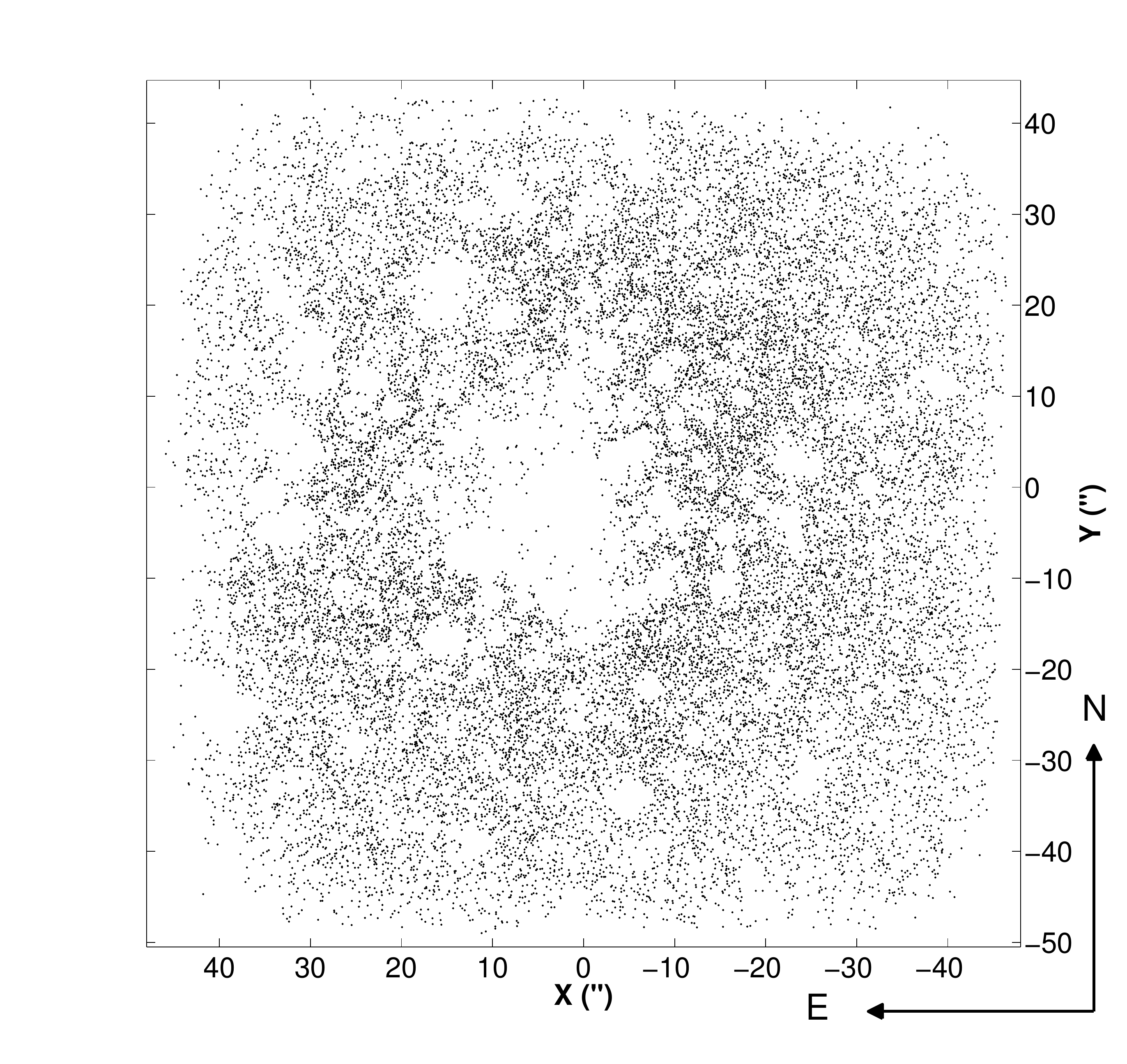}
\caption{Positions of the stars in the final J-K$_\mathrm{s}$ CMD of Figure~\ref{fig:cmds}.\label{fig:pos}}
\end{figure}

Figure~\ref{fig:pos} shows the spatial positions of the stars in the J-K$_\mathrm{s}$ CMD. Holes correspond to regions of extreme crowding, as well as the positions of several very bright and saturated stars that prevent accurate photometry from being obtained for neighboring objects.

\section{Discussion and conclusion}\label{sec:conclusion}

In this paper, we have presented a detailed analysis of the photometric performance of GeMS/GSAOI using data obtained during its science verification phase. In particular, we have focused on developing the optimal strategies for obtaining precise stellar photometry from MCAO images using science-based metrics to quantify the success or otherwise of our approaches for crowded fields. Our strategy is based on the well known technique of PSF fitting, specifically with the DAOPHOT II suite of programs, but modified to account for the challenges posed by MCAO data, especially the significant temporal and spatial variation in the PSF. Efficient exploitation of MCAO imaging in the NIR is absolutely essential to the future of ground-based astronomy given the development of MCAO instruments on the TMT and E-ELT. The following is a list of our primary conclusions:

\begin{itemize}
\item MCAO delivers a more uniform correction than classical AO over the moderately wide field of view. However, the shape of the PSF can still vary significantly in a complex manner, both spatially and temporally (i.e., between exposures; Section~\ref{sec:performance}), as also found by \cite{bib:ascenso15}. It is possible that some of this variation has been removed since GeMS has undergone several realignments of the optics, but residual variability is still expected. As a consequence, this should be reflected in the creation of the PSF model by allowing it to have a high degree of spatial variability, and by treating all exposures independantly.

\item The MCAO correction is applied on a scale equal to the ``control radius'' of the system. The majority of the spatial variability of the PSF is  found within this radius. We demonstrate that we obtain the most precise photometry when the radius of the PSF model is set approximately equal to the control radius of the AO system (Section~\ref{sec:psfrad}). If the PSF radius is considerably smaller than the control radius, part of the variable PSF controlled by the MCAO system is not measured, causing a systematic error dependent on the position of the stars. If the PSF radius is much larger than the control radius, then the resulting model attempts to include the halo of the PSF. This region is generally not varying and contains a fixed fraction of the flux in the PSF, and can therefore be accounted for during the calibration procedure. If the PSF model includes these outer regions, then the low flux levels mean that other features, including noise artifacts and faint stars, are included and identified as intrinsic features of the PSF. This results in an incorrect PSF model for these regions and a reduction in the precision of the photometry.

\item There is an appreciable improvement in photometric precision when the PSF model is based on a second iteration of the derivation procedure, i.e., derived from an image which has had non-PSF stars subtracted (Section~\ref{sec:psfclean}). This practice is likely less important with sparse fields.

\item During the profile fitting, the sky flux under every star is measured interior to the PSF radius. In exposures with a large FWHM, we have noticed that using an area significantly smaller than the PSF radius increases the precision of the photometry (Section~\ref{sec:psfsky}). We postulate that this is because a larger sky region is more likely to have contamination from unidentified stars, leading to inaccuracies in the sky measurements.

\item Simultaneous profile fitting of all the observations, such as is done with ALLFRAME, should be avoided because of the difficulty in finding the proper geometric transformation for the edges of some of the images. Treating the images independently leads to more robust photometric/astrometric solutions  (Section~\ref{sec:psfindep}).

\item Calibration of MCAO photometry is complex. Standard star observations are insufficient to estimate the zeropoint since the zeropoint changes for each exposure as a result of the MCAO correction. It is therefore advisable to match every exposure and chip independently to a reference catalog of the same field, that is itself calibrated to the NIR 2MASS system. The measurement of the zeropoints must account for the different spatial resolutions of the catalogs (Section~\ref{sec:calibration}). For GeMS/GSAOI, we do not measure a significant color term.
\end{itemize}

The procedures described in this paper have allowed us to derive extremely deep and precise NIR photometry for the globular cluster NGC 1851. Comparison to isochrones suggest remarkable agreement between our observational data and the theoretical expectations for the color and magnitude distribution of the stars (from the main sequence knee, through the main sequence turn-off to the subgiant and giant branch) given our understanding of the age, metallicity, and distance of this cluster as derived from optical data. When combined with optical HST ACS data, the (V-J) and (V-K$_\mathrm{s}$) CMDs allow for the identification of the split subgiant branch (already demonstrated for the K$_\mathrm{s}$ band in \cite{bib:turri15}. With precision photometry and high quality PSF modeling in hand, a future contribution will discuss the astrometric performance and precision of our data using a similar approach of the identification of science-based metrics.

MCAO is a key technology for the future of ground-based NIR astronomy, especially as we approach the era of ELTs, where the large apertures will provide diffraction limits that will significantly surpass even the James Webb Space Telescope. GeMS/GSAOI allows us to explore the utility of this technology for science observations and develop solutions to the new challenges that are encountered. For example, NFIRAOS, the MCAO system  on TMT \citep{bib:herriot14}, is still expected to present noticeable PSF variability even though it will be considerably more uniform and stable than GeMS. The instrument will be thermally controlled, on a Nasmyth platform, and equipped with internal calibration tools. The geometry of the system will also improve the quality of the correction thanks to the use of more laser guide stars, a larger overlapping of the volumes probed by the guide stars, and a smaller scientific field of view. A comparable fraction of the flux will still reside outside the control radius compared to GeMS, since the density of actuators on the deformable mirrors will be similar. Our current analysis suggests that the photometric calibration of TMT observations of crowded fields will be a significant challenge. Our present technique of using a calibrated reference catalog of the field would require relatively deep, relatively high spatial resolution photometry if we are to identify large numbers of stars in common with ultra-deep, ultra-high resolution imaging from TMT. By exploring and solving these issues now, before the new generation of MCAO instruments are on-sky, we can expect to improve significantly their science return at first light.

\acknowledgments

We thank Tim Davidge at NRC Herzberg for his comments on the manuscript that helped to greatly improve it. Thanks also to Gaetano Sivo, Vincent Garrel and Rodrigo Carrasco who provided insight into the GeMS and GSAOI instruments at the Gemini South telescope.

Based on observations for the program GS-2012B-SV-406 (P.I.: McConnachie) at the Gemini Observatory, which is operated by the Association of Universities for Research in Astronomy, Inc., under a cooperative agreement with the NSF on behalf of the Gemini partnership: the National Science Foundation (United States), the National Research Council (Canada), CONICYT (Chile), Ministerio de Ciencia, Tecnolog\'{i}a e Innovaci\'{o}n Productiva (Argentina), and Minist\'{e}rio da Ci\^{e}ncia, Tecnologia e Inova\c{c}\~{a}o (Brazil).

The GeMS data were acquired through the Gemini Science Archive and processed using the Gemini IRAF package.

This publication makes use of data products from the ACS Survey of Galactic Globular Clusters (P.I.: Sarajedini).

The data for the isochrones are from the Dartmouth Stellar Evolution Database.

G. Fiorentino has been supported by the FIRB 2013 (MIUR grant RBFR13J716).

\facility{Gemini:South (GeMS, GSAOI)}

\software{IRAF, DAOPHOT II}

\end{document}